\documentclass[12pt, draftclsnofoot, onecolumn]{IEEEtran}

\usepackage{etoolbox}
\usepackage{amsmath, amssymb, cite}
\usepackage{amsthm,bm,bbm}
\usepackage{mathtools}
\usepackage{caption}
\usepackage{multirow,color,amsfonts}
\usepackage{tabulary}
\usepackage{subfigure}
\usepackage{graphicx}
\usepackage{setspace}
\usepackage{enumerate}
\usepackage[]{algorithm2e}
\usepackage{comment}
\usepackage{pdfpages}
\usepackage{cleveref} 
\usepackage{enumitem}

\usepackage{xr}

\usepackage{soul}
\usepackage[normalem]{ulem}
\newcommand{\stkout}[1]{\ifmmode\text{\sout{\ensuremath{#1}}}\else\sout{#1}\fi}

\newcommand{\chasenoma}{{\sf SE}_{\sf sum}^{\sf CC,NOMA}}
\newcommand{\chasetin}{{\sf SE}_{\sf TIN}^{\sf CC,NOMA}}

\newcommand{\chaseoma}{{\sf SE}_{\sf sum}^{\sf CC,OMA}}
\newcommand{\chasetinort}{{\sf SE}_{\sf TIN}^{\sf CC,OMA}}%orthogonal TIN for chase

\newcommand{\MA}{{grant-based access }}%random access vs grant-based access

\newcommand{\clasicaltin}{{\sf SE}_{\sf TIN}^{\sf Clas.}}
\newcommand{\clasicalsumoptimal}{{\sf SE}_{\sf sum}^{\sf Clas.}}

\newcommand{\IRoma}{{\sf SE}_{\sf sum}^{\sf IR,OMA}}
\newcommand{\IRtinort}{{\sf SE}_{\sf TIN}^{\sf IR,OMA}}

\newcommand{\Cb}{C_{\sf buf}}

\newcommand{\SE}{{\sf SE}}

\newcommand{\Jnomaframe}{JT}

\newcommand{\Jnomaslot}{J}
\newcommand{\Jomaslot}{(J/T)}

\newtheorem{defi}{Definition}

\newtheorem{prop}{Proposition}

\newtheorem{cor}{Corollary}

\newcommand\norm[1]{\left\lVert#1\right\rVert}

\def\compactify{\itemsep=0pt \topsep=0pt \partopsep=0pt \parsep=0pt}
\let\latexusecounter=\usecounter
\newenvironment{CompactItemize}
  {\def\usecounter{\compactify\latexusecounter}
   \begin{itemize}}
  {\end{itemize}\let\usecounter=\latexusecounter}

\title{The Interplay of Spectral Efficiency, User Density, and Energy in Grant-based Access Protocols}
\author{
   Derya Malak\thanks{D. Malak is with the Commun. Systems Dept., EURECOM, Biot Sophia Antipolis, 06904 FRANCE (email: derya.malak@eurecom.fr).}\thanks{The material in this paper was presented in part at the 20th Int. Symp. Modeling and Optim. in Mobile, Ad Hoc, and Wireless Netw. (WiOpt 2022), Turin, Italy, and received the Best Paper Award \cite{MalakThroughputWiopt2022}.}\thanks{Derya Malak’s research is partially supported from a Huawei France-funded Chair towards Future Wireless Networks. This work was supported in part by the National Science Foundation under Grant CNS 2008639.}
}

\begin{document}
\maketitle
\begin{abstract}
We employ \MA with retransmissions for multiple users with small payloads, particularly at 
low spectral efficiency (SE). 
The radio resources are allocated  
via non-orthogonal multiple access (NOMA) in the time into $T$ slots and frequency dimensions, with a measure of non-orthogonality~$\eta$.  
Retransmissions are stored in a receiver buffer with a finite size $\Cb$ and combined via Hybrid Automatic Repeat reQuest (HARQ), using Chase Combining (CC) and Incremental Redundancy (IR).   
We determine the best scaling for the SE (bits/rdof) and for the user density $J/n$, for a given number of users $J$ and a blocklength $n$, versus signal-to-noise ratio (SNR, $\rho$) per bit, i.e., the ratio $E_b/N_0$,  
for the sum-rate optimal regime and when the interference is treated as noise (TIN), using a finite blocklength analysis.  
Contrasting the classical scheme (no retransmissions) with CC-NOMA, CC-OMA, and IR-OMA strategies in TIN and sum-rate optimal cases, the numerical results on the SE demonstrate that CC-NOMA outperforms, almost in all regimes, the other approaches. For high $\Cb$ and small $\eta$, IR-OMA could surpass CC-NOMA. 
At low $E_b/N_0$, the SE of CC-OMA with TIN, as it exploits CC and offers lower interference, can approach the trend of CC-NOMA and outperform the other TIN-based methods.  
In the sum-rate optimal regime, the scalings of $J/n$ versus $E_b/N_0$ deteriorate with $T$, yet from the most degraded to the least, the ordering of the schemes is as (i) classical, (ii) CC-OMA, (iii) IR-OMA, and (iv) CC-NOMA, demonstrating the robustness of CC-NOMA. 
Contrasting TIN models at low $\rho$, the scalings of $J/n$ for CC-based models improve the best,  whereas, at high $\rho$, the scaling of CC-NOMA is poor due to higher interference, and CC-OMA becomes prominent due to combining retransmissions and its reduced interference. 
The scaling results are applicable over a range of $\eta$, $T$, $\Cb$, and $J$, at low received SNR. The proposed analytical framework provides insights into resource allocation in \MA and specific 5G use cases for massive ultra-reliable low-latency communications (URLLC) uplink access.
\end{abstract}


%%%%%%
\section{Introduction}
\label{sect:intro}

The fifth-generation (5G) communication networks will support a wide range of use cases beyond high data rate applications, including Ultra-reliable, low-latency communication (URLLC) settings with small payload sizes transmitted by a large number of users with stringent power requirements. 4G LTE cannot effectively handle the heterogeneity because it ensures interference-free transmission via scheduled access %to individual users via incorporating an access overhead 
and is designed to support fewer devices with large payloads. On the other hand, the overhead of scheduled access in 4G LTE is not desirable in URLLC applications.

Motivated by the challenges in scheduled access, we consider a wireless multiple access %communication 
channel (MAC) model where a set of users sends their fixed payloads (in bits) given a preallocation of uplink resources. A given set of shared spectral resources of bandwidth $\omega$ Hertz (Hz) is partitioned into $B$ non-overlapping frequency bins shared by the users via non-orthogonal multiple access (NOMA), and the time of duration $\tau$ second (sec) is divided into $T$ transmit opportunities. Note that in HARQ, erroneous packets are not discarded and are instead stored in a buffer with a finite size $\Cb$ and combined with retransmitted packets \cite{danieli2010maximum}. Keeping this situation in mind, we consider different forms of Hybrid Automatic Repeat reQuest (HARQ): (i) HARQ with Chase Combining (CC) of NOMA transmissions, CC-NOMA, (ii) HARQ with Chase combining of OMA transmissions, CC-OMA, and (iii) HARQ with Incremental Redundancy (IR), IR-OMA. The general challenge is to design a random access protocol to maximize the scaling of the density of users versus the SNR per bit.

Via the proposed retransmission-based \MA scheme, we aim to address the following central questions for 5G wireless networks and beyond:
\begin{itemize}[leftmargin=*]
\item The spectral efficiency (SE, which is the total number of data bits per total real number of degrees of freedom, rdof) versus signal-to-noise ratio (SNR) per bit (or equivalently $E_b/N_0$) tradeoff for different HARQ schemes with retransmissions via Chase combining or Incremental redundancy. What are the gains in the sum-rate optimal technique\footnote{The sum-rate optimal rate is achieved via successive interference cancellation (SIC) by treating interference as noise (TIN)  
\cite{polyanskiy2018information}.} that describes an upper bound to the sum rate of the users, %and is attainable via SIC, 
versus the per-user rate approach based on TIN only?

\item How sensitive is the scaling of the SE versus $E_b/N_0$ to the number of retransmissions, $T$, different SNR, $\rho$, regimes, different uplink load $J$ regimes, where we keep the total power  fixed?

\item The blocklength, $n$, versus the number of retransmissions, $T$. We assume that %the time is slotted and 
each time slot $t\in\mathcal{T}=\{1,\dots, T\}$ accommodates the transmission of %a block of 
$m$ symbols. How does signal-to-interference-plus-noise ratio (SINR) change with $n=mT$, where $m$ is the blocklength per retransmission? %Why is it important, e.g., in (\ref{SINR_NOMA_retransmission})?

\item The impact of a finite HARQ buffer size on the throughput of HARQ. The size of the buffer available at the receiver, denoted by $\Cb$, to store previously received packets impacts the throughput of retransmission-based schemes. %The buffer capacity is $m\Cb$, where $\Cb$ is the buffer size normalized with the packet lengths. 
How does $\Cb$ affect the scaling performances?

\item NOMA-based signalling and the effect of non-orthogonal user signature %(random and non-orthogonal) 
correlations, denoted by the non-orthogonality factor $\eta$, on the scaling of the user density $J/n$ (users/rdof) versus $E_b/N_0$ given a number of users, $J$, and as a function of $T$. 
How should we design\footnote{$\eta$ can be made sufficiently small for large blocklengths. For a blocklength $m=\frac{n}{T}$ per transmission, $\eta\approx \frac{1}{\sqrt{m}}$  \cite{polyanskiy2018information}.} $\eta$ for the conventional matched filter receiver (MFR) for single-user detection (SUD) for decoding of random signatures? %How does the scaling behavior of $J/n$ change as function of $T$?

\end{itemize}

We next review the connections to the state-of-the-art and summarize the bottlenecks.

%%%%%  
\subsection{Related Work}
\label{sect:relatedwork}

\emph{Channel access models and throughput scaling.}
Random-access protocols have been pioneered with the emergence of ALOHA \cite{abramson1970aloha} and slotted or reservation-based ALOHA \cite{roberts1975aloha}, \cite{crowther1973system} schemes, which later yielded the development of  carrier sense multiple access.  
However, these contention-based schemes do not have desirable throughput and delay performances and do not guarantee a deterministic load. 
Recently, different uplink schemes have been proposed to accommodate massive access \cite{madueno2014efficient}. 
%https://web.stanford.edu/~dntse/papers/tutorial_v2.pdf
In general, the resource being shared is on a time-frequency grid, and each transmission costs one time-frequency slot (TFS). %and the signal strengths provide a new dimension at the receiver. 
%What is the resource being shared? The number of resources at the receiver equals $\#of resolvable bits per sample \times bandwidth \times time$
The throughput -- %via 
incorporating the user identification -- has been characterized for massive user connectivity with orthogonal access in \cite{yu2017fundamental} from a DoF perspective. %where the users are assigned orthogonal TFSs.  %time-frequency resource blocks. %the BS can schedule the successfully identified users to the time-frequency resource blocks for transmission. The transmit signals from different users can be assured to be orthogonal. However, to perform scheduling, the BS must inform the users that they are correctly identified. Further, the BS must provide the index of the time-frequency resources to the users.
Other models include sparse code multiple access for grant-free %multiple 
access
%https://www.arxiv-vanity.com/papers/1910.06529/
\cite{bayesteh2014blind}, multi-user detectors (MUDs) to improve performance of random-CDMA \cite{verdu1999spectral}, \cite{tse1999linear} for spread spectrum systems, %(in different bases), 
e.g., orthogonal multiple access (OMA), %(TDMA, FDMA, orthogonal CDMA, OFDM), %(orthogonal subcarrier signals), 
coded OFDM, %or CDMA with random non-orthogonal spreading, 
and %generally 
NOMA \cite{vaezi2019non}. %https://arxiv.org/pdf/1809.07224.pdf
%%%
Others have focused on the capacity of Gaussian MACs  %multiple access channels (MAC) 
\cite{kowshik2021fundamental}, %many-access channels 
MAC with user identification \cite{chen2017capacity}, %Many access channel model where the number of users scales unboundedly with the blocklength for MUD (where the number of users scales with the blocklength. 
quasi-static fading MAC \cite{kowshik2019quasi}, Gaussian MAC with feedback \cite{ozarow1984capacity}. %,tandon2009capacity}. %noiseless feedback \cite{ozarow1984capacity}, and with noisy feedback \cite{tandon2009capacity}. %and computing sum of sources over classical-quantum MAC \cite{sohail2022unified}.  %https://arxiv.org/pdf/2202.10403.pdf
Finite blocklength (FBL) achievability bounds for the Gaussian MAC and random access channel under average-error and maximal-power constraints have been devised in \cite{yavas2021gaussian}, with single-bit decoder feedback. %to instruct the encoder when to stop transmitting. %https://arxiv.org/pdf/2001.03867.pdf
\begin{comment}
Instead of orthogonal multiple access models, such as time-division multiple access (TDMA), frequency-division multiple access (FDMA), orthogonal code-division multiple access (orthogonal CDMA), or orthogonal frequency-division multiplexing (OFDM), which provide orthogonality in different bases, we allow sharing of resource blocks via non-orthogonal multiple access (NOMA). 
%There are different models, TDMA, FDMA, CDMA (in different bases). 
\end{comment}

\emph{Random-access versus grant-based access.} We have studied grant-free access in \cite{malak2018throughput} and \cite{malak2017fundamental} to maximize the rate of users simultaneously accessing the channel given a common outage constraint, with no upper bound on the transmitted energy. %While grant-free access protocols are better suited for scenarios where only a fraction of users are simultaneously active, e.g., the Internet of Things (IoT), sideline, or 5G New Radio applications, where a specific subset of users share the resources at any given time, and in particular to eMTC for a broad range of IoT use cases, e.g., asset trackers and wearables, and narrowband-IoT technologies \cite{eMTC_NBIoT2017}, the proposed approach allows the grouping of multiple users to share the resources and tolerates interference and collisions between the users. 
Grant-free access protocols are better suited for scenarios where only %a fraction of users are simultaneously active, 
a specific subset of users share the resources at any given time, e.g., applications of the Internet of Things (IoT), sideline, or 5G New Radio. These protocols are convenient for a broad range of IoT use cases of enhanced Machine-Type Communication (eMTC) %, e.g., asset trackers and wearables, 
and narrowband-IoT %technologies 
\cite{eMTC_NBIoT2017}. On the other hand, our proposed approach focuses on the scaling of the user density while allowing the grouping of multiple users to share the resources and at the same time tolerating interference and collisions between the users. 
%Hence, it lies between grant-free and grant-based techniques and is better tailored to capture the tradeoff between SE and SNR per bit for eight distinct transmission models.
Hence, our %the proposed 
protocol %in this paper 
lies between grant-free and grant-based techniques because it allows (i) multiple users to share the resources, %as in grant-free techniques, 
and (ii) a collision-aware resource allocation, %as in grant-based protocols, 
by grouping users to maximize the scaling of the user density per SNR per bit. Therefore, it %our protocol 
is better tailored to capture the tradeoff between SE and SNR per bit for eight %distinct 
(re)transmission models.

\emph{Interference management and resource sharing.}
Different interference management techniques have been studied under different spectral efficiency models. To accommodate massive random access, interference cancellation \cite{dovelos2017finite}, collision resolution \cite{madueno2014efficient}, load control \cite{koseoglu2016pricing}, and interference cancellation given a target outage rate \cite{dhillon2013power} have been proposed. 
From the perspective of fundamental limits, the best achievable rate region for two user Gaussian interference channel is given by Han-Kobayashi \cite{han1981new}. 
While interference alignment is a good technique for specific channel parameters \cite{cadambe2008interference,huang2012interference,etkin2009degrees,cadambe2008interferencealign}, the capacity region for a large number of users is unknown because ideal interference cancellation is not practical. In \cite{malak2018throughput}, we characterized the scaling of throughput (user density) with a deadline for a suboptimal but practical random access system where the time and frequency domains are slotted, and the receiver uses conventional SUD under an SINR-based outage constraint. 
%The receiver uses conventional SUD, which decodes a desired user’s data by treating other users’ interference as noise, subject to an SINR-based outage constraint. 
%In \cite{malak2018throughput}, the main limitation is fixed per-user power $E_b/N_0$ needs to go to $\infty$ as the number of users scale (or the blocklength).
However, the fixed per-user power in \cite{malak2018throughput} causes a linear scaling between the received SNR and the number of users.

\emph{Critical performance metrics.} 
Using different power levels to reduce $E_b/N_0$ has been considered in \cite{ahmadi2021random}. %https://arxiv.org/pdf/2104.12186.pdf
Random linear coding with approximate message passing %(AMP) 
decoding for many-user Gaussian MAC has been studied in \cite{hsieh2021near}, where the authors derive the asymptotic error rate achieved for a given user density, user payload in bits, and user energy. %(and energy per bit versus user density tradeoff). %https://arxiv.org/pdf/2102.04730.pdf 
%It is demonstrated that in the large system limit, a spatially coupled coding scheme with AMP decoding achieves near-optimal tradeoffs for a wide range of user densities. Furthermore, when the user payload is large, the tradeoff between energy-per-bit and spectral efficiency is studied and methods to reduce decoding complexity have been discussed.
%
Cognitive radio and NOMA have been blended to maximize the achievable rate of the secondary user without deteriorating the outage performance of primary user \cite{liu2022new}.  %https://arxiv.org/pdf/2204.06157.pdf
Dynamic power allocation and decoding order at the base station for two-user uplink cooperative NOMA-based cellular networks has been studied in \cite{elhattab2022optimizing}, where the authors demonstrated the superior performance over traditional two-user uplink NOMA (without cooperation). % In addition, they demonstrated that in UL C-NOMA, decoding the far NOMA user first at the BS provides the best performance. %https://arxiv.org/pdf/2203.13100.pdf

\emph{HARQ models and generation of %(orthogonal or decorrelated) 
coding sequences.} HARQ is a combination of Automatic Repeat reQuest (ARQ) and forward error correction (FEC) \cite{sesia2005techniques}. %https://pastel.archives-ouvertes.fr/file/index/docid/500058/filename/Thesis3.pdf 
In particular, there are three models known as HARQ with Selective Repeat, HARQ with CC, and HARQ with IR \cite{lee2015harq}, %https://arxiv.org/pdf/1501.06166.pdf
\cite{yafeng2003performance}, \cite{rowitch2000performance}.  %https://web.archive.org/web/20050301214127id_/http://cwc.ucsd.edu:80/~drowitch/papers/rcpt_jun2000.pdf
This one is a salient variant of HARQ that captures puncturing via parity bits, e.g., puncturing with Turbo codes \cite{frenger2001performance}, and effects of different HARQ buffer sizes 
%There also exist memory efficient HARQ implementations 
\cite{danieli2010maximum}. %https://backend.orbit.dtu.dk/ws/portalfiles/portal/4564313/Danieli.pdf
In general, it is well known that successive refinement of information can provide an optimal description from a rate-distortion perspective \cite{equitz1991successive}, \cite{kostina2019successive}, %https://arxiv.org/pdf/1707.09567.pdf
and incremental refinements and multiple descriptions with feedback have been explored \cite{ostergaard2020incremental}. %https://arxiv.org/pdf/2011.02747.pdf

%%%%%
Coding sequences have been devised for massive access, including Walsh sequences and decorrelated sequences \cite{helleseth2020resolution}, %https://arxiv.org/pdf/2006.12239.pdf
%or almost affinely disjoint subspaces \cite{liu2021almost}, %https://arxiv.org/pdf/2007.01792.pdf
%
%where it is possible to characterize the Hamming weight distribution of pseudorandom sequences \cite{shirvanimoghaddam2021hamming},  %https://arxiv.org/pdf/2107.05156.pdf
%
and Khachatrian-Martirossian construction to enable $K>n$ users signal in $n$ dimensions simultaneously, where $K \approx \frac{1}{2}n \log_2 n$ is the optimal scaling \cite[Slides 57-59]{polyanskiy2018information}. %http://people.lids.mit.edu/yp/homepage/data/NASIT18-MAC-tutorial.pdf 
Furthermore, it has been shown that when the inputs are constrained to $\pm 1$, it is possible to have $K\gg n$. %Total information grows with $K$: $H(X_1+\dots+X_K)\sim \frac{1}{2} \log K$. Similar to $\frac{1}{2} \log (1+KP)$ in GMAC.
Zadoff-Chu sequences provide low complexity and constant-amplitude output signals, and have been widely used in 3GPP LTE air interface, including the control and traffic channels \cite{zepernick2013pseudo}. %Zadoff-Chu sequences with higher preamble capacities than state-of-the-art sequences have been proposed in \cite{pawar2022high}. %https://arxiv.org/pdf/2204.04604.pdf
Multi-amplitude sequence design for grant-free MAC has been contemplated in \cite{yu2021uniquely}. %where the authors devised uniquely-decodable multi-amplitude sequences.  %https://arxiv.org/pdf/2110.11827.pdf 
However, inducing a high $E_b/N_0$, this approach is not desirable in a practical massive access scenario.

%%%%%%%%
\subsection{Overview, Contributions and Organization}
\label{sect:overview}

The goal of this paper is to analyze a retransmission and %based general 
\MA framework that unifies the properties of NOMA-based transmissions with HARQ-based protocols that rely on CC and IR to provide insights on uplink resource allocation strategies for future 5G wireless communication networks. In Section \ref{CommsModel}, we detail the system model for \MA and the key performance metrics, SE (bits/rdof), the SNR per bit ($E_b/N_0$), and user density (users/rdof) for a given blocklength, total received power constraint, and a total number of retransmissions. We delineate the retransmission and \MA schemes in Section \ref{NOMA_retransmission_no_fading} and analyze their SE and the SNR per bit for the sum-rate optimal and TIN cases. 
More specifically, we consider retransmission-based models where the receiver jointly decodes transmissions via (i) the classical transmission scheme with no retransmissions and the retransmission-based schemes with combining, namely (ii) CC-NOMA, (iii) CC-OMA, and (iv) IR-OMA. 
Our analysis incorporates channel power gains and the capacity for the FBL channel model. In Section \ref{sect:numericalresults}, we numerically evaluate the SE versus SNR per bit tradeoff and the user density versus SNR per bit tradeoff, and show their behaviors with respect to the number of transmissions $T$, received SNR $\rho$, HARQ buffer size $\Cb$, non-orthogonality factor $\eta$, and the total number of users $J$.

The key design insights for the proposed \MA framework are as follows: 
\begin{CompactItemize}[leftmargin=*]
\item {\bf \emph{The low $\rho$ regime is relevant.}} We exploit the conventional MFR for SUD suitable at low SNRs $\rho$. 
We show that the user density of NOMA-based models scales significantly better at low $\rho$ versus high $\rho$. The interference cannot be exploited at high $\rho$, degrading the performance of TIN-based models. The minimum SNR per bit to achieve a non-zero user density grows with $\rho$.

\item {\bf \emph{The SE of the sum-rate optimal strategy improves with NOMA.}} The scalings of SE versus $E_b/N_0$ for various schemes show that for any given value of $E_b/N_0$, the best performance is attained by $\chasenoma$, and mainly for small $T$. Compared to OMA-based transmissions, CC-NOMA has a better SE versus $E_b/N_0$ performance. 
The performance of IR-OMA approaches that 
of the classical model as $\Cb$ at the receiver increases.  
The numerical results indicate that $\chasenoma$ outperforms the other strategies almost in all regimes. While $\chasenoma$ significantly improves with increasing $J$, and $\chaseoma$ is less sensitive to $\Cb$, at high $\Cb$, $\IRoma$ performs, in general, better than $\chaseoma$, 
and it could outperform $\chasenoma$ for small $\eta$.

\item {\bf \emph{The SE of the TIN strategy is optimal at low $\rho$.}} Provided that $\Cb$ is sufficiently large, TIN is good at low SE. If not, a higher $T$ is required. The scaling results are sensitive to $\eta$ for CC-NOMA, and a codebook with a smaller $\eta$ can significantly improve the SE of TIN.   
At low $E_b/N_0$, the performance of $\chasetinort$ can %approach or 
outperform $\chasenoma$  
and  
other TIN-based methods because it exploits CC and offers lower interference than $\chasetin$.  
%Furthermore, for 
At large $\Cb$ and $T$, $\chasetinort$ can be superior to $\clasicaltin$.

\item {\bf \emph{User density is sensitive to retransmissions.}} For the sum-rate optimal model, although the performances of CC-OMA, IR-OMA, and the classical techniques degrade with %increasing 
$T$, CC-NOMA does not sacrifice the number of users per rdof as much.  
A higher number of users $J$, under fixed per-user power, %when the per-user power is kept fixed, 
results in a lowered received SNR $\rho$ per user, which improves the SE for the sum-rate optimal CC-NOMA model, yet for the TIN-based model, the SINR drops due to the higher %effective 
interference.  
For TIN, CC-OMA, and CC-NOMA perform well for low $\rho$, and CC-OMA can effectively combine retransmissions even for high $\rho$. However, the SNR per bit demand for CC-NOMA is sensitive to $\rho$, deteriorating the performance at high $\rho$. The SE of the classical model and IR-OMA do not scale as well as CC-OMA because the former models cannot compensate for the interference at high $\rho$ and hence cannot leverage retransmissions. The $J/n$ versus $E_b/N_0$ performances of sum-rate optimal models deteriorate in $T$. The ordering of the %different 
models in the sum-rate optimal regime, from the most to the least sensitive to degradation, as an increasing function of $T$, is (i) classical, (ii) CC-OMA, (iii) IR-OMA, and (iv) CC-NOMA, demonstrating the robustness of CC-NOMA to retransmissions.

\item {\bf \emph{User density scales up with SNR per bit.}} The user density $J/n$ can superlinearly scale with $E_b/N_0$ (where the scaling does not necessarily degrade with $T$ in the case of CC-NOMA versus the %other models that rely on OMA
OMA-based models) in the FBL %finite (FBL) 
and the infinite blocklength (IBL) regimes, where IBL gives an upper bound to the %actual user density 
scaling, which becomes tighter as $n$ increases. Both for sum-rate optimal and TIN-based models, the scaling of $J/n$ %the user density 
versus SNR per bit does not improve with $\rho$ due to the increase in the SNR per bit.  
Comparing different TIN models, at low $\rho$, the scalings of $J/n$ versus $E_b/N_0$ for CC-NOMA and CC-OMA improve similarly, whereas the schemes that do not promote CC do not perform as well. %However, 
With increasing $\rho$, the scaling of CC-NOMA deteriorates due to high interference, whereas CC-OMA performs the best among all as it combines retransmissions and provides reduced interference.  
In the TIN-based CC-NOMA and CC-OMA models, the scalings of $J/n$ improve with increasing $T$, and the scalings for the IR-OMA and the classical models are not sensitive to $T$. 

\end{CompactItemize}

Our insights could be applied to 5G wireless system design with delay and resource-constrained communications, which is critical in use cases such as URLLC or mMTC. Nevertheless, the scaling results in our framework provide an upper bound on the achievable SE and the user density because of the following additional assumptions: ideal negative acknowledgment with no error or delay, the IBL regime capacity-achieving encoding, perfect power control, perfect synchronization among users, and decoding via a suboptimal receiver, through matched filtering and SUDs, versus MUD, which could strictly improve performance of random-CDMA \cite[slide 146]{polyanskiy2018information}. A more general framework that allows studying the above key points as well as path loss  
and outage capacity-based models 
both for the IBL and the FBL regimes 
will be considered as a future work, as detailed in Section \ref{sect:conclusion}.

%%%%%%%
\section{System Model}
\label{CommsModel}

\begin{figure}[t!]
    \centering
    \includegraphics[width=0.75\textwidth]{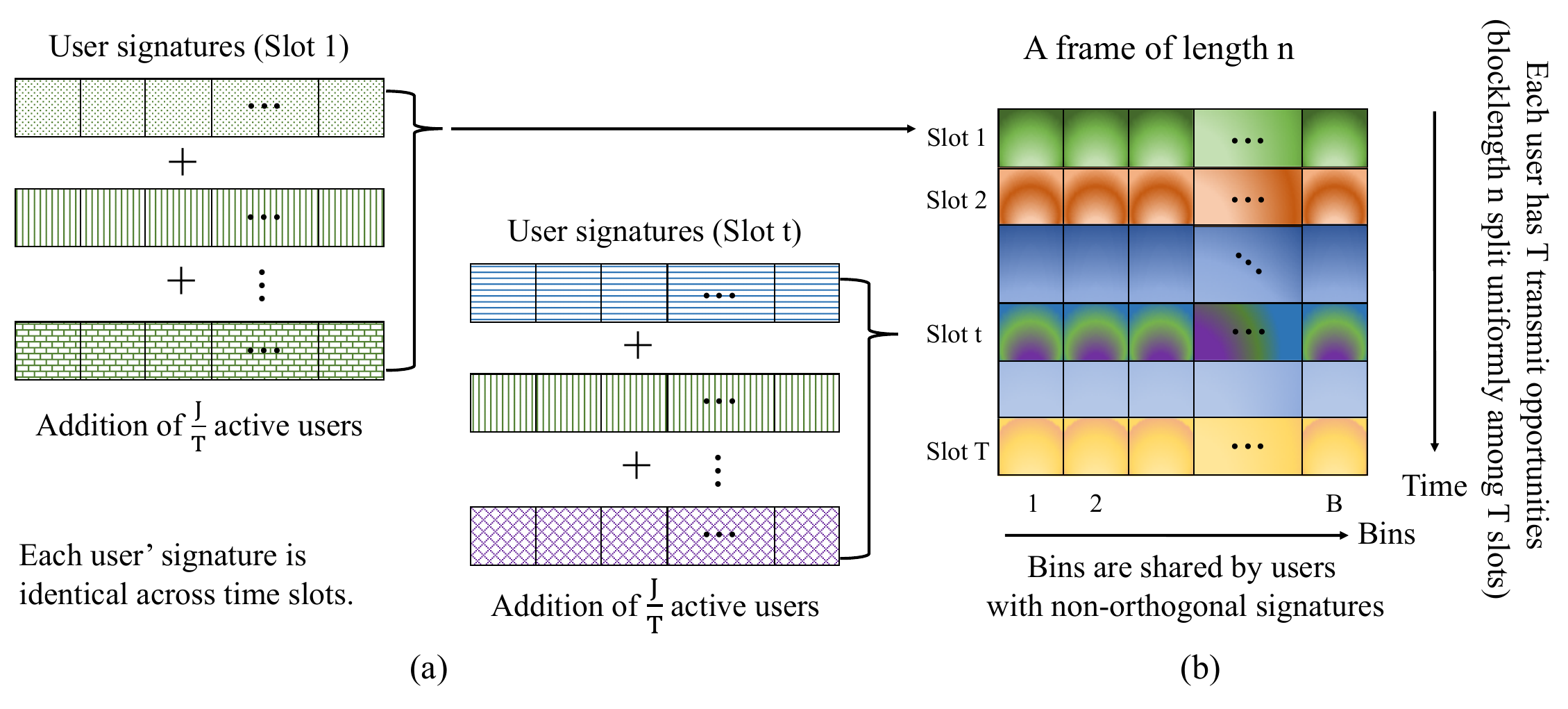}
    \caption{(a) Non-orthogonal user signatures at time slots $1$ and $t$. Each user uses the same signature across all time-frequency resources. The second user in slot $1$ is repeated in slot $t$ (same signature). (b) The frame structure where time is partitioned into $T$ transmit opportunities, and the time-frequency resources are shared in a non-orthogonal manner by the users. }
    \label{fig:frame}
\end{figure}

We consider a wireless \MA communication model where a collection of users transmits over shared radio resources to a common receiver. The goal of each user is to transmit its payload of fixed size ($L$ bits) within a latency constraint (blocklength $n$). A user is granted $T$ retransmission attempts, i.e., time slots, to communicate its payload. The users use non-orthogonal signatures to transmit their payloads, as shown in Figure \ref{fig:frame}-(a). The signatures are kept identical at each attempt.

\paragraph{Frame structure} A frame has a total bandwidth of $\omega$ Hz and the time of duration $\tau$ sec, and is partitioned into $B$ frequency bins of equal width, and $T$ time slots, i.e., transmit opportunities, of equal duration. We refer to a given time slot and frequency bin as a  
TFS. For the proposed frame structure, the total number of resources or real degrees of freedom (rdof) in a frame is $N=\omega \tau$, which is evenly split into $T$ retransmissions. The TFSs  
in a frame are shared by a collection of users in a non-orthogonal manner. While in OMA-based approaches, the rdof is split orthogonally among the users,  
in NOMA-based transmissions, the TFSs in a frame are shared by a collection of users in a non-orthogonal manner. Each user attempts to transmit its payload of fixed-size $L$ bits over shared resources. Given $\omega$, $\tau$, $m$, and $T$, the number of symbols in a TFS is $\omega \tau/(B T)$. Under the orthogonal division of the resources, the coding rate is $LBT/(\omega \tau)$ bits per transmitted symbol.

\paragraph{User (source) model} Given $T$ (re)transmission attempts, the total blocklength $n$ per-user is split uniformly across $T$ attempts to accommodate the retransmission of a packet. Hence, the blocklength per transmission at each time slot  
is $m=n/T$. Let $J_t$ and $\mathcal{J}_t$ be the number and set of users at slot $t\in\mathcal{T}$, respectively, such that $J=\sum_{t=1}^T J_t$, and $\mathcal{J}$ be the set of all users in the frame. 

Let ${\bf U}_j=(U_{j1},U_{j2},\dots,U_{jK})$ be the $K$ dimensional source vector corresponding to user $j\in\mathcal{J}$. In the case of no feedback, let $\phi_{tji}:\mathcal{U}_j^K \to \mathcal{V}_{j}$ for $i\in\{1,\dots,m\}$ be the encoder function for $j\in\mathcal{J}$ that captures the mapping from ${\bf U}_{j}$ to the channel input ${\bf V}_{tj}=\phi_{tji}({\bf U}_{j})=b_{tj}{\bf S}_{j}$, i.e., the product of the complex amplitude $b_{tj}\in\mathbb{C}$ of the transmitted symbol and selected signature sequence ${\bf S}_{j}$, for %retransmission 
attempt $t\in\mathcal{T}$, where each retransmission ${\bf V}_{tj}=(V_{tj1},V_{tj2},\dots,V_{tjm})$ from user $j$ has a blocklength $m$.

\paragraph{User signatures} The number of rdof $N$ in a frame can be thought of as the total length of the signature sequences of the active users over $B$ frequency bins. 
Each user has the same signature across all time-frequency resources. The TFSs are shared in a non-orthogonal manner, where each waveform at a given time slot is a sum of non-orthogonal signatures, which is shown 
in Figure \ref{fig:frame}-(a). We assume that the signature sequences ${\bf S}_{j}$ are unitary, $\norm{{\bf S}_{j}}=1$, i.e., each signature has unit variance, $\mathbb{E}[{\bf S}_{j}^{\intercal}{\bf S}_{j}]=1$, and $|\langle{\bf S}_{j},{\bf S}_{j'}\rangle|=|\mathbb{E}[{\bf S}_{j}^{\intercal}{\bf S}_{j'}]|=\eta$ for any $\{(j,j')\in\mathcal{J}_t:\ j\neq j'\}$. The maximum value of $J_t$ to ensure that all $j\in\mathcal{J}_t$ is decoded with zero-error is given by the Khachatrian-Martirossian 
construction \cite{polyanskiy2018information} %Slide 139 YP
allows $J_t>m$ users. 
Under this setup, when ${\bf S}_{j}$'s are random and $m$ is large, $\eta\approx \frac{1}{\sqrt{m}}$ with high probability. 
We sketch the frame structure with overlapping NOMA traffic in Figure \ref{fig:frame}-(b).  

%%%
\paragraph{Received signal and conventional matched filter decoding} 
The transmitted signal from user $j\in \mathcal{J}_t$ multiplied by the channel gain determines the received signal, given by ${\bf X}_{tj}=a_{tj}{\bf S}_{j}$, where $a_{tj}\in\mathbb{C}$ is the complex amplitude of the product of the values of the transmitted symbol $b_{tj}$, and the channel gain $H_{tj}$ of user $j$ at slot $t$, accounting for fading. Hence,  $|a_{tj}|^2=|b_{tj}|^2\cdot  |H_{tj}|^2$ represents the transmitted signal power times channel power gain variable (see Appendix~\ref{App:Chase_SINR_nonorthogonal_sumoptimal}).  
We denote the received signal vector during transmission $t\in\mathcal{T}$ by ${\bf Y}_t=(Y_{t1},Y_{t2},\dots,Y_{tm})$. The channel is additive such that the received signal vector during transmission $t\in \mathcal{T}$ is\footnote{In the case with feedback, $V_{tji}=\phi_{tji}({\bf U}_{j},{\bf Y}_t^{i-1})$ is the channel input from user $j\in\mathcal{J}$ at time $i\in\{1,\dots,m\}$ for attempt $t\in\mathcal{T}$, where ${\bf Y}_t^{i}=(Y_{t1},Y_{t2},\dots,Y_{ti})$, and $\phi_{tji}:\mathcal{U}_j^K \times \mathcal{Y}^{i-1}\to \mathcal{V}_{j}$ is the encoder function. We leave the feedback setting for future work.} 
\begin{align}
\label{retransmissions_combining_chase}
{\bf Y}_t={\bf X}_{tj}+\sum\limits_{j'\in \mathcal{S}_{t,-j}}{\bf X}_{tj'}+{\bf Z}_t=a_{tj}{\bf S}_{j}+\sum\limits_{j'\in \mathcal{S}_{t,-j}}a_{tj'}{\bf S}_{j'}+{\bf Z}_t,    \quad t\in \mathcal{T} \ ,
\end{align}
where $\mathcal{S}_{t,-j}$ is the collection of the interferers of $j\in \mathcal{J}_t$ in the same time slot $t$, i.e., $\mathcal{S}_{t,-j}=\{j'\in\mathcal{J}_t:\,j'\neq j\}$, and ${\bf Z}_t\sim \mathcal{CN}(0,\sigma_t^2 I_m)$ is a complex Gaussian random variable. We assume perfect channel knowledge at the receiver, whereas the receiver has no access to ${\bf S}_{j}$.

%Justification for matched filter decoding.
We consider the {\em conventional matched filter receiver} (MFR) for decoding, which performs approximately optimal when the target SINR is low. In this case, the effective bandwidth required by the conventional approach is small versus %MMSE and 
the linear decorrelator receiver, which %is desired as it 
allows many users per DoF, where the other users' signals are treated as additive white Gaussian noise (AWGN) \cite{tse1999linear}. %On the other hand, when
If the target SINR is high, both the linear minimum mean-square error (MMSE) and the linear decorrelator receiver decorrelate a user from the rest, yielding no more than one DoF per interferer \cite{tse1999linear}. When %In the case that 
$\{{\bf S}_{j}\}_{j\in\mathcal{J}}$ are known to the receiver, an MMSE-based receiver provides a better signal-to-interference ratio (SIR) per-user via exploiting the structure of the interference \cite{tse1999linear}. 
The maximum number of supported users for %the decorrelator and the 
MFR %in the asymptotic regime is %also 
derived in \cite{tse1999linear} as a function of a target SIR, %${\rm SIR}^*$, 
and the received power %, $P$, 
under power control. 
%It is given as $\frac{1}{{\rm SIR}^*}-\frac{\sigma^2}{P}$, which increases in $P$, and is equal to $\frac{1}{{\rm SIR}^*}$ when $P$ tends to infinity. 
In \cite{tse1999linear}, different from our approach, the characterization of the scaling results for the users is based on (i) a target SIR requirement at the receiver for all users, and (ii) the asymptotic regime in the number of users, contrary to the FBL regime analyzed in the current work.

\paragraph{Maximum ratio combining} We assume that the receiver's HARQ buffer size equals the number of coded symbols per coded packet, where the retransmitted packets are summed up with previously received erroneous packets via maximum ratio combining (MRC) of retransmissions prior to decoding.

The common receiver has the decoder function $\Phi_T: \mathcal{Y}^{n}\to \{\mathcal{U}_j^K\}_j$ that combines $T$ retransmissions to decode the individual source vectors $\{{\bf U}_j\}_j$ from the received signal vectors ${\bf Y}_t$, $t\in\mathcal{T}$.  
Using (\ref{retransmissions_combining_chase}), the MRC of $T$ transmissions results in the following combined signal:
\begin{align}
\label{retransmissions_combining_chase_decoder}
{\bf Y}={\bf U}_j+\sum\limits_{t=1}^T a^*_{tj} \sum\limits_{j'\in \mathcal{S}_{t,-j}}a_{tj'}{\bf S}_{j'}+{\bf Z}\ ,
\end{align}
where ${\bf Y}=\sum\limits_{t=1}^T a^*_{tj} {\bf Y}_t$, and ${\bf U}_j=\sum\limits_{t=1}^T |a_{tj}|^2 {\bf S}_{j}$,  
and ${\bf Z}=\sum\limits_{t=1}^Ta^*_{tj}{\bf Z}_t$ are $m$ dimensional vectors. 
We assume that the coefficients $a_{tj}$ are known. These coefficients can be estimated using the least mean square (LMS) algorithm and then utilized by the MRC for generating the decision variable.

\paragraph{Per-user received SNR} The noise power each user sees is assumed to be additive and constant with value $\sigma_t^2$, $t\in\mathcal{T}$ per dimension, i.e., $\langle {\bf Z}_{t},{\bf Z}_{t}\rangle=m\sigma_t^2$, where $m\sigma_t^2$ is the total noise power across the number of frequency bins, which is $B$. The average received power of user $j\in\mathcal{J}$ during transmission $t\in\mathcal{T}$, which is the total noise power times the received SNR, under unit channel power gain, is 
\begin{align}
\sum\limits_{i=1}^m\mathbb{E}[X^2_{tji}]=m\mathbb{E}[X^2_{tj1}]=\mathbb{E}[{\bf X}_{tj}^{\intercal}{\bf X}_{tj}]=|a_{tj}|^2 \mathbb{E}[{\bf S}_{j}^{\intercal}{\bf S}_{j}]=|a_{tj}|^2=|b_{tj}|^2=m\sigma_t^2\rho_{tj} \ , \nonumber
\end{align}
where $\rho_{tj}$ denotes the received SNR from $j\in\mathcal{J}$ during transmission $t\in\mathcal{T}$, noting that $\mathbb{E}[{\bf S}_{j}^{\intercal}{\bf S}_{j}]=1$.

The energy constraint for each transmitted symbol $j\in\mathcal{J}$ at any given $t\in\mathcal{T}$ is
\begin{align}%{\bf X}_{tj} is length m
\label{PowerConstraint}
\mathbb{E}[{\bf X}_{tj}^{\intercal}{\bf X}_{tj}]=m\sigma_t^2\rho_{tj}\leq \frac{K E_j}{T}    \ ,
\end{align}
where the parameter $K$ is the message size of any source in bits, and $E_j$ denotes an upper bound on the received energy of user $j\in\mathcal{J}$ per source dimension. 
In (\ref{PowerConstraint}), the total power of channel input linearly scales with the message size $K$, yielding a maximum total energy of $K E_j$ per message of user $j\in\mathcal{J}$, and an upper bound on the received energy per slot, denoted by $\frac{K E_j}{T}$. We assume that $\mathbb{E}[X_{tji}]=0$ and $X_{tji}$'s across $j\in\mathcal{J}$ are not independent such that $%\sum\limits_{i=1}^m\mathbb{E}[X_{tji}X_{tj'i}]=
\mathbb{E}[{\bf X}_{tj}^{\intercal}{\bf X}_{tj'}]=a^*_{tj}a_{tj'} \mathbb{E}[{\bf S}_{j}^{\intercal}{\bf S}_{j'}]=a^*_{tj}a_{tj'}\eta\leq \frac{K E_{jj'}}{T}$ for $\{(j,j'):\ j\neq j'\}$, noting that the non-orthogonal user signatures satisfy $|\langle{\bf S}_{j},{\bf S}_{j'}\rangle|=\eta$.
%$a^*_{tj}\cdot a_{t'j}=0$ for $t'\neq t$

Assuming that $\mathbb{E}[X_{tji}^2]$ does not change with $i\in\{1,\dots,m\}$, and $\sigma_t^2=\sigma^2$, from (\ref{PowerConstraint}) we have
\begin{align}
\rho_{tj}=\frac{\mathbb{E}[{\bf X}_{tj}^{\intercal}{\bf X}_{tj}]}{m\sigma^2}=\frac{1}{m\sigma^2}\sum\limits_{i=1}^m \mathbb{E}[X_{tji}^2]=\frac{\mathbb{E}[X_{tji}^2]}{\sigma^2}=\frac{|a_{tj}|^2}{m\sigma^2} \ ,\quad j\in\mathcal{J} \ .\nonumber
\end{align}

%%%%%%%%%%%%%%%%%%%%%%%%%%%%%%%

We  
assume that $\rho_{tj}$ are identical and denoted by $\rho$. 
Given a constant received 
power of $m\sigma^2\rho $, the received SNR is $\rho=\frac{\mathbb{E}[{\bf X}_{tj}^{\intercal}{\bf X}_{tj}]}{m\sigma^2}$. 
For NOMA-based transmissions, the total power spent by all users is
\begin{align}
\label{transmit_power}
P_{tot}=\frac{\Jnomaframe m\sigma^2\rho}{n}=\Jnomaslot\sigma^2\rho \ ,    
\end{align}
or equivalently, the total energy spent for a given %code 
blocklength $n$ is $nP_{tot}$. %Does this differ for NOMAversus OMA?
For OMA-based transmissions, the number of users per slot is $J/T$ %$\Jomaslot$ 
(versus $\Jnomaslot$ for NOMA-based), and $P_{tot}=\Jomaslot\sigma^2\rho$ \cite[Ch. 4-6]{tse2005fundamentals}.

The overall problem is to determine some key performance metrics, which are the spectral efficiency, the SNR per bit, and the user density, and their joint behavior, which we describe in the sequel.

%%%
\paragraph{The spectral efficiency (SE)}
It is the maximum number of bits per channel use (bits/s/Hz):
\begin{align}
\label{SE_defn}
\SE=\frac{\mbox{Total number of data bits}}{\mbox{rdof}}\ ,    %Total real DoF
\end{align}
where rdof represents the total number of real DoF, denoted by $n$. %The number of real degrees of freedom (rdof) $n=\omega \tau$ given $\omega$ hertz (Hz) bandwidth and duration $\tau$ second (sec).

\begin{defi}
\label{def:IBL}
%https://www.inc.cuhk.edu.hk/InformationTheory/files/PDF/7_6.pdf
{\bf (Achievable channel coding rate in the IBL regime \cite{yeung2008information}.)} A rate $R$ is achievable %with complete feedback 
for a discrete memoryless channel (DMC) %$p(y\,|\,x)$ 
if for rates below capacity $C$ such that 
\begin{align}
%\frac{1}{n}\log M>R-\epsilon\ .\nonumber   
R=\frac{1}{n}\log M  < C \ , 
\end{align}
%for any $\epsilon > 0$, 
there exists for sufficiently large $n$ an $(M,\, n)$ code with complete feedback, with maximal probability of error $\epsilon\to 0$. Conversely, for a sequence of codes $(M,\, n)$, if $\epsilon>0$, %$\epsilon$ is bounded away from $0$, 
then it must hold that $R > C$.
\end{defi}

Assume that a user attempts to transmit a payload of fixed size $L$ bits over the channel. Hence, the relation between the required codebook size $M$ and $L$ is $L=\log M$. %within a latency constraint of $T$ seconds 
Hence, the blocklength $n$ should be chosen sufficiently large so that the achievable transmit rate, $\frac{L}{n}$, satisfies:
\begin{align}
\label{blocklength_vs_SE}
\frac{L}{n} \leq C=\frac{1}{2}\log_2(1+{\rm SINR})\,\, bit/rdof, \quad n\leq N \ , %n\leq \frac{\tau \omega}{T m} \ ,
%C=\frac{1}{2}\log(1+{\rm SINR}),\, {\text{nats/channel use}}
\end{align}
where $C$ is %Shannon's 
channel capacity, and SINR represents the signal-to-interference-plus-noise ratio, for an AWGN channel where interference is treated as noise (TIN). The capacity is achievable at an arbitrarily low error rate in the IBL regime, i.e., as $n\to\infty$. However, since $N$ is finite, the ratio $L/N$ is always finite. Hence, given $L$, the IBL scheme gives an upper bound on $R$, and a lower bound on $n$.

In the FBL regime, let $M(n,\epsilon)$ be the maximal code size achievable with a given finite blocklength $n$, and average error probability $\epsilon$. Then, the maximal rate achievable is approximated by \cite{polyanskiy2010channel}. 

\begin{defi}
\label{def:FBL}
{\bf (Achievable channel coding rate in the FBL regime \cite{polyanskiy2010channel}.)} A rate $R$ is achievable with complete feedback for a DMC %$p(y\,|\,x)$ 
if for any $\epsilon > 0$, there exists an $(M,\, n)$ code such that 
\begin{align}
\label{FBL_capacity}
R(n,\epsilon)=\frac{1}{n}\log M(n,\epsilon)\approx C - \sqrt{\frac{V}{2n}} Q^{-1}(\epsilon) \ ,
\end{align}
for sufficiently large $n$, 
%with complete feedback,  
where $M(n,\epsilon)$ is the maximal code size achievable with a given blocklength $n$ and average error probability $\epsilon$, and $Q(x) = \frac{1}{\sqrt{2\pi}} \int\nolimits_{x}^{\infty}e^{-u^2/2}\,{\rm d}u$ is the tail probability of the standard normal distribution where $Q^{-1}$ is the inverse $Q$-function. Furthermore, in (\ref{FBL_capacity}), $V=1-\frac{1}{(1+{\rm SINR})^2}$ is the channel dispersion, and $C=\frac{1}{2}\log(1+{\rm SINR})$ is the capacity in the units of nats per channel use. %which satisfies (\ref{blocklength_vs_SE}) for the AWGN channel model by TIN. 
\end{defi}

While the Khachatrian-Martirossian construction is designed for the noiseless adder channel \cite{khachatrian1998code}, it achieves a sum rate that approximates the sum rate of a Gaussian MAC (i) under the assumption of perfect channel inversion power control, such that $\rho=1$, and (ii) when $J$ is high, justifying (\ref{FBL_capacity}) in Definition~\ref{def:FBL}. 
Furthermore, the Gaussian approximation to the FBL regime is tight \cite{polyanskiy2010channel}.

In the following, we instead express (\ref{FBL_capacity}) as  $R(n,\epsilon)\approx C-\Delta(C,n,\epsilon)$, where the channel dispersion $V$ can be written as function of  ${\rm SINR}=\exp(2C)-1$, and hence, the term $\Delta(C,n,\epsilon)=\sqrt{\frac{V}{2n}} Q^{-1}(\epsilon)=\sqrt{\frac{1}{2n}(1-\frac{1}{\exp(2C)^2})} Q^{-1}(\epsilon)$ captures the joint behavior of $C$, $n$, and $\epsilon$.

We note that at low $\rho$, the value of SINR is small and the channel dispersion in the FBL regime becomes negligible, yielding from (\ref{FBL_capacity}) that the IBL approximation is good in the TIN regime. %For more details on the FBL regime, we refer the reader to \cite{polyanskiy2010channel}.

%%% $N_0$ is the noise spectral density, the noise power in a 1 Hz bandwidth, measured in watts per hertz or joules.
%
%But when the signal bandwidth is well defined, $E_b/N_0$ is also equal to the SNR in that bandwidth divided by the gross link spectral efficiency in bits/s/Hz, where the bits in this context again refer to user data bits, irrespective of error correction information and modulation type \cite{heegard1999turbo}.
\paragraph{The SNR per bit, $E_b/N_0$}
It represents the ratio of the energy-per-bit to the noise power spectral density, which is a normalized SNR measure:
\begin{align}
\label{EbN0_defn}
\frac{E_b}{N_0}=\frac{\mbox{Total energy spent}}{2\times \mbox{Total number of bits}} \ ,   
\end{align}
which is dimensionless, and usually expressed in decibels (dB). We note that the scaling $2$ in the denominator captures the total number of bits over the entire bandwidth, which is $2\times rdof$.%We note that the scaling $2$ in the denominator captures the total spectral efficiency over the entire bandwidth.

%%%
\paragraph{User density (users/rdof)} %real degrees of freedom (rdof)
Given a total count of users $J=\sum_{t=1}^T J_t$, where $J_t$ is the count of users active in slot $t\in\mathcal{T}$, and a total blocklength $n$ %per-user 
given a frame duration $T$, {\em user density}, $J/n$, gives the total number of users per rdof that can transmit within the same frame. 
For the OMA-based schemes, where the blocklength per retransmission is $m$, the density of users in slot $t$ is $J_t/m$. 
For NOMA-based schemes, the ratio $J_t/n$ denotes the maximum density of users that can simultaneously transmit in  $t\in\mathcal{T}$. From (\ref{SE_defn}), (\ref{blocklength_vs_SE}), and (\ref{EbN0_defn}), the achievable $J/n$ is affected by the SE versus SNR per bit tradeoffs of the retransmission-based protocols for uplink access, which we detail in Section \ref{NOMA_retransmission_no_fading}.

\begin{comment}
We provide the notation used in the current paper in Table \ref{table:notation}.
\begin{table*}[h!]\small%\footnotesize
\setlength{\extrarowheight}{2pt}
\begin{center}
\begin{tabular}{l | l }
{\bf Description} & {\bf Symbol}\\ 
\hline
Maximal code size with a given error probability $\epsilon$ and blocklength $n$ & $M(n,\epsilon)$\\
Set of all users & $\mathcal{J}$\\
User $j\in\mathcal{J}$ message & $U_j$\\
Per-user received SNR & $\rho$\\
Noise power & $\sigma^2$\\
A measure of non-orthogonality of the user signatures & $\eta$\\
Input (source) dimension (same for each user $j$) & $K$\\ 
Blocklength per transmission & $m$\\
%Encoder$\times$channel function for user $j\in\mathcal{J}$ at time $i\in\{1,\dots,n\}$ & $\phi_{ji}$\\
%Decoder function & $\Phi$\\
Channel input of user $j\in\mathcal{J}$ at time $i\in\{1,\dots,n\}$ & $X_{ji}$\\
%Rate-distortion parameters & ?\\
%Error parameters & ?\\
%Energy parameters & ?\\
Total number of retransmissions & $T$\\
Blocklength per transmission & $m=n/T$\\
Set of users at slot $t\in\mathcal{T}$ & $\mathcal{J}_t$\\
Collection of the interferers of $j\in \mathcal{J}_t$ in the same time slot $t$ & $\mathcal{S}_{t,-j}$ \\
%%%
HARQ buffer size for IR-OMA & $\Cb$\\%We assume that the buffer size is exact for perfect recovery in the cases of CC-NOMA and CC-OMA.
Quantization noise for IR-OMA at slot $t$ given $T$ & $\sigma_{q}^2(t,T)$\\
\hline
\end{tabular}
\end{center}
\caption{Notation.}
\label{table:notation}
\end{table*}
%The capacity notations
\end{comment}

%%%
\section{Combining NOMA-based Retransmissions in Uplink}
\label{NOMA_retransmission_no_fading}

We focus on the scaling behaviors of the SE, the SNR per bit, and the user density for the retransmission-based \MA schemes. The senders must contend not only with the receiver noise but also with interference from each other. To that end, we next analyze the behavior of the SE and SNR per bit performances of the HARQ-based schemes for first, the {\em sum-rate optimal regime} that is attainable via SIC, and then the achievable data rate of a single user, i.e., per-user rate via treating the total interference from all other users as noise, i.e., {\em TIN}. 
However, our analysis does not capture the joint decoding of the intended user and the strongest interferers, which we leave as future work.

%%%%%
\subsection{The Classical Transmission Scheme with No Multiplexing of Retransmissions}
\label{no_retransmissions}

We commence with the classical interference-based model with no multiplexing across different time slots. Each user selects one slot to transmit its message given a  blocklength $n$. The time resources are split uniformly across $T$ slots. There are $J_t=J/T$ users per slot sharing the frame resources. 
In general, transmissions are exposed to different channel conditions, more specifically, the fading (e.g., Rayleigh fading) or path loss. Incorporating the channel gains $|H_{tj}|^2$, $t\in\mathcal{T}$, $j\in\mathcal{J}$, and assuming that $|H_{tj}|^2$ has unit power and is independent across the slots %\cite{lee2015harq}
with a known cumulative distribution function, $F_{|H|^2}$, %Incorporating the channel gains, 
we can express the SE of the classical sum-rate optimal transmission approach as
\begin{align}
\label{SE_no_retransmissions_sum_optimal_fading}
\clasicalsumoptimal=\frac{1}{2T}\sum\limits_{t\in\mathcal{T}}\log_2\Big(1+\rho \sum\limits_{j\in \mathcal{J}_t}|H_{tj}|^2\Big)  \,\, bit/rdof\ .  
\end{align}

The SNR per bit of the classical sum-rate optimal transmission model is equal to
\begin{align}
\label{EbN0_no_retransmissions_sum_optimal}
\frac{E_b}{N_0}%=\frac{J\sigma^2\rho n}{m\log_2\big(1+\rho \frac{J}{T}\big)}=\frac{J\sigma^2\rho T}{\log_2\big(1+\rho \frac{J}{T}\big)} \ .  \\
=\frac{\Jomaslot\sigma^2\rho n}{2m(\clasicalsumoptimal-\Delta(\clasicalsumoptimal,n,\epsilon))} \ ,
\end{align}
where the total energy spent for a given blocklength $n$ is $nP_{tot}=\Jomaslot\sigma^2\rho n$, which is adapted for classical OMA-based transmissions from the relation in (\ref{transmit_power}) for NOMA-based transmissions.

The SE of the classical model via TIN for decoding $j\in\mathcal{J}_t$, where $J_t=J/T$ users per slot, it holds that $\rho=0$ for the remaining $T-1$ slots for which $j\notin\mathcal{J}_{t'}$, $t'\in\mathcal{T}\backslash \{t\}$, is expressed as 
\begin{align}
\label{SE_no_retransmissions_TIN_fading}
\clasicaltin=\frac{1}{2T}\sum\limits_{t\in \mathcal{T}}\sum\limits_{j\in \mathcal{J}_t}\log_2\Big(1+\frac{\rho |H_{tj}|^2}{\rho  \sum\limits_{j'\in\mathcal{S}_{t,-j}} |H_{tj'}|^2+1}\Big)  \,\, bit/rdof \ ,
\end{align}
where the summation  $\sum\limits_{j\in\mathcal{J}_t}$   in the front of the logarithm in (\ref{SE_no_retransmissions_TIN_fading}) denotes the achievable SE per time slot for the TIN (versus no coefficient for the sum-rate optimal model in (\ref{SE_no_retransmissions_sum_optimal_fading})).

Similarly, the SNR per bit for the classical transmission model with TIN is given as
\begin{align}
\label{EbN0_no_retransmissions_TIN}
\frac{E_b}{N_0}%=\frac{J\sigma^2\rho n }{n\frac{J}{T}\log_2\big(1+\frac{\rho }{\rho  (J-1)+1}\big)}=\frac{\sigma^2\rho T}{\log_2\big(1+\frac{\rho }{\rho  (\frac{J}{T}-1)+1}\big)} \ ,\\
=\frac{\Jomaslot\sigma^2\rho n}{2n(\clasicaltin-\Delta(\clasicaltin,n,\epsilon))} \ .
\end{align}

We next provide two results on $E_b/N_0$ for the IBL regime, for the classical transmission model.

\begin{cor}
\label{EbN0_no_multiplexing}
{\bf The classical transmission model.} 
For the classical transmission model %when $T=1$ 
in the IBL regime, for $|H_{tj}|=1$, $\forall$ $t\in\mathcal{T}$, $j\in\mathcal{J}_t$, exploiting the relation between $\SE$ and $E_b/N_0$, we next provide the relations between %the measures 
$\SE$, $P_{tot}$, and $J$, for the sum-rate optimal and TIN %-based 
regimes.

(i) \underline{The classical sum-rate optimal approach.} 
The measures $\SE$ and $P_{tot}$ satisfy the relation 
\begin{align}
%\frac{(2^{2\SE}-1)\sigma^2}{2\SE}=\frac{P_{tot}}{\log_2(1+{P_{tot}}/{\sigma^2})}\ ,\nonumber\\
(2^{2\clasicalsumoptimal}-1)\sigma^2=P_{tot} \ .\nonumber
\end{align}

(ii) \underline{The classical transmission approach with TIN.} 
The measures
$\SE$, %as given in (\ref{SE_no_retransmissions_TIN}), 
$P_{tot}$, and $J$ satisfy 
%the SNR per bit as a function of SE in (\ref{SE_no_retransmissions_TIN}) is given as 
%$\frac{E_b}{N_0}=\frac{(2^{\frac{2\SE}{J}}-1)J\sigma^2}{2\SE\big[1-(J-1)(2^{\frac{2\SE}{J}}-1)\big]}=\frac{P_{tot}}{\log_2(\frac{\rho J+1}{\rho J-\rho+1})}$, 
\begin{align}
%\frac{(2^{\frac{2\SE}{J}}-1)}{2\SE\big[1-(J-1)(2^{\frac{2\SE}{J}}-1)\big]}=\frac{\rho}{\log_2\big(1+\frac{\rho}{P_{tot}/\sigma^2-\rho+1}\big)} \ ,\nonumber\\
\frac{J}{T}\sigma^2{(2^{2\clasicaltin\cdot\frac{T}{J}}-1)}\Big/{\Big(1-\Big(\frac{J}{T}-1\Big)(2^{2\clasicaltin\cdot\frac{T}{J}}-1)\Big)}=P_{tot}\ ,\nonumber
\end{align}
where $\SE$ is given in (\ref{SE_no_retransmissions_TIN_fading}). We note under fixed total power $P_{tot}$ that  $\lim\limits_{\rho\to 0}\, \frac{E_b}{N_0}= \log 2\cdot  T\sigma^2$ from (\ref{EbN0_no_retransmissions_TIN}). 
\end{cor}

From Cor. \ref{EbN0_no_multiplexing}, we note that both $\clasicalsumoptimal$ and $\clasicaltin$ increase in $P_{tot}$. Hence, we can determine the common $E_b/N_0$ value that leads to the classical sum-rate optimal and the classical TIN-based models to achieve ${\rm SE}\to 0$ (which is attained when $J\to\infty$ \cite[Slide 69]{polyanskiy2018information}). The convergence behavior for different $T$ values can be observed in Section \ref{sect:numericalresults} (see e.g., Figures \ref{fig:SE_EbN0_eta1_J10} and \ref{fig:SE_EbN0_eta_J_varying}).

In the case of no retransmissions, TIN is essentially optimal for low SE \cite{polyanskiy2018information}. 
However, for strategies combining the retransmissions, TIN may not be optimal even at low SE, see e.g., \cite{polyanskiy2018information} and \cite{geng2015optimality}. We will next analyze the SE and the SNR per bit by incorporating the channel gains 
(to accurately capture the SINR) for the HARQ models in Sections \ref{chase_retransmissions}, \ref{chase_orthogonal_retransmissions}, and \ref{IR_retransmissions}, which will be followed by numerical simulations in Section \ref{sect:numericalresults} to contrast the various HARQ schemes and demonstrate that sum-rate optimal  
schemes could be more energy efficient via combining of retransmissions versus TIN.

%%%%%
\subsection{Chase Combining with NOMA-based Retransmissions}
\label{chase_retransmissions}

For a given payload $L$,  
a user transmits $T$ times within a frame of duration $\tau$ sec. At the receiver, Chase combining is a common form of HARQ that is used to combine signal energy for a given user’s transmissions over $T$ slots. Chase combining has been shown to increase throughput in relatively poor channel conditions  
\cite{malak2018throughput}. 
We assume that the transmission is successful, i.e., a user can have its payload decoded, at the end of $T$  
attempts when the Chase-combined SINR exceeds the critical threshold. 
Once a user's transmission is successfully decoded after $T$ attempts, it stops transmitting.  
The duration of each slot is $\tau/T$ seconds, to ensure that the user will meet the latency constraint $\tau$.

In CC-HARQ, each transmission contains the same data and parity bits. 
The receiver's HARQ buffer size for CC-HARQ equals the number of coded symbols per coded packet, where the retransmitted packets are summed up at the receiver with previously received erroneous packets via MRC of retransmissions prior to decoding. 
In Figure \ref{fig:CC-HARQ} (left), we sketch CC-HARQ. 
We next derive the SE for the Chase combining of NOMA-based retransmissions (CC-NOMA) for the sum-rate optimal model. 
\begin{prop}
\label{Chase_SINR_nonorthogonal_sumoptimal}
{\bf Sum-rate optimal model --- Chase combining of non-orthogonal transmissions.} The SE of CC-NOMA for the sum-rate optimal model incorporating channel power gains is given as 
\begin{align}
\label{SE_CC_NOMA_sum_fading}
\!\!\!\!\chasenoma=\frac{1}{2}\log_2\Big(1+\rho \sum\limits_{t\in\mathcal{T}}|H_{tj}|^2\Big[1+\eta^2\Big(\sum\limits_{t\in\mathcal{T}}|H_{tj}|^2\Big)^{-2}\Big|\sum\limits_{t\in \mathcal{T}} \sum\limits_{j'\in \mathcal{S}_{t,-j}}H_{tj}H^*_{tj'}\Big|^2\Big]\Big)\,\, bit/rdof\ ,    
\end{align}
where $|H_{tj}|^2$ is the channel power gain of user $j\in \mathcal{J}$ (the one with the largest SINR) at slot $t\in\mathcal{T}$. 

The SNR per bit for CC-NOMA for the sum-rate optimal model, using $\chasenoma$ in (\ref{SE_CC_NOMA_sum_fading}), equals 
\begin{align}
\label{Eb_N0_CC_NOMA_sum}
\frac{E_b}{N_0}
=\frac{J\sigma^2\rho n}{2n(\chasenoma-\Delta(\chasenoma,n,\epsilon))}\ .
\end{align}
\end{prop}

\begin{proof}
See Appendix \ref{App:Chase_SINR_nonorthogonal_sumoptimal}.
\end{proof}

We next provide two results on $E_b/N_0$ for CC-NOMA under the sum-rate optimal IBL model.

%Unit channel gains
\begin{cor}
\label{Sensitivity_EbN0_CC_NOMA_sum}
{\bf CC-NOMA under the sum-rate optimal model.} %we next provide two results on 
In the IBL regime for unit channel power gains, the following relations hold in the limit as $\rho\to 0$ for $|H_{tj}|=1$, $\forall$ $t\in\mathcal{T}$, $j\in\mathcal{J}_t$.

(i) \underline{A lower bound on $E_b/N_0$.} 
The SNR per bit 
satisfies 
\begin{align}
\label{Eb_N0_CC_NOMA_sum_LB}
\frac{E_b}{N_0}
\geq -1.59 dB+10\log_{10}J\sigma^2-10\log_{10}\Big(T\Big[1+\eta^2\Big(\frac{J}{T}-1\Big)^2\Big]\Big) \ .
\end{align}

(ii) \underline{Sensitivity of the $E_b/N_0$ limit versus $J$.}  
The SNR per bit 
in the limit as $\rho\to 0$, approaches
\begin{align}
\label{EbN0_limit_CC_NOMA_sum}
\lim\limits_{\rho\to 0}\,\frac{E_b}{N_0}=\log 2 \cdot P_{tot}\ .
\end{align}

\end{cor}

\begin{proof}
For Part (i) of the corollary, from (\ref{SE_CC_NOMA_sum_fading}) and (\ref{Eb_N0_CC_NOMA_sum}), we have
\begin{align}
\frac{E_b}{N_0}&=J\sigma^2\cdot \frac{(2^{2\SE}-1)}{2 \SE} \cdot{1}\Big/{\Big(T\Big[1+\eta^2\Big(\frac{J}{T}-1\Big)^2\Big]\Big)}\nonumber\\
&\geq -1.59 dB+10\log_{10}J\sigma^2-10\log_{10}\Big(T\Big[1+\eta^2\Big(\frac{J}{T}-1\Big)^2\Big]\Big) \ , \nonumber
\end{align}
where the inequality is due to $\frac{2^{2\SE}-1}{2 \SE}\geq -1.59 dB$ as $\SE\to 0$. 

For Part (ii), taking the limit of (\ref{Eb_N0_CC_NOMA_sum}) as $\rho\to 0$, or as $J\to\infty$ for a given finite $P_{tot}$, we obtain
\begin{align}
\lim\limits_{J\to\infty}\frac{E_b}{N_0}&=\lim\limits_{J\to\infty}\log 2 \cdot{\sigma^2\rho\Big(1+\rho T\Big[1+\eta^2\Big(\frac{J}{T}-1\Big)^2\Big]\Big)}\Big/{\Big(\rho T\eta^2 2\Big(\frac{J}{T}-1\Big)\frac{1}{T}\Big)}\nonumber\\
&=\lim\limits_{J\to\infty}\log 2\cdot \sigma^2\rho T \Big(\frac{J}{T}-1\Big)
\ ,  \nonumber
\end{align}
where the first step follows from L'H\^{o}pital's rule, and the last step 
from $P_{tot}=J\sigma^2\rho$.
\end{proof}

Cor. \ref{Sensitivity_EbN0_CC_NOMA_sum} (Part (ii)) implies that if $P_{tot}=J\sigma^2\rho$ scales by a factor of $A$, then the SE curve for the sum-rate optimal model moves to the left by $10\log_{10}A$ dB, as indicated in Section \ref{sect:numericalresults} (see %e.g., 
Figure \ref{fig:SE_EbN0_eta_J_varying}).

\begin{prop}
\label{Chase_SINR_nonorthogonal_TIN}
{\bf TIN model --- Chase combining of non-orthogonal transmissions.} 
The SE for CC-NOMA with TIN incorporating channel power gains is 
\begin{align}
\label{SE_CC_NOMA_TIN_fading}
\chasetin=\frac{1}{T}\sum\limits_{t\in\mathcal{T}}\sum\limits_{j\in\mathcal{J}_t}\frac{1}{2}\log_2\Big(1+\frac{\rho\big(\sum\limits_{t=1}^T  |H_{tj}|^2  \big)^2}{\sum\limits_{t=1}^T |H_{tj}|^2+\rho\eta^2 \big|\sum\limits_{t=1}^T \sum\limits_{j'\in \mathcal{S}_{t,-j}}  H_{tj}H^*_{tj'}\big|^2}\Big) \,\, bit/rdof \ ,      
\end{align}
where $|H_{tj}|^2$ is the channel power gain of user $j\in \mathcal{J}$ (the one with the largest SINR) at slot $t\in\mathcal{T}$.

The SNR per bit of CC-NOMA under TIN, using $\chasetin$ in (\ref{SE_CC_NOMA_TIN_fading}) is given as
\begin{align}
\label{EbN0_TIN}
 \frac{E_b}{N_0}
 =\frac{J\sigma^2\rho n}{2n(\chasetin-\Delta(\chasetin,n,\epsilon))} \ .
\end{align}
\end{prop}

\begin{proof}
See Appendix \ref{App:Chase_SINR_nonorthogonal_TIN}.
\end{proof}

We next provide two lower bounds on $E_b/N_0$ for CC-NOMA under TIN for the IBL model. 

%Unit channel gains
\begin{cor}
\label{Sensitivity_EbN0_CC_NOMA_TIN}
{\bf CC-NOMA under TIN.} In the IBL regime, the followings hold in the limit as $\rho\to 0$.

(i) \underline{A lower bound on $E_b/N_0$.} 
%In the limit as $\rho\to 0$, 
The SNR per bit for $|H_{tj}|=1$, $\forall$ $t\in\mathcal{T}$, $j\in\mathcal{J}_t$ %of CC-NOMA under TIN 
satisfies 
\begin{align}
\label{EbN0_TIN_LB}
 \frac{E_b}{N_0}\geq-1.59 dB+10\log_{10}\sigma^2 \ .
\end{align}

(ii) \underline{Sensitivity of $E_b/N_0$ limit versus $J$ for CC-NOMA under TIN.} %$\chasetin$
For a given finite $J$, the SNR per bit for $|H_{tj}|=1$, $\forall$ $t\in\mathcal{T}$, $j\in\mathcal{J}_t$  
approaches the following lower bound: 
\begin{align}
\label{EbN0_for_high_J_limit_CC_NOMA_TIN}
\lim\limits_{\rho\to 0}\,\frac{E_b}{N_0}=\log 2\cdot \sigma^2\ .
\end{align} 
\end{cor}

\begin{proof}
For Part (i) of the corollary, from (\ref{SE_CC_NOMA_TIN_fading}) %(\ref{SE_TIN}) 
and (\ref{EbN0_TIN}), we have
\begin{align}
 \frac{E_b}{N_0}={J\sigma^2\rho n}\Big/{\Big(n\frac{J}{T}\log_2\Big(1+\frac{\rho T^2}{T+\rho\eta^2(J-T)^2}\Big)\Big)}
 =\frac{J\sigma^2}{2\SE}\cdot \frac{\frac{1}{T}(2^{\frac{2T}{J} SE}-1)}{1-\eta^2\big(\frac{J}{T}-1\big)^2 (2^{\frac{2T}{J} \SE}-1)} 
 \ ,\nonumber
\end{align}
where the last step follows from using $\frac{2^{2\SE}-1}{2 \SE}\geq 10\log_{10}(\log 2)=-1.59 dB$ as $\SE\to 0$.

For Part (ii), taking the limit of (\ref{EbN0_TIN}) as $\rho\to 0$, and incorporating that $P_{tot}=J\sigma^2\rho$, we obtain
\begin{align}
\label{EbN0_limit_CC_NOMA_TIN_as_function_of_Ptot}
\lim\limits_{\rho\to 0}\,\frac{E_b}{N_0}&=\lim\limits_{\rho\to 0}\,{J\sigma^2\rho n}\Big/{\Big(n\frac{J}{T}\log_2\Big(1+\frac{\rho T^2}{T+\rho\eta^2(J-T)^2}\Big)\Big)}\nonumber\\  
&=\lim\limits_{\rho\to 0}\,\log 2\cdot {T\sigma^2\rho}\Big/{\Big(\frac{\rho T^2}{T+\rho\eta^2\big(\frac{P_{tot}}{\sigma^2\rho}-T\big)^2}\Big)} 
=\lim\limits_{\rho\to 0}\,\log 2\cdot \frac{\sigma^2}{T}\Big(T+\rho\eta^2\Big(\frac{P_{tot}}{\sigma^2\rho}-T\Big)^2\Big) \ .
\end{align}
For a given finite $J$, in the limit as $\rho\to 0$, $E_b/N_0$ goes to $\log 2\cdot \sigma^2$, where $P_{tot}$ goes to $0$.
\end{proof}

From (\ref{EbN0_limit_CC_NOMA_TIN_as_function_of_Ptot}), for the IBL model, under a given finite $P_{tot}$,  
the value of $\rho$ is inversely proportional to  
$J$, and when $P_{tot}$ is held fixed, the $E_b/N_0$ limit scales with $\rho^{-1}$. When $P_{tot}$ scales with $J$, Cor.~\ref{Sensitivity_EbN0_CC_NOMA_TIN} implies that the SNR per bit limit as ${\rm SE}\to 0$ for CC-NOMA under TIN is not sensitive to  
$J$ and $T$, whereas from (\ref{SE_CC_NOMA_TIN_fading}),  
SE for a given $E_b/N_0$ improves with $T$ and a lower $E_b/N_0$ is indeed achievable.

%%%%
\subsection{Chase Combining with OMA-based Retransmissions}
\label{chase_orthogonal_retransmissions}

In this model, the retransmissions of each user are combined to enhance its received SNR. This scheme is a simplified version of CC-NOMA where the users have orthogonal messages, namely OMA with Chase combining or CC-OMA, which was introduced in \cite{malak2018throughput}. We next provide its SE. 

\begin{prop}
\label{Chase_SINR_orthogonal_sumoptimal}%TWC model with Howard
{\bf Sum-rate optimal model --- Chase combining of orthogonal transmissions.} 
The SE of CC-OMA for the sum-rate optimal model {\em incorporating channel power gains} is 
\begin{align}
\label{SE_OMA_fading}
\chaseoma=\frac{1}{2}\log_2\Big(1+\rho \sum\limits_{t\in\mathcal{T}} |H_{tj}|^2\Big[1+\Big(\sum\limits_{t\in\mathcal{T}} |H_{tj}|^2\Big)^{-2} \Big(\sum\limits_{t\in\mathcal{T}} \sum\limits_{j'\in \mathcal{S}_{t,-j}} H_{tj} H^*_{tj'} \Big)\Big]\Big) \,\, bit/rdof \ ,
\end{align}
where user $j$ is the one with the largest SINR.

The SNR per bit of CC-OMA for the sum-rate optimal model, using $\chaseoma$ in (\ref{SE_OMA_fading}) is given as
\begin{align}
\label{EbN0_CC_OMA_sum}
 \frac{E_b}{N_0}%=\frac{J\sigma^2\rho n}{n\log_2\Big(1+\rho T\Big[1+\frac{1}{T}\Big(\frac{J}{T}-1\Big)\Big]\Big)} \ .\\
 =\frac{\Jomaslot\sigma^2\rho n}{2n(\chaseoma-\Delta(\chaseoma,n,\epsilon))} \ .
\end{align}
\end{prop}

\begin{proof}
See Appendix \ref{App:Chase_SINR_orthogonal_sumoptimal}.
\end{proof}

\begin{prop}
\label{Chase_SINR_orthogonal_TIN}
{\bf TIN model --- Chase combining of orthogonal transmissions.} 
The SE of CC-OMA with TIN incorporating channel power gains is given as 
\begin{align}
\label{SE_CC_OMA_TIN_fading}
\chasetinort = \frac{1}{T}\sum\limits_{t\in\mathcal{T}}\sum\limits_{j\in\mathcal{J}_{t}}\frac{1}{2}\log_2\Big(1+\frac{\rho \sum\limits_{t\in\mathcal{T}}|H_{tj}|^2}{1+\rho\sum\limits_{t\in\mathcal{T}}  \sum\limits_{j'\in \mathcal{S}_{t,-j}} |H_{tj}|^2   |H_{tj'}|^2 / \sum\limits_{t\in\mathcal{T}} |H_{tj}|^2  }\Big) \,\, bit/rdof \ . 
\end{align}

The SNR per bit of CC-OMA with TIN, using $\chasetinort$ given in (\ref{SE_CC_OMA_TIN_fading})  is given as
\begin{align}\frac{E_b}{N_0}=\frac{\Jomaslot\sigma^2\rho n}{2n(\chasetinort-\Delta(\chasetinort,n,\epsilon))} \ .
\end{align}
\end{prop}

\begin{proof}
See Appendix \ref{App:Chase_SINR_orthogonal_TIN}.
\end{proof}

In the limit as $J\to \infty$, it holds that $\chasetinort\leq \frac{T}{2\log 2}$. %This follows from using (\ref{SE_CC_OMA_TIN}) and taking the limit as $J\to \infty$ and letting $P_{tot}=J\sigma^2\rho$ be fixed.
The subsequent result follows using the definition (\ref{EbN0_defn}) for SNR per bit and the SE given in (\ref{SE_CC_OMA_TIN_fading}) for CC-OMA under the TIN model.

\begin{cor}
\label{cor:CC_OMA_TIN}
{\bf CC-OMA under TIN.} At IBLs, when $|H_{tj}|=1$ for all $t\in\mathcal{T}$, $j\in\mathcal{J}_t$, the SNR per bit $E_b/N_0$ of $\chasetinort$ satisfies the lower bound given as
\begin{align}
 \frac{E_b}{N_0}\geq -1.59 dB+10\log_{10}\sigma^2 \ .
\end{align}
\end{cor}

\begin{proof}
The SNR per bit of $\chasetinort$ is given as
\begin{align}
\label{EbN0_CC_OMA_TIN}
 \frac{E_b}{N_0}&=\frac{\Jomaslot\sigma^2\rho n}{2n \chasetinort}={\Jomaslot\sigma^2\rho n}\Big/{\Big(n\frac{J}{T}\log_2\Big(1+\frac{\rho T}{1+\rho \big(\frac{J}{T}-1\big)}\Big)\Big)}\nonumber\\
 &=\frac{\Jomaslot\sigma^2}{2\SE}\cdot {\frac{1}{T}(2^{\SE\frac{2T}{J}}-1)}\Big/{\Big(1- \frac{1}{T}\big(\frac{J}{T}-1\big)(2^{\SE\frac{2T}{J}}-1)\Big)}%\nonumber\\
 %&=\sigma^2\frac{(2^{SE\frac{2T}{J}}-1)}{1- \frac{1}{T}\big(\frac{J}{T}-1\big)(2^{SE\frac{2T}{J}}-1)}\frac{1}{\frac{2T}{J}SE}\nonumber\\
 %&\geq 10\log_{10}(\log 2)+10\log_{10}\sigma^2,
 \geq -1.59 dB+10\log_{10}\sigma^2 \ ,
 \end{align}
 where the second equality follows from using (\ref{SE_CC_OMA_TIN_fading}) which yields $\rho=\frac{(2^{\SE\frac{2T}{J}}-1)}{T- \big(\frac{J}{T}-1\big)(2^{\SE\frac{2T}{J}}-1)}$, and the last step follows from the same intuition as in (\ref{Eb_N0_CC_NOMA_sum_LB}) and (\ref{EbN0_TIN}).
\end{proof}

From Cor. \ref{cor:CC_OMA_TIN}, it is clear that the SNR per bit limit as ${\rm SE}\to 0$ for CC-OMA under TIN is not sensitive to the parameters $J$ or $T$. We refer the reader to Section \ref{sect:numericalresults} (see e.g., Figures \ref{fig:SE_EbN0_eta1_J10} and \ref{fig:SE_EbN0_eta_J_varying}).

%%%%
\subsection{Incremental Redundancy with OMA-based Retransmissions}
\label{IR_retransmissions}

%The outage characterization is inspired from: https://arxiv.org/pdf/1501.06166.pdf
%the probability of an unsuccessful transmission up to the T-th retransmission (the failure probability) can be characterized, and also for the finite blocklength regime using the Gaussian approximation in \cite{devassy2014finite}.

We next consider an incremental redundancy model with OMA  (IR-OMA). 
From Sections~\ref{chase_retransmissions} and \ref{chase_orthogonal_retransmissions}, due to the finite HARQ buffer size, the throughput of CC-NOMA is determined by the addition of all active users' signals at any given time slot. %Hence, a finite HARQ buffer has an impact on the throughput of HARQ. 
Unlike for CC, where the buffer size is the same as the number of packets per transmission, in IR-OMA, also known as HARQ Type~III, the buffer size is equal to the number of coded bits of the total transmitted coded packets, where each retransmitted packet is self-decodable, and contains different information than the previous one.

\begin{figure*}[t!]
\begin{center}
\includegraphics[width=0.95\textwidth]{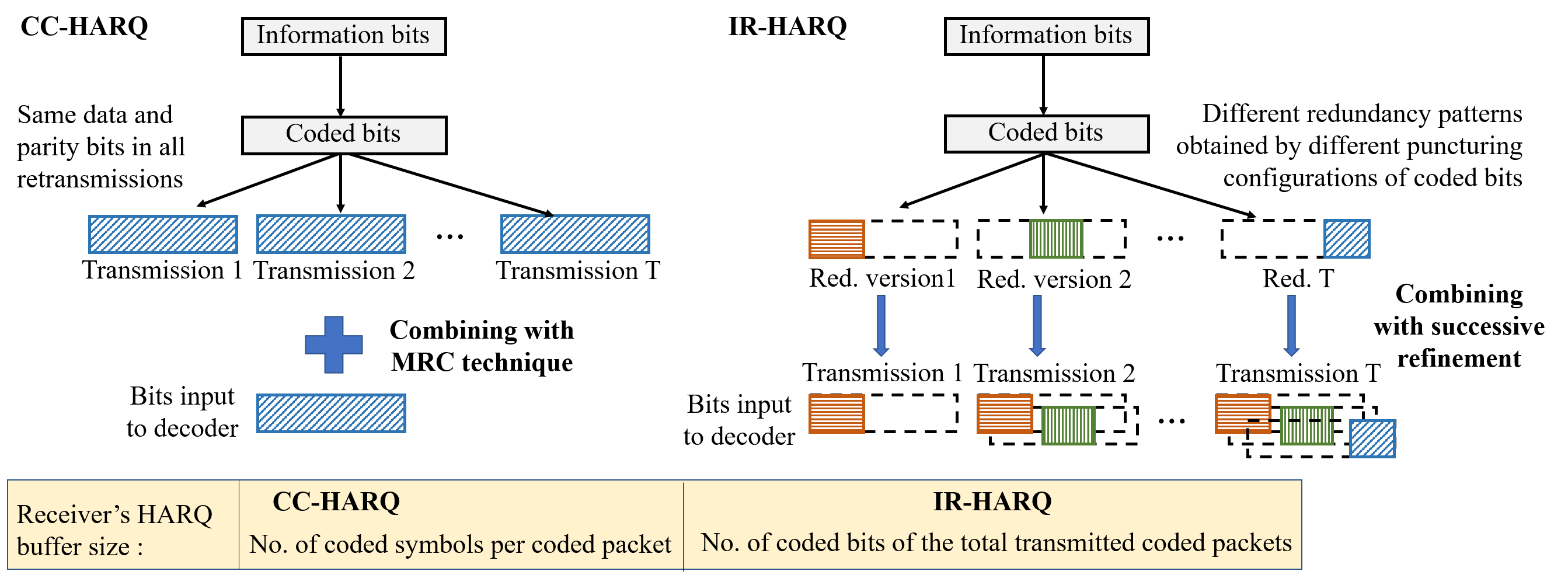}
\caption{\small{(Left) CC-HARQ, where the retransmissions contain the same data and parity bits, which are summed at the receiver prior to decoding. (Right) IR-HARQ, where each retransmission provides some additional bits, and is self-decodable.}}
\label{fig:CC-HARQ}
\end{center}
\end{figure*}

In IR-HARQ, each transmission consists of new redundancy bits from the channel encoder, which enables IR-OMA to achieve a superior performance over CC-OMA. However, IR has higher complexity due to additional signaling of retransmission numbers and a larger receiver buffer size \cite{frenger2001performance}.  %https://citeseerx.ist.psu.edu/document?repid=rep1&type=pdf&doi=0fca7038e12142b9bae9ab8bcc648507fb3d884f
IR is better suited for highly time-varying channels where the rate of the error control code is adapted to the current channel state. In IR-HARQ, multiple distinct sets of code bits are generated for the same information bits used in a packet, and %. Each set of coded bits is 
transmitted under different channel conditions. These sets consist of distinct redundancy patterns  
obtained by different puncturing configurations of a common code. The rate adaptation achievable via puncturing reduces the decoder complexity. The transmitter and receiver only share a series of puncturing tables to specify which code bits are to be transmitted for a specific code rate \cite{uhlemann2006puncturing}. The receiver then simply inserts erasures for all code bits that are not received. %Rate-compatible 
Punctured turbo codes are used for unequal error protection \cite{barbulescu1995rate} and IR-HARQ, % schemes, 
e.g., \cite{jung1997pragmatic,li1997performance,rowitch1997rate}. %cut sentence before
In Figure \ref{fig:CC-HARQ} (right), we sketch IR-HARQ, where each retransmitted packet provides some additional information bits, and is self-decodable, i.e., it provides successive refinement \cite{equitz1991successive}  
by iteratively improving the rate-distortion tradeoff as more information is transmitted. 
Since we focus on the fixed-access strategy, the analysis of IR-NOMA would be relatively simple. Due to limited space, we only detail IR-OMA here.

\paragraph{Expected quantization distortion} Using the refinement-based approach in \cite{equitz1991successive}, the average quantization distortion is characterized as the mean squared error distortion between the quantized signal ${\bf \hat{Y}}_{t,T}$ and the received signal ${\bf Y}_t$. 
The quantized $m$ dimensional signals are given by ${\bf \hat{Y}}_{t,T}={\bf Y}_t+{\bf Q}_{t,T}$. The quantization noise satisfies ${\bf Q}_{t,T}\sim\mathcal{CN}\big(0,\frac{2\sigma_{q}^2(t,T)}{m} I_m\big)$, where $\sigma_{q}^2(t,T)$ represents the total quantization noise power per rdof (the quantization distortion per frequency bin is $\sigma_{q}^2(t,T)/B$) for IR-OMA at slot $t$ given a total number of $T$ retransmissions, where attempt $t$ is unsuccessful if $1\leq t<T$ and the retransmission is successful at attempt $T$. From (\ref{retransmissions_combining_chase}), ${\bf Y}_t$ has a dimension $m=n/T$.

\begin{prop}
\label{IR_SINR}%The sum rate of IR-OMA
{\bf Sum-rate optimal model --- Incremental redundancy of orthogonal transmissions.} The SE of IR-OMA for the sum-rate optimal model {\em incorporating channel power gains} is given as
\begin{align}
\label{SE_IR_sum}
\IRoma
=\sum\limits_{t=1}^{T}\frac{B}{2}\log_2\Big(1+\frac{\rho \sum\limits_{j\in\mathcal{J}_{t}}|H_{tj}|^2/B}{1+\sigma_{q}^2(t,T-1)/(Bm\sigma^2)}\Big)\,\, bit/rdof/(T\mbox{ slots}) \ ,
\end{align}
where the following relation holds between the quantization noise $\sigma_{q}(t,T)$ and the buffer size $\Cb$:
\begin{align}
\label{distortion_IR_sum_optimal}
\sigma_{q}^2(t,T)=\frac{B(J\rho/B+1)m\sigma^2}{2^{\frac{2\Cb}{TB}}-1},\, \quad t<T\ , 
\end{align}
and   
$\sigma_{q}^2(T,T)=0$, i.e., at retransmission $T$, the received signal ${\bf Y}_T={\bf \hat{Y}}_T$, i.e., the receiver recovers ${\bf Y}_T$. 
Furthermore, $\sigma^2_q(t,T-1)$ for $t\leq T-1$ can be derived from (\ref{distortion_IR_sum_optimal}), and  
$\sigma^2_q(T,T-1)=0$.

%%%%%%
The SNR per bit of IR-OMA for sum-rate optimal case, using $\IRoma$ given in (\ref{SE_IR_sum}), is given as
\begin{align}
\label{EbN0_IR_sum_optimal}
\frac{E_b}{N_0}
=\frac{J\sigma^2\rho n}{2n(\IRoma-\Delta(\IRoma,n,\epsilon))/T} \ .
\end{align}
\end{prop}

\begin{proof}
See Appendix \ref{App:IR_SINR}.
\end{proof}

%vector optimization
Note that in Prop. \ref{IR_SINR}, $\log_2\Big(1+\frac{\rho J\sigma^2}{\sigma^2}\Big)\leq \frac{B}{2}\log_2\Big(1+\frac{\rho J\sigma^2/B}{\sigma^2}\Big)$, which follows from employing $\sum\limits_{i=1}^n\log(1+x_i)\leq n\log(1+\frac{1}{n}\sum\limits_{i=1}^n x_i)$ with $x_1=1$ and  
$x_i=0$, $i\neq 1$. 
In this paper, we do not optimize $B$ and the division of total transmit power across $B$ bins, which is left as future work. Instead, in Section \ref{sect:numericalresults}, we assume $B=1$ to provide lower bounds on the performance tradeoffs.

As the buffer size $\Cb\to\infty$, $\frac{E_b}{N_0}\to \frac{J\sigma^2\rho }{\log_2(1+\rho J)}$. Similarly, for smaller $\Cb$, $\frac{E_b}{N_0}>\frac{J\sigma^2\rho }{\log_2(1+\rho J)}$. 
Hence, it is easy to note that as $\Cb$ increases the IR-OMA sum SE matches the sum SE %(black curve) 
for the classical problem without combining transmissions (sum-rate optimal case). However, when $\Cb$ is small the gap between the SE for the classical transmission model and the IR-OMA sum SE grows as $T$ increases.

\begin{prop}
\label{SNR_IR_OMA_TIN}
{\bf TIN model --- Incremental redundancy of orthogonal transmissions.} 
The SE of IR-OMA with TIN {\em incorporating channel power gains} is 
\begin{align}
\label{SE_IR_OMA_TIN_fading}
\IRtinort
=\sum\limits_{t=1}^{T}\sum\limits_{j\in\mathcal{J}_{t}}\frac{B}{2}\log_2\Big(1+\frac{\rho |H_{tj}|^2 /B}{\rho /B \sum\limits_{j'\in \mathcal{S}_{t,-j}}|H_{tj'}|^2+1+\frac{\sigma_{q}^2(t,T-1)}{Bm\sigma^2}}\Big)\,\, bit/rdof/(T\mbox{ slots}) \ ,
\end{align}
where the following relation between the quantization noise $\sigma_{q}(t,T)$ and the buffer size $\Cb$:
\begin{align}
\label{distortion_IR_TIN}
\sigma_{q}^2(t,T)=\frac{B(\rho/B+1)m\sigma^2}{2^{\frac{2\Cb}{TB}}-1}-\Big(\frac{J}{T}-1\Big)\rho m\sigma^2 \ .
\end{align}

The SNR per bit of IR-OMA for TIN, using $\IRtinort$ given in (\ref{SE_IR_OMA_TIN_fading}), is given as
\begin{align}
\label{EbN0_IR_TIN}
\frac{E_b}{N_0}
=\frac{J\sigma^2\rho n}{2\frac{n}{T}(\IRtinort-\Delta(\IRtinort,n,\epsilon))}\ .
\end{align}
\end{prop}

\begin{proof}
See Appendix \ref{App:SNR_IR_OMA_TIN}.
\end{proof}

We next provide 
two results on $E_b/N_0$ for the IBL regime for unit channel power gains.

\begin{table*}[t!]\small%\footnotesize
\setlength{\extrarowheight}{4pt}
\begin{center}
\begin{tabular}{l | l | l}
{\bf SE} & $\frac{\partial }{\partial T}$SE & $\frac{\partial }{\partial \eta}$SE\\ 
\hline
$\clasicalsumoptimal
=\frac{1}{2}\log_2\Big(1+\rho \frac{J}{T}\Big)$ & $<0$ & NA\\ 
\hline
$\clasicaltin
=\frac{J}{2T}\log_2\Big(1+\frac{\rho}{\rho  (\frac{J}{T}-1)+1}\Big)$ & $<0$ & NA\\ 
\hline
$\chasenoma=\frac{1}{2}\log_2\Big(1+\rho T\Big[1+\eta^2\Big(\frac{1}{T}\sum\limits_{t=1}^T J_t-1\Big)^2\Big]\Big)$ & $>0$ & $>0$ 
\\
\hline
$\chasetin=\frac{J}{2T}\log_2\Big(1+\frac{\rho T^2}{T+\rho\eta^2(J-T)^2}\Big)$ & $<0$ at high SE, $>0$ at low SE   & $<0$\\
\hline
$\chaseoma =\frac{1}{2}\log_2\Big(1+\rho T\Big[1+\frac{1}{T}\Big(\frac{J}{T}-1\Big)\Big]\Big)$ & $>0$ at high SE, $<0$ at low SE & NA
\\
\hline
$\chasetinort = \frac{J}{2T}\log_2\Big(1+\frac{\rho T}{1+\rho \big(\frac{J}{T}-1\big)}\Big)$ & $>0$ & NA
\\
\hline
$\IRoma=\frac{B}{2}\log_2(1+\rho J/B)+\sum\limits_{t=1}^{T-1}\frac{B}{2}\log_2\Big(1+\frac{\rho J/B}{1+\sigma_{q}^2(t,T-1)/(Bm\sigma^2)}\Big)$ & $<0$  & NA\\
\hline
$\IRtinort=\frac{JB}{2T}\log_2\Big(1+\frac{\rho/B}{\rho/B(J/T-1)+1}\Big)+\frac{JB(T-1)}{2T}\log_2\Big(1+\frac{\rho/B}{1+\rho/B\zeta_t}\Big)$ & $>0$ at high $\Cb$, $<0$ at low $\Cb$ & NA
\\
\hline
\end{tabular}
\end{center}
\caption{\small{The SE of the different retransmission-based models with combining under unity channel power gain.}}
\label{table:SE}
\end{table*}

%Unit channel gains
\begin{cor}
\label{cor:IR_OMA_TIN}
{\bf IR-OMA under TIN.} At IBLs, the followings hold for $|H_{tj}|=1$, $\forall$ $t\in\mathcal{T}$, $j\in\mathcal{J}_t$.

(i) \underline{A limit on $E_b/N_0$ as $\Cb\to\infty$.}
The SNR per bit  
in the limit as $\Cb\to\infty$, approaches 
\begin{align}
\label{SNR_per_bit_IR_OMA_TIN}
\lim\limits_{\Cb\to\infty}\frac{E_b}{N_0}=\frac{T\sigma^2\rho }{\log_2\big(1+\frac{\rho }{\rho  (J/T-1)+1}\big)}\ ,
\end{align}
which  
is the SNR per bit for the classical transmission model with TIN in (\ref{EbN0_no_retransmissions_TIN}) for the IBL regime. 

(ii) \underline{A limit on $E_b/N_0$ as $\rho\to 0$.} For large buffer sizes $\Cb$, as $\rho\to 0$, it holds 
that
\begin{align}
\label{EbN0_TIN_IR_OMA_limit_J_high}
\lim\limits_{\rho\to 0}\,\frac{E_b}{N_0}\approx 
\log 2 \cdot P_{tot} \ .
\end{align}

\end{cor}

\begin{proof}
For Part (i) of the corollary, from Prop.~\ref{SNR_IR_OMA_TIN}, as $\Cb\to\infty$, $\zeta_t=(J/T-J/(T-1))$ for $t<T$, and $\zeta_T=(J/T-1)$, and the SNR per bit for IR-OMA with TIN approaches 
\begin{align}
\label{SNR_per_bit_IR_OMA_TIN_proof}
\lim\limits_{\Cb\to\infty}\frac{E_b}{N_0}= {T\sigma^2\rho }\Big/{\Big(\frac{1}{T}\sum\limits_{t=1}^{T}\log_2\big(1+\frac{\rho}{\rho\zeta_t+1}\big)\Big)}
\leq {T\sigma^2\rho }\Big/{\log_2\big(1+\frac{\rho }{\rho  (J/T-1)+1}\big)}\ ,
\end{align}
where the inequality in the second step is indeed an equality because $\sigma_{q}^2(t,T)\to 0$ as $\Cb\to\infty$, where (\ref{TIN_Cb}) no longer holds, and the desired bit-error-rate (BER) is met 
from (\ref{SE_IR_OMA_TIN_fading}).

Part (ii) of the corollary is immediate from Prop. \ref{SNR_IR_OMA_TIN}.
\end{proof}

From Cor. \ref{cor:IR_OMA_TIN}, the behavior of SE versus $E_b/N_0$ is highly affected by $\Cb$, and to achieve the same SE, it is required to have a larger $E_b/N_0$ when $\Cb$ is low. In the regime as $\Cb\to\infty$, we can observe from (\ref{SNR_per_bit_IR_OMA_TIN}) that the SNR per bit $\frac{E_b}{N_0}$ for IR-OMA under TIN behaves similarly as the classical transmission model with TIN in (\ref{EbN0_no_retransmissions_TIN}). For smaller $\Cb$, the ratio $\frac{E_b}{N_0}$ is typically higher than the SNR per bit for the classical TIN case. We will illustrate this behavior in Section \ref{sect:numericalresults} (see Figures \ref{fig:SE_EbN0_eta1_J10} and \ref{fig:SE_EbN0_eta_J_varying}).
  
In Table \ref{table:SE}, we summarize the SE of different retransmission-based models (with unit channel power gain), with two additional columns describing the behavior of SE with respect to $T$ and $\eta$. Note also that $\frac{\partial }{\partial \rho}$SE$>0$ for all models. 
For the different HARQ models in hand, next in Section \ref{sect:numericalresults}, we will study the SE versus $E_b/N_0$ and $J/n$ versus $E_b/N_0$ tradeoffs by exploiting the joint behavior of SE, $E_b/N_0$, and $\rho$, for the FBL regime (for fixed $n$ and $\epsilon$).  
For the IBL models, we refer the reader to \cite{MalakThroughputWiopt2022}.

%%%
\section{Numerical Evaluation of Scaling Results}
\label{sect:numericalresults}

\begin{figure}[t!]
\begin{center}
\includegraphics[width=0.32\textwidth]{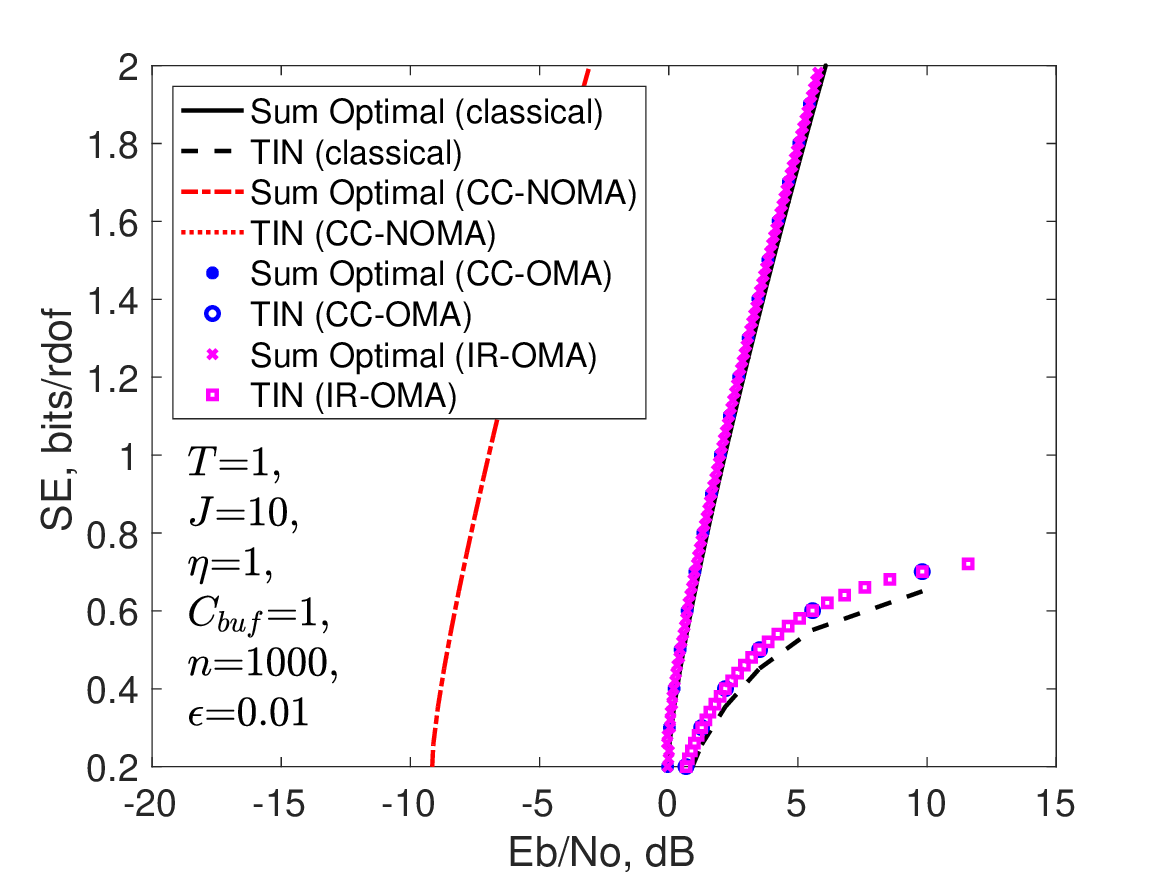}
\includegraphics[width=0.32\textwidth]{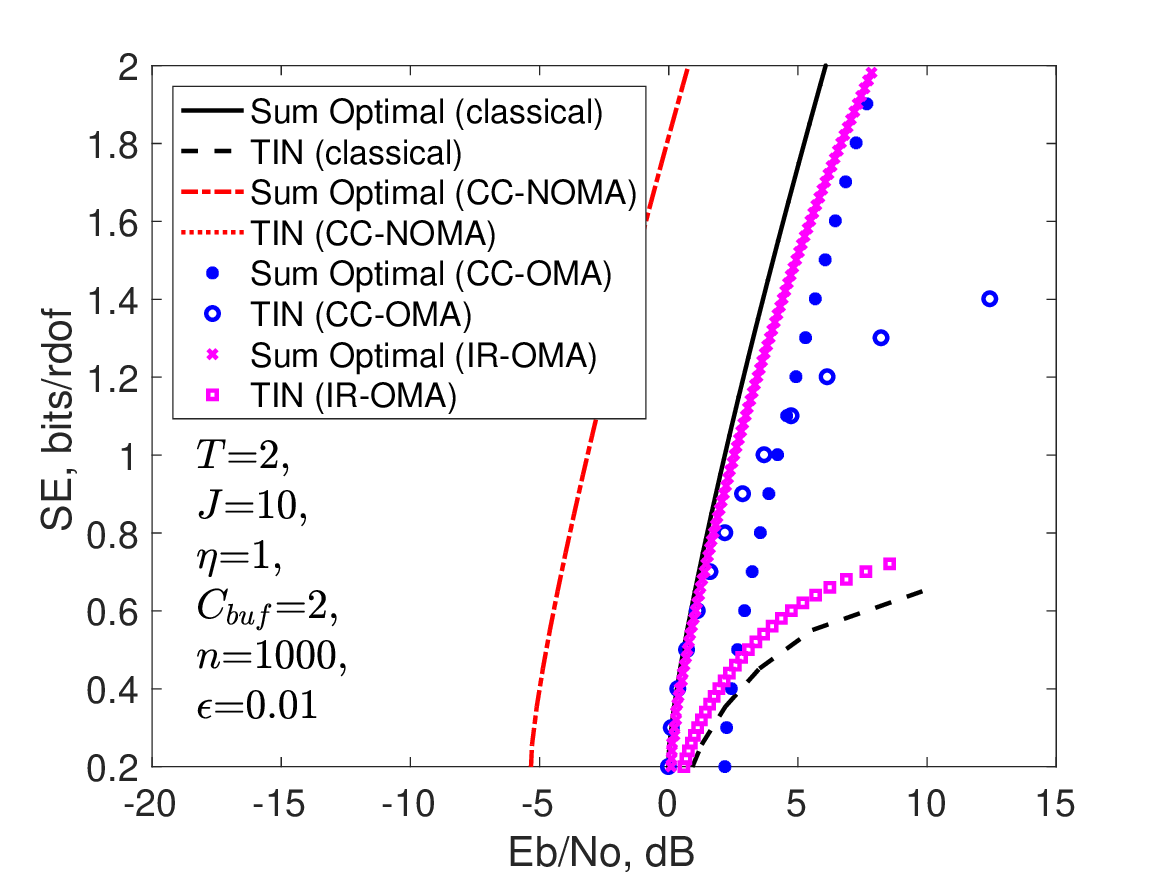}
\includegraphics[width=0.32\textwidth]{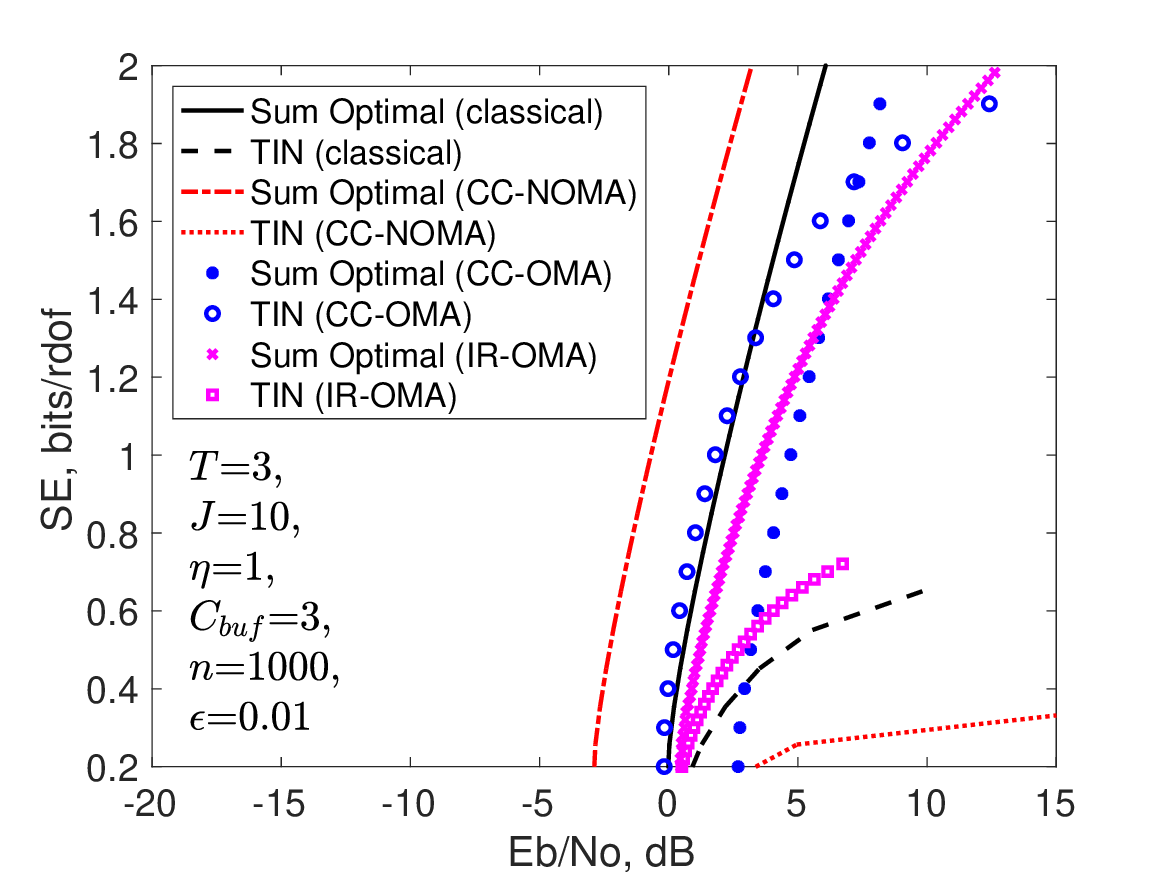}
\includegraphics[width=0.32\textwidth]{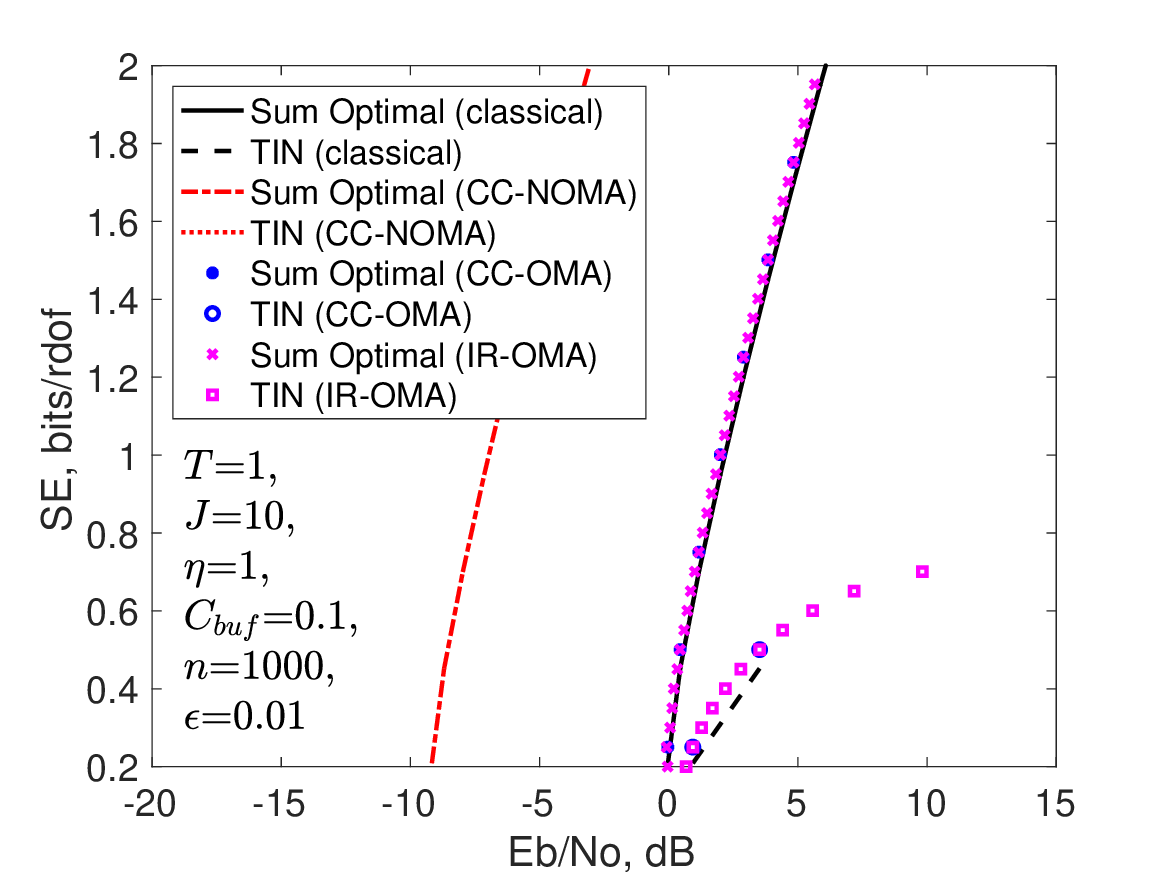}
\includegraphics[width=0.32\textwidth]{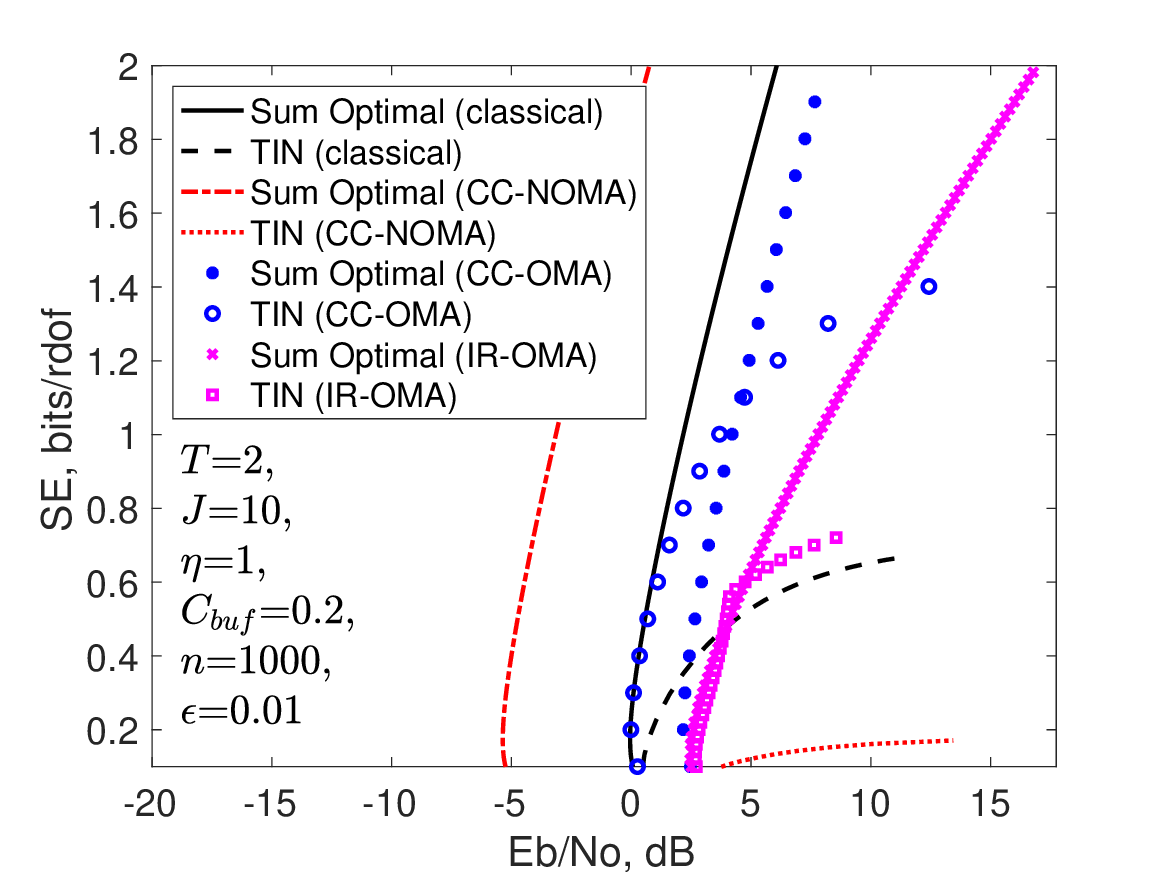}
\includegraphics[width=0.32\textwidth]{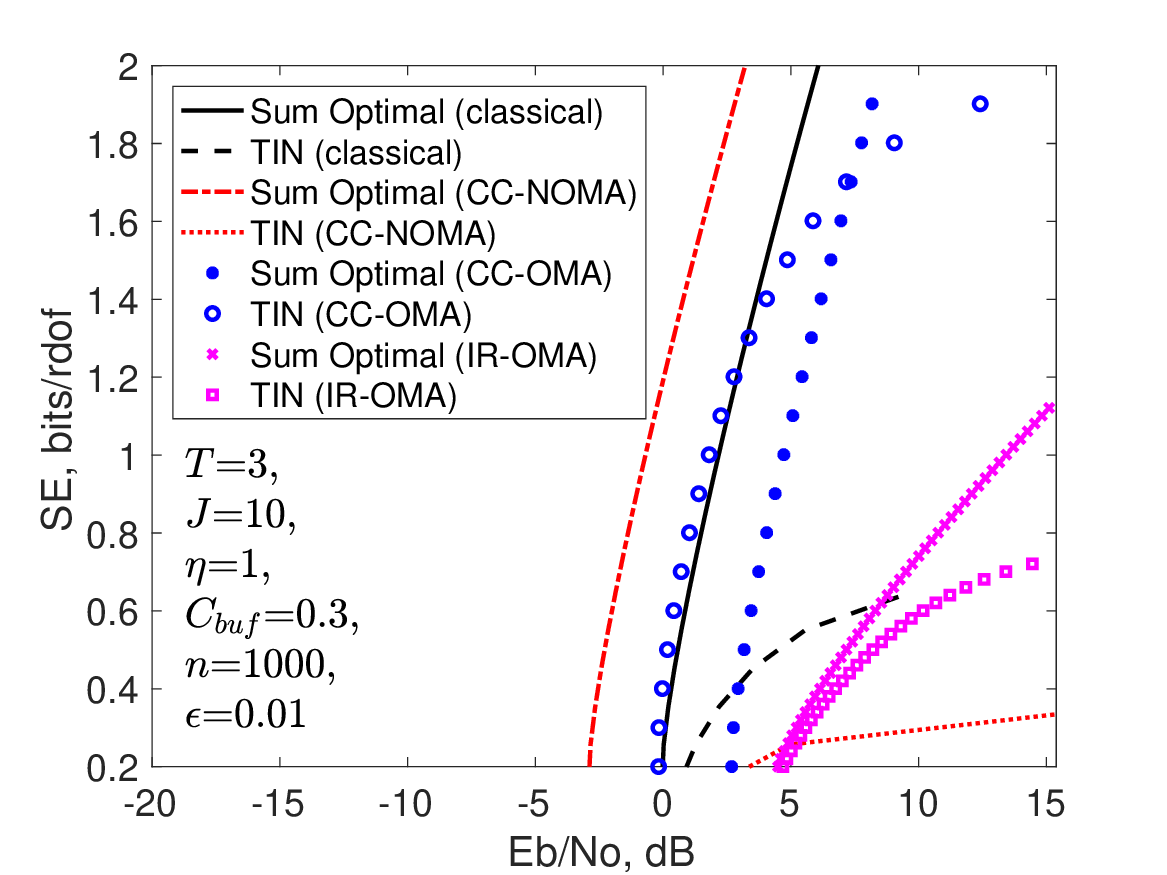}
\end{center}
\caption{\small{Scaling of SE versus $E_b/N_0$ for varying $T$ for $\eta=1$ and $J=10$. %(Row I)  large buffer size, $\Cb=10T$. 
(Row I) moderate buffer size, $\Cb=T$. (Row II) small buffer size, $\Cb=0.1T$.}}
\label{fig:SE_EbN0_eta1_J10}
\end{figure}

\begin{figure}[t!]
\begin{center}
\includegraphics[width=0.32\textwidth]{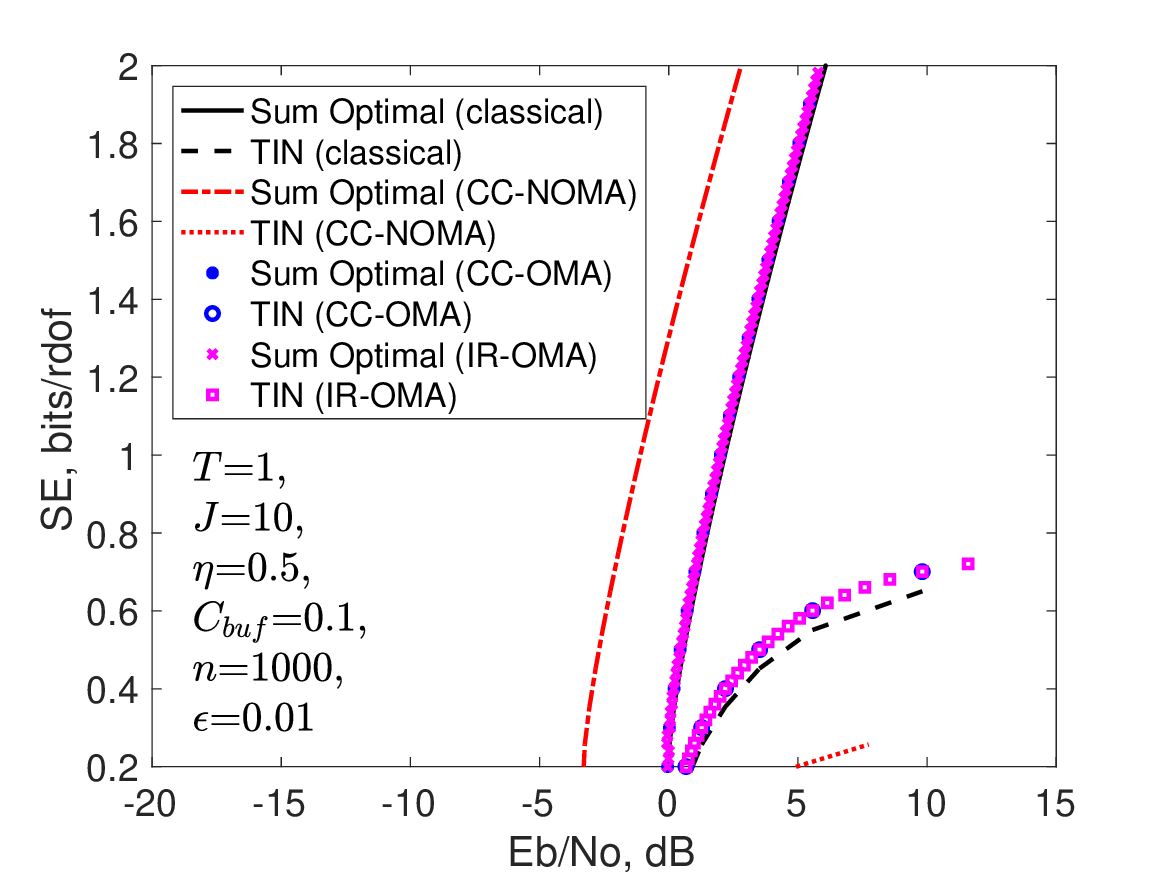}
\includegraphics[width=0.32\textwidth]{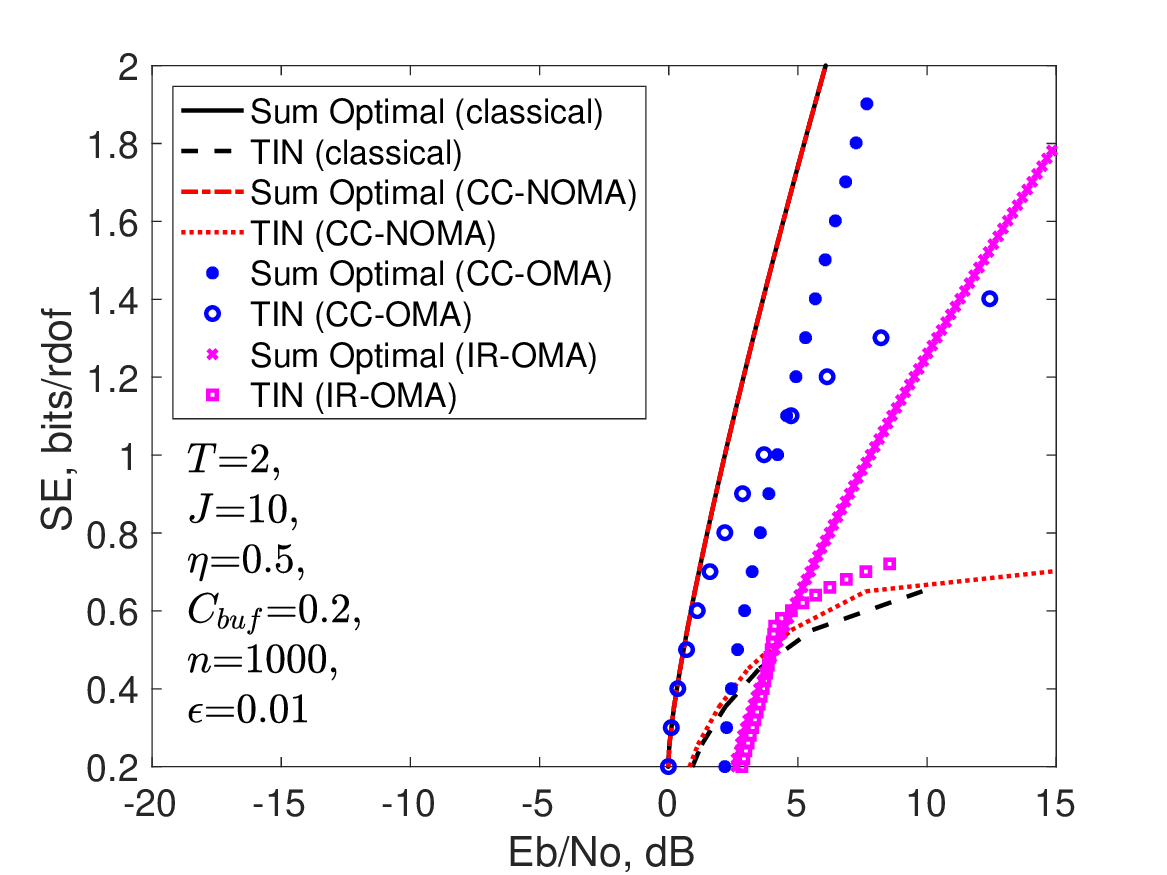}
\includegraphics[width=0.32\textwidth]{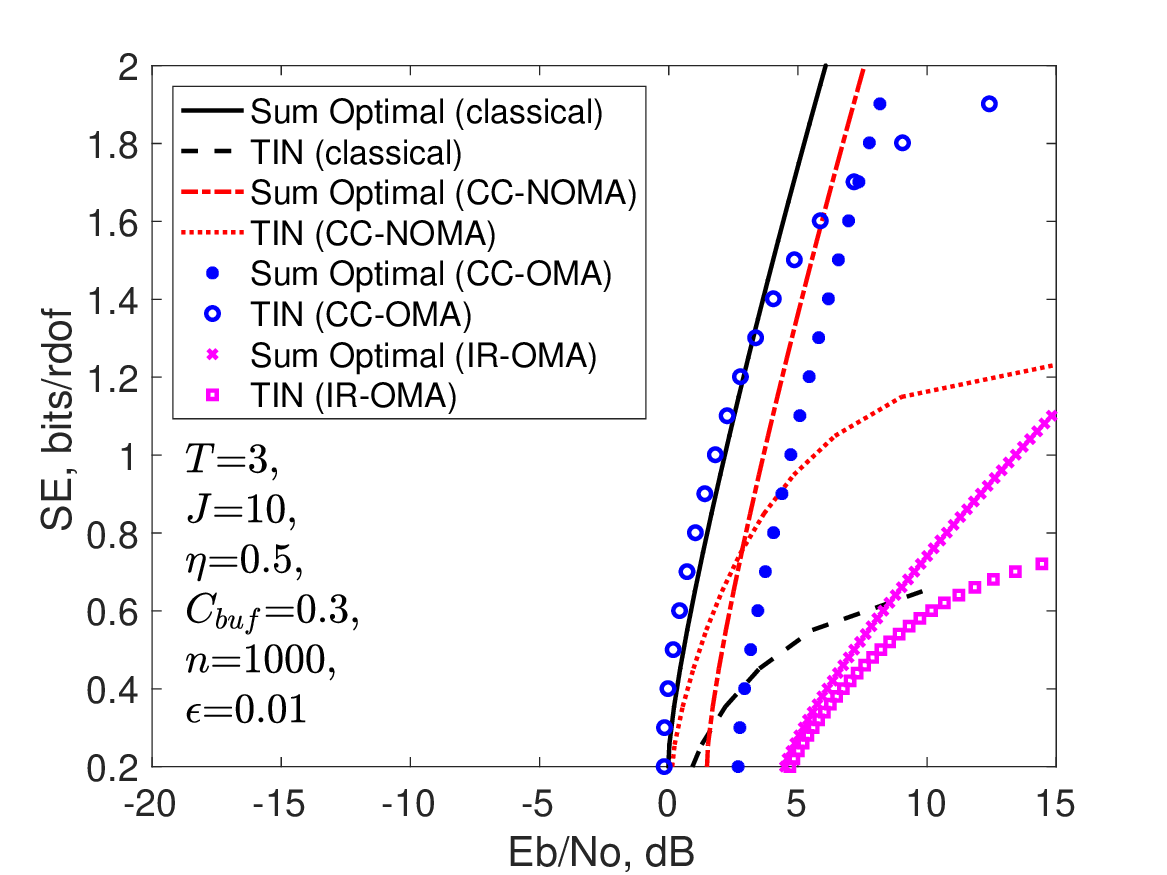}
\includegraphics[width=0.32\textwidth]{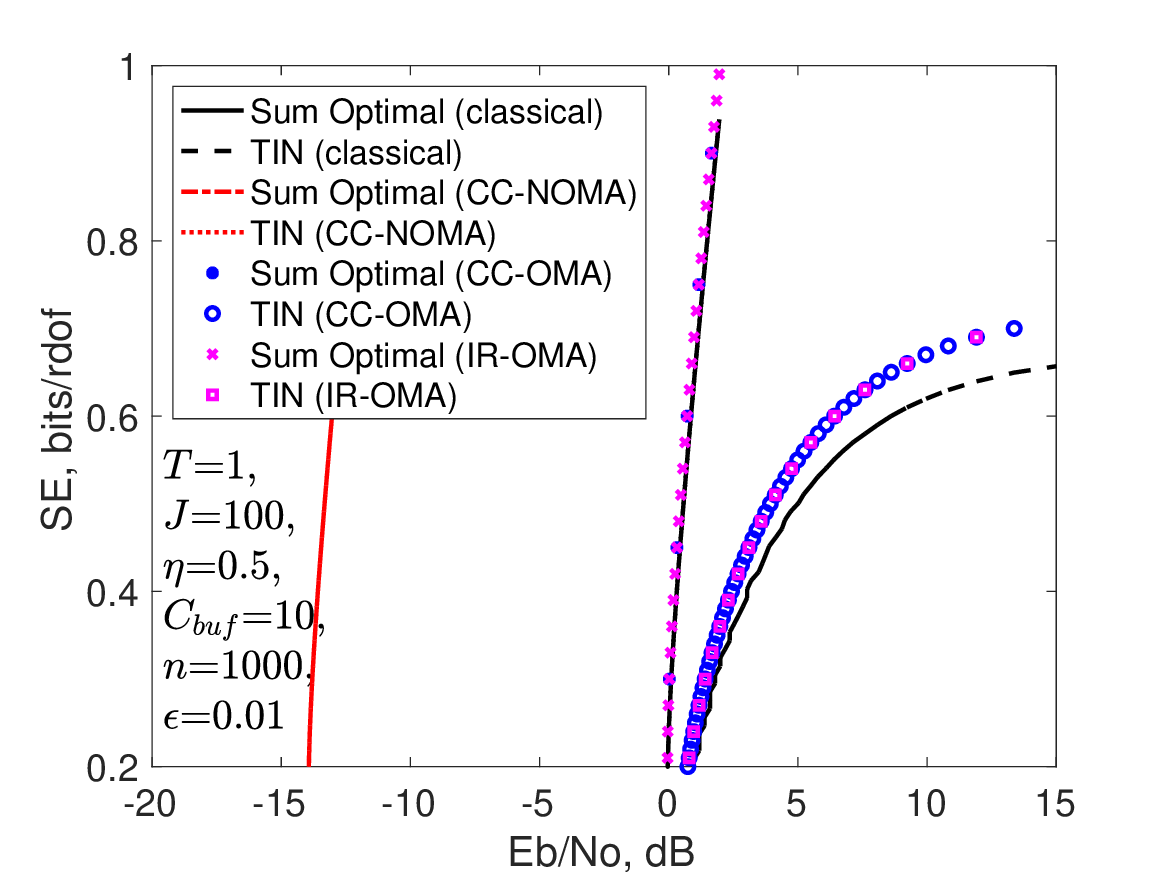}
\includegraphics[width=0.32\textwidth]{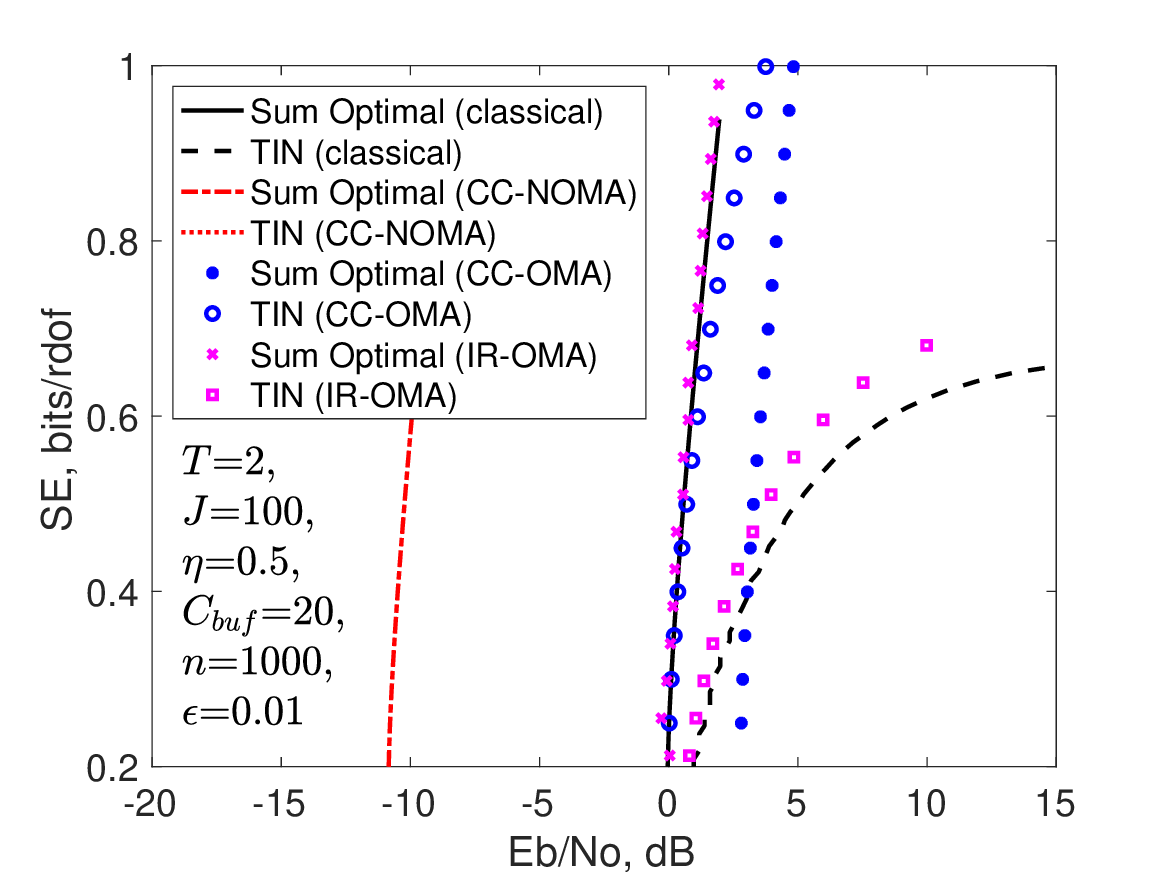}
\includegraphics[width=0.32\textwidth]{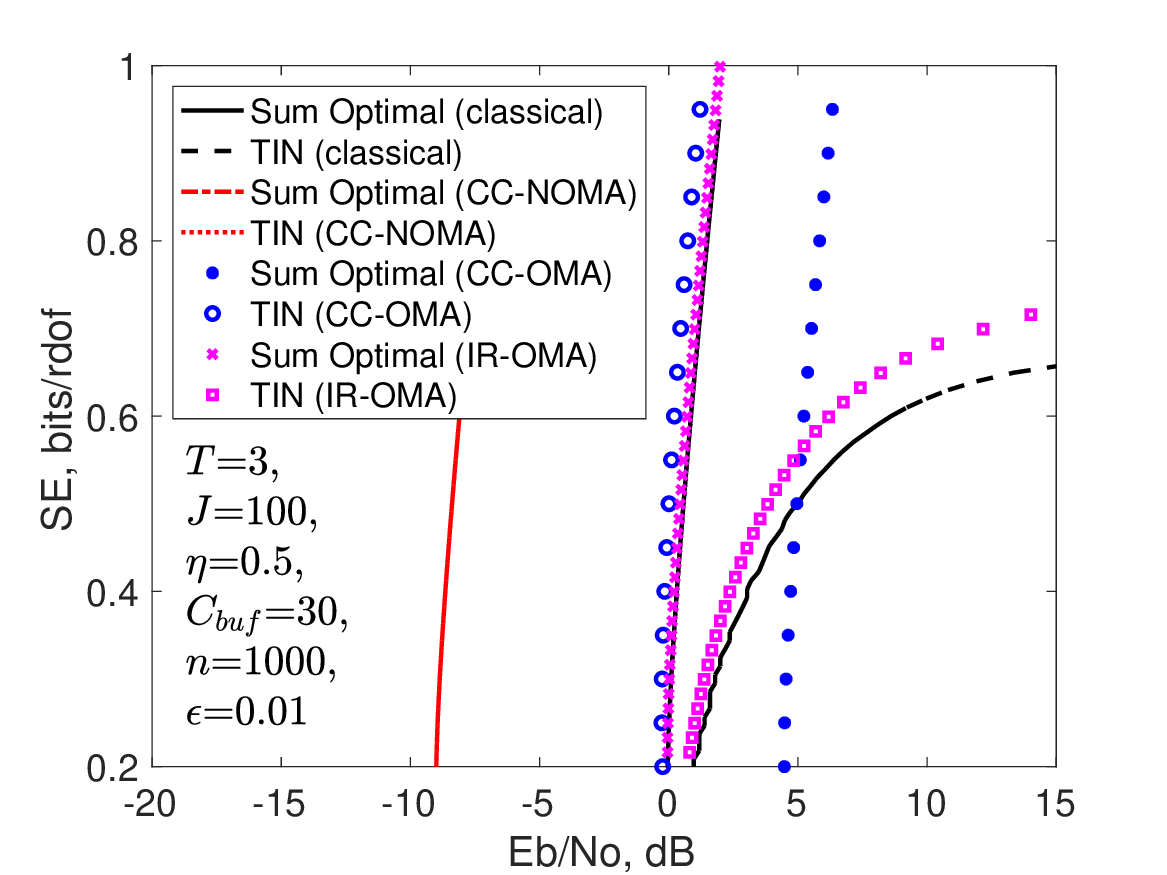}
\end{center}
\caption{\small{Scaling of SE versus $E_b/N_0$. (Row I) $\eta=0.5$, $J=10$, $\Cb=0.1T$. (Row II) $\eta=0.5$, $J=100$, $\Cb=10T$. %(Row III) $\eta=1$, $J=100$, $\Cb=10T$.
}}
\label{fig:SE_EbN0_eta_J_varying}
\end{figure}

\begin{figure}[t!]
\begin{center}
\includegraphics[width=0.32\textwidth]{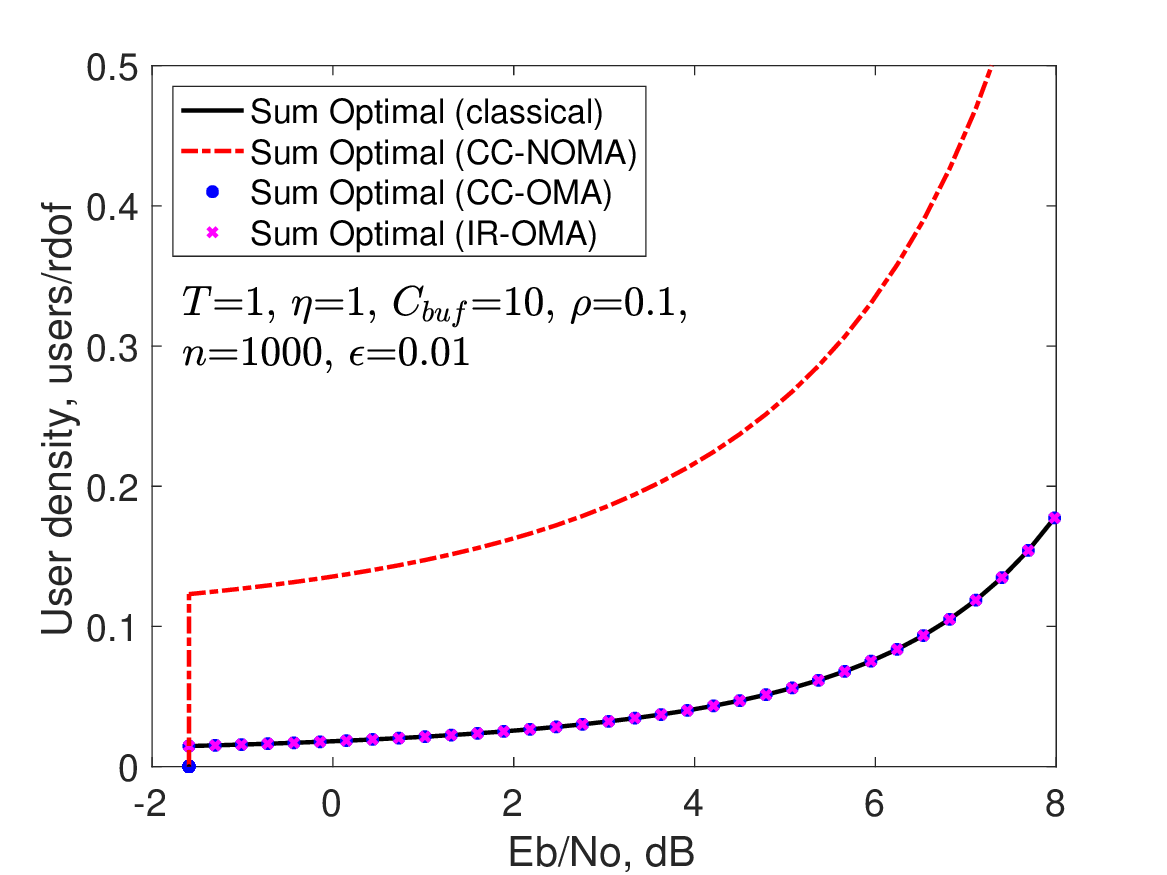}
\includegraphics[width=0.32\textwidth]{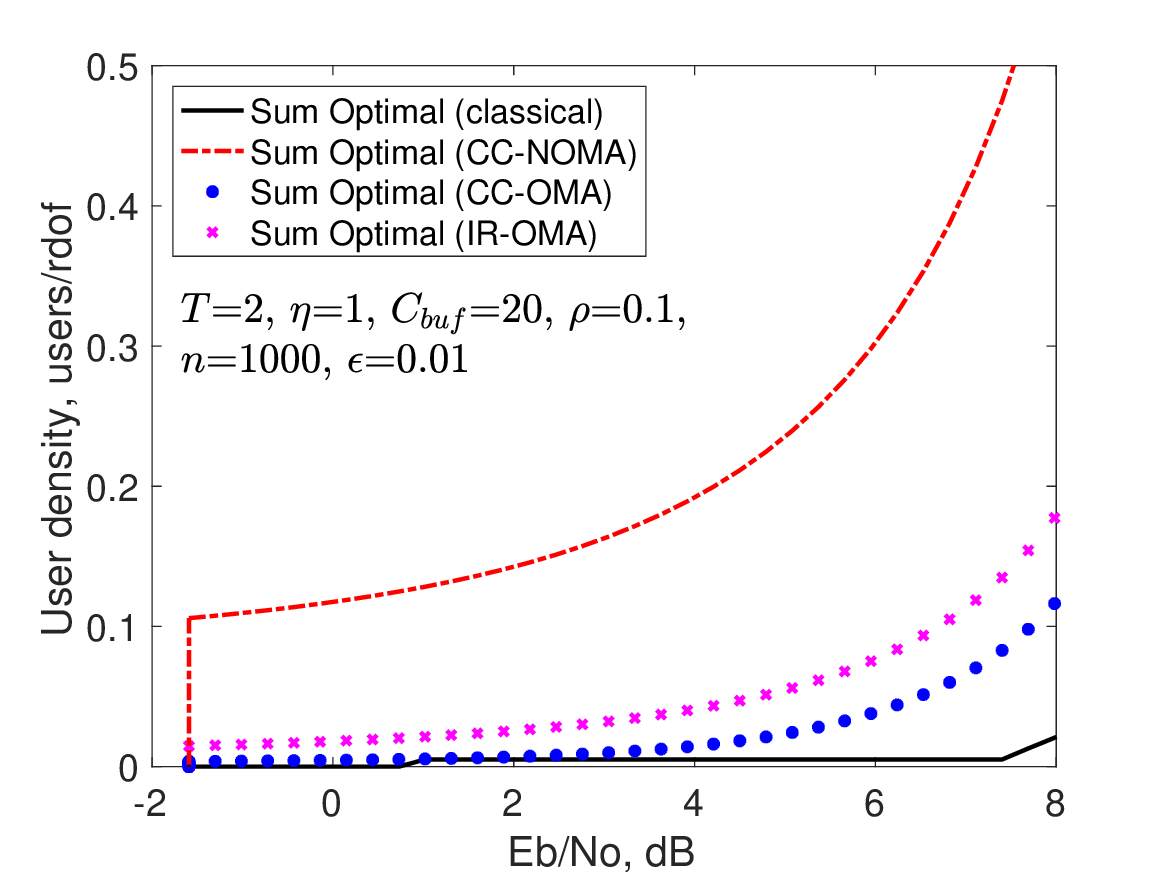}
\includegraphics[width=0.32\textwidth]{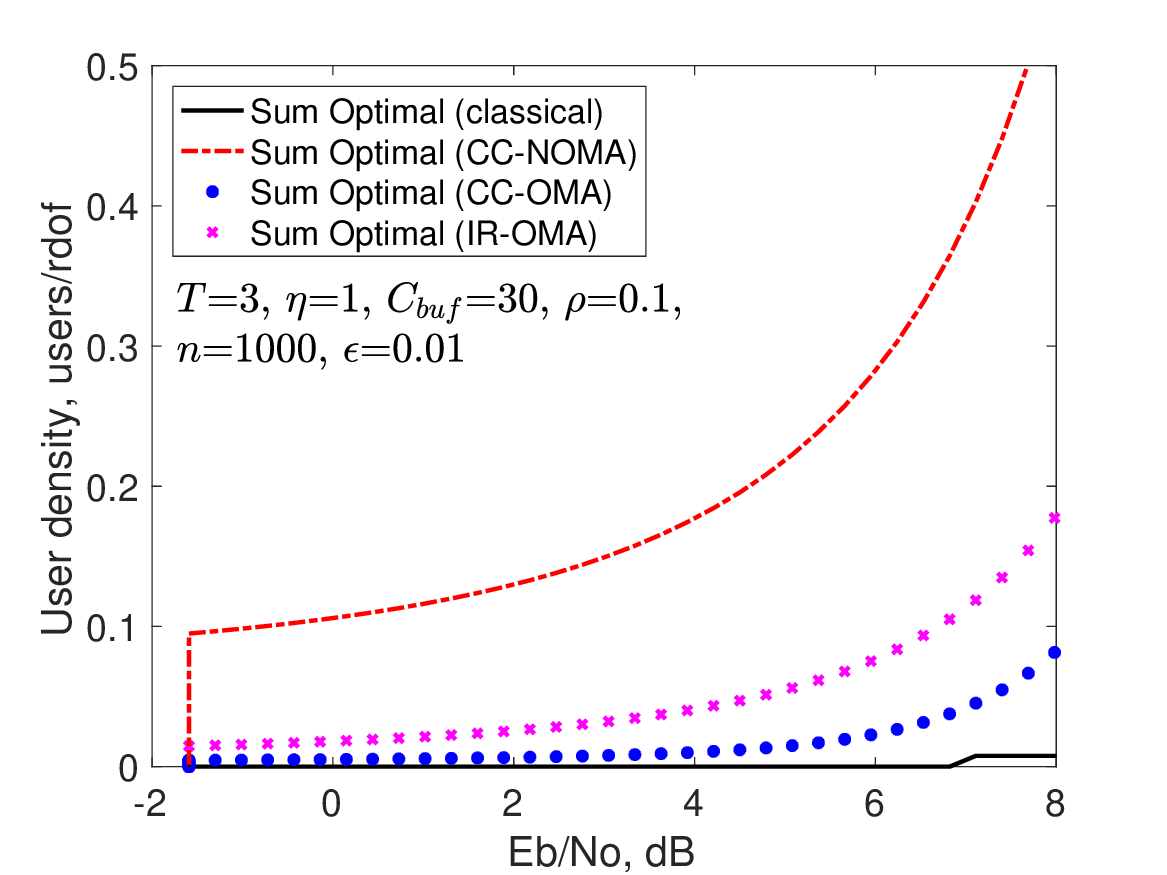}
\includegraphics[width=0.32\textwidth]{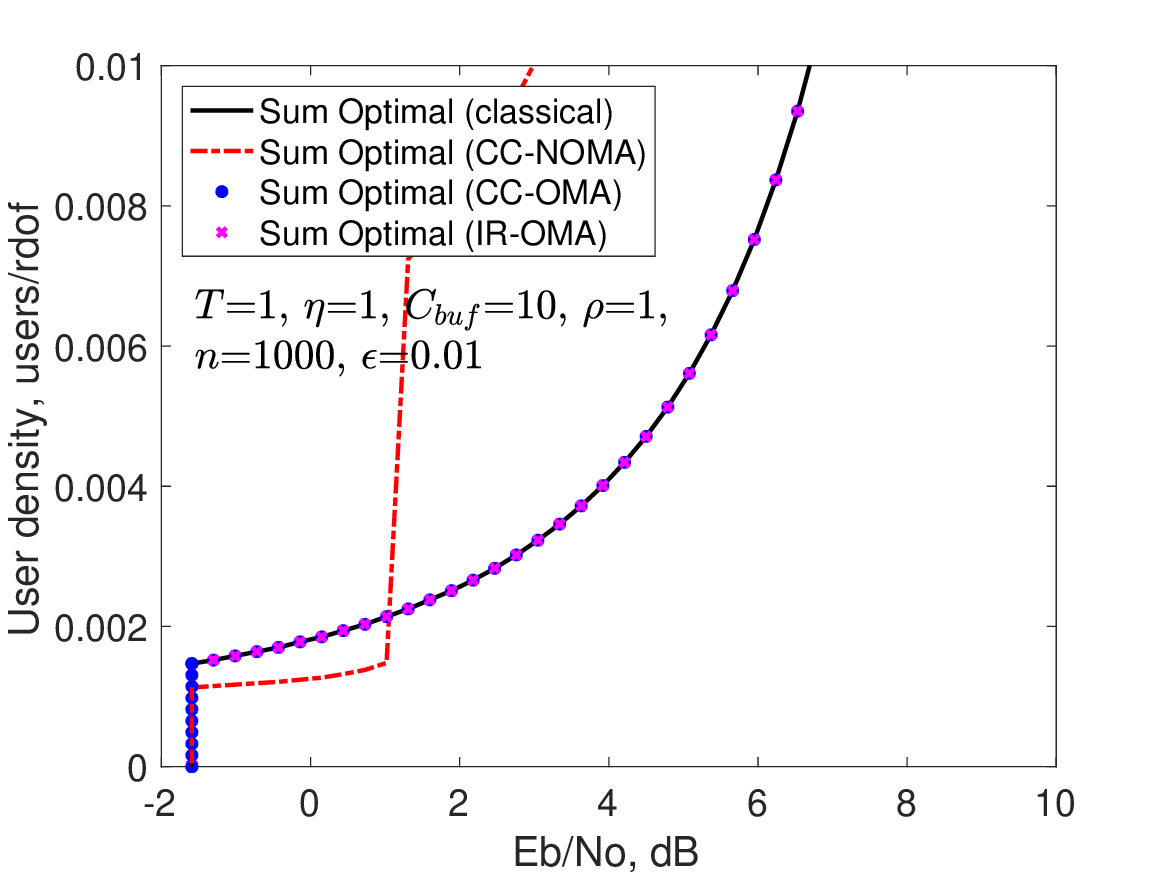}
\includegraphics[width=0.32\textwidth]{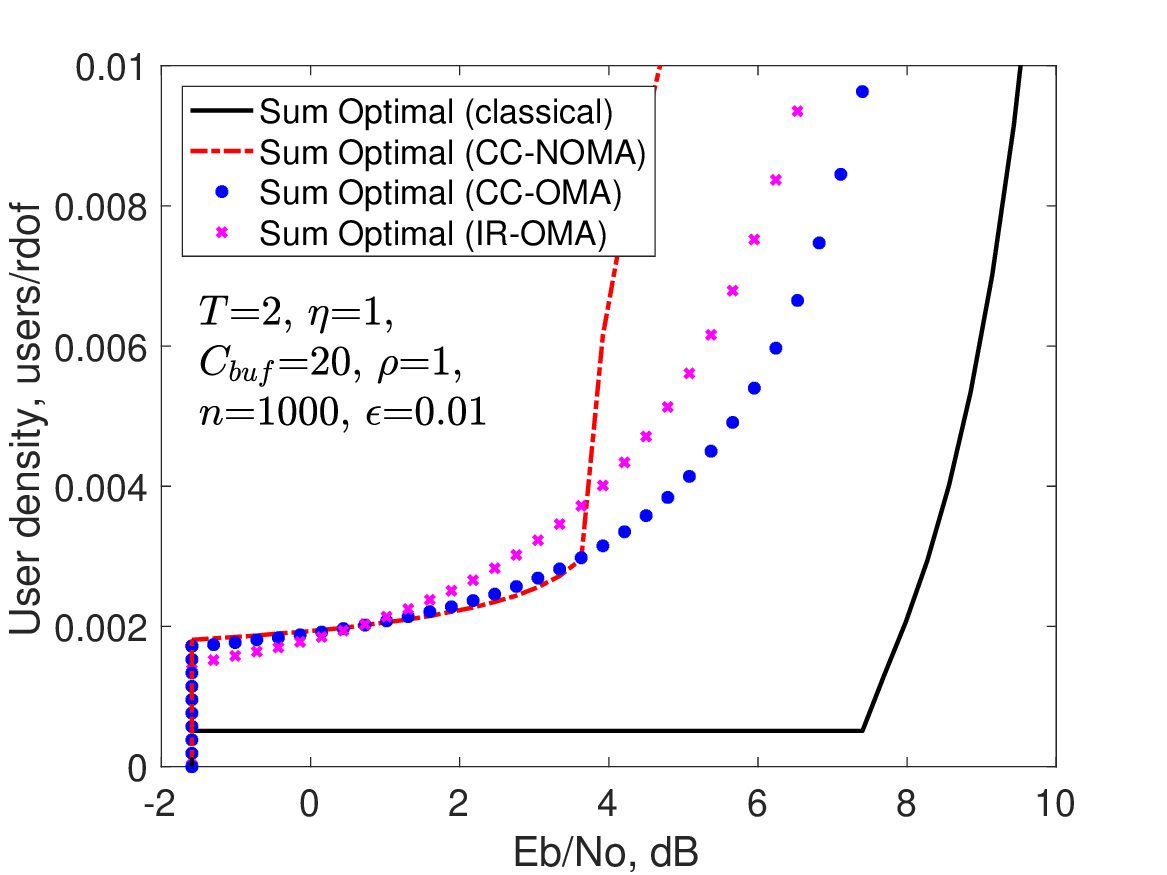}
\includegraphics[width=0.32\textwidth]{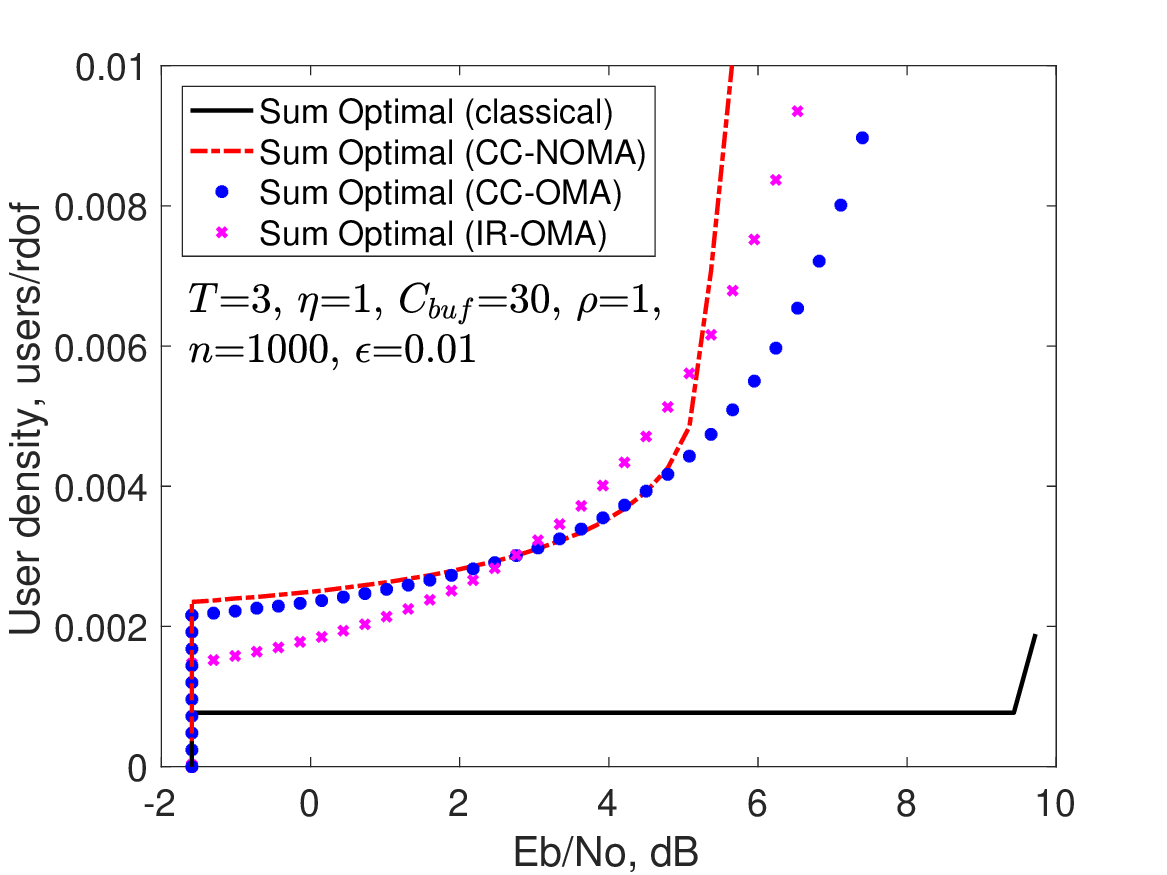}
\end{center}
\caption{\small{(Sum-rate optimal) Scaling of $J/n$ versus $E_b/N_0$ for varying $\rho$ and $\Cb=10T$. (Rows I-II) $\rho=0.1$, and $\rho=1$.}}
\label{fig:SO_UserDensity}
\end{figure}

\begin{figure}[t!]
\begin{center}
\includegraphics[width=0.32\textwidth]{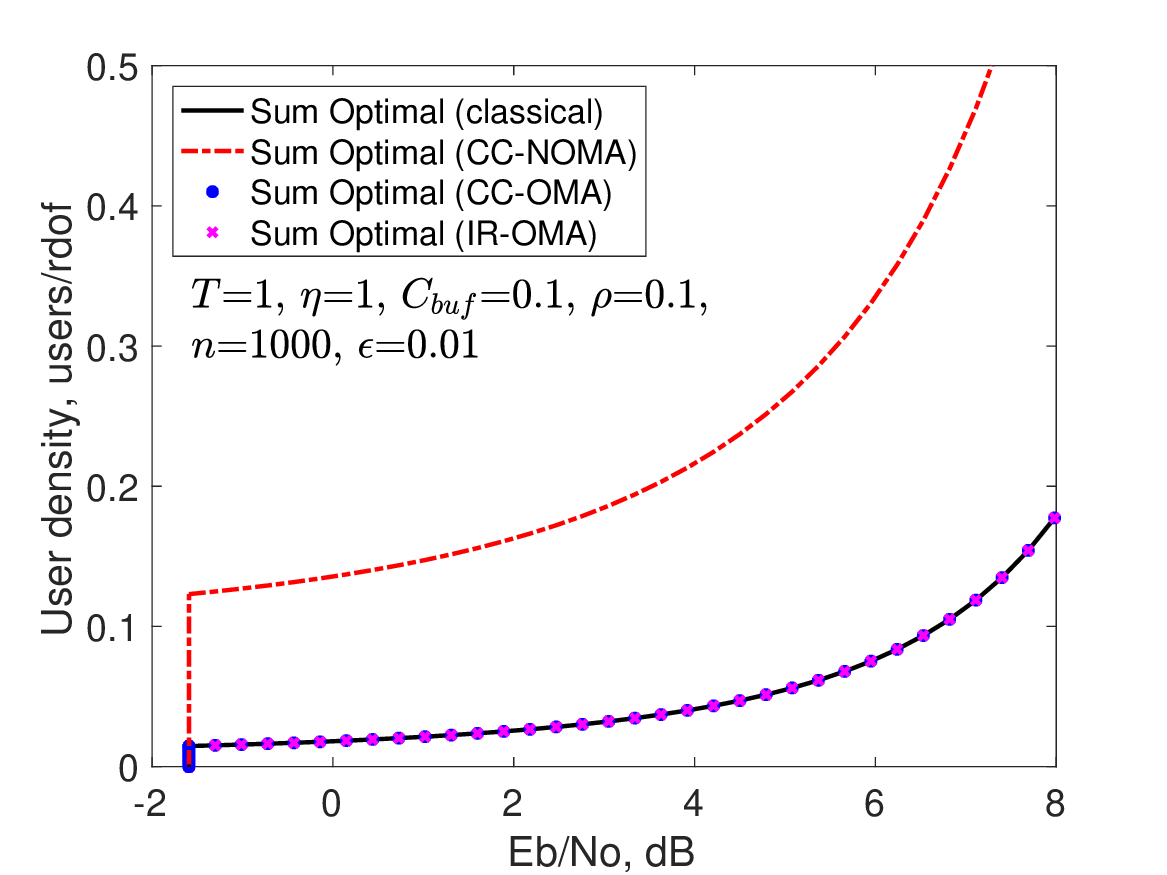}
\includegraphics[width=0.32\textwidth]{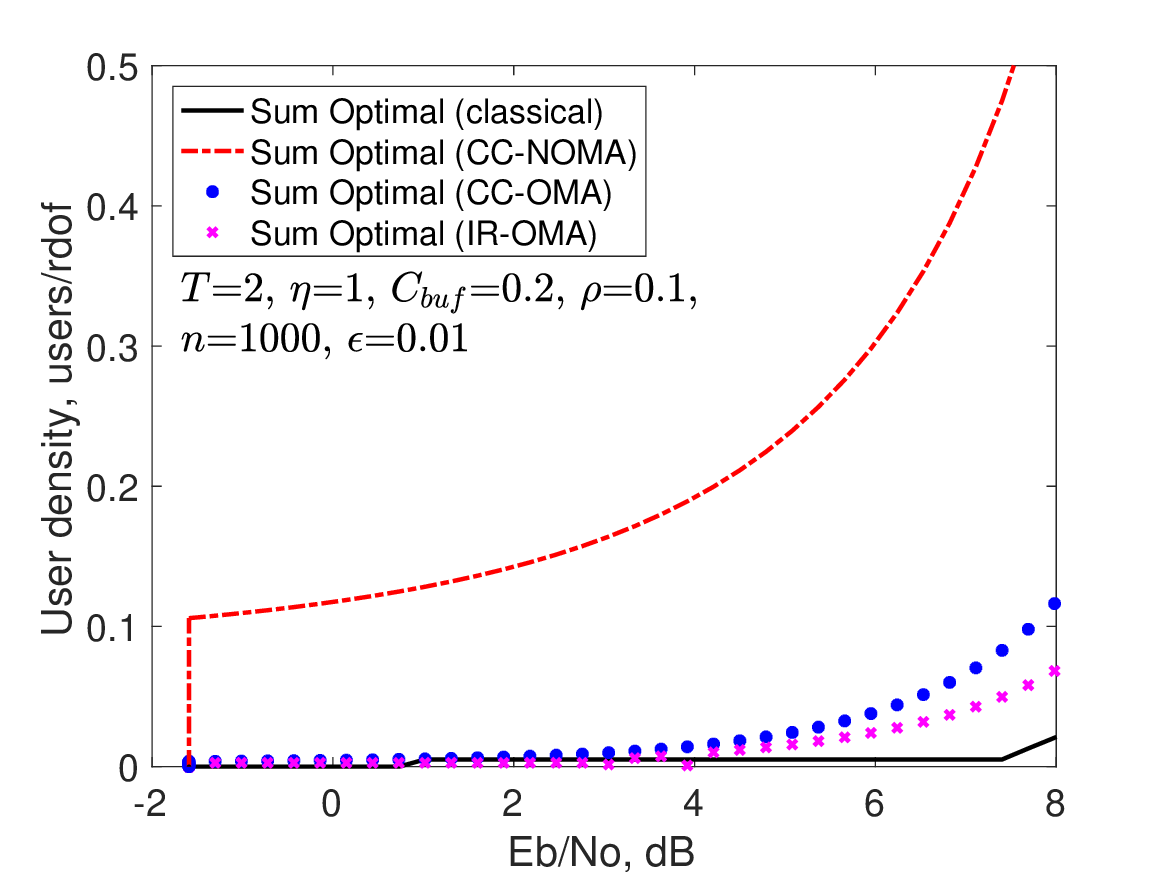}
\includegraphics[width=0.32\textwidth]{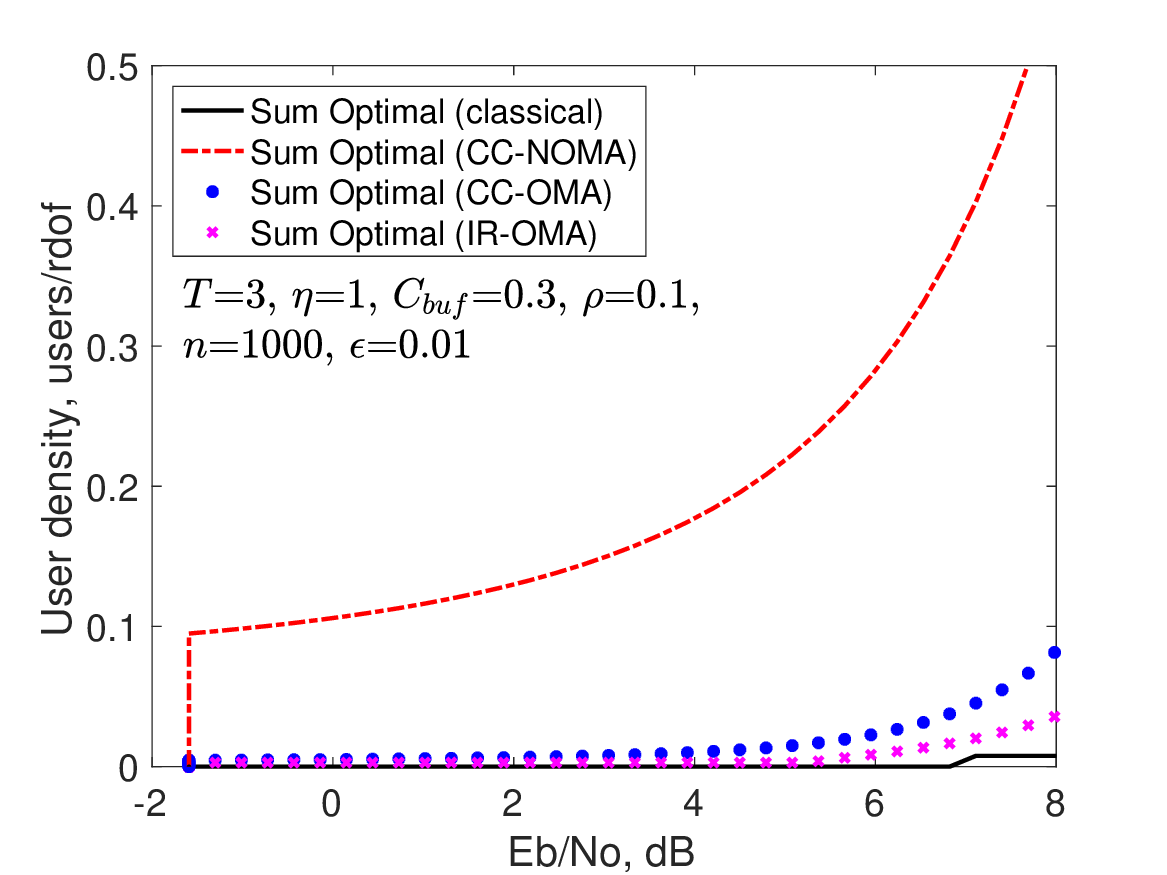}
\includegraphics[width=0.32\textwidth]{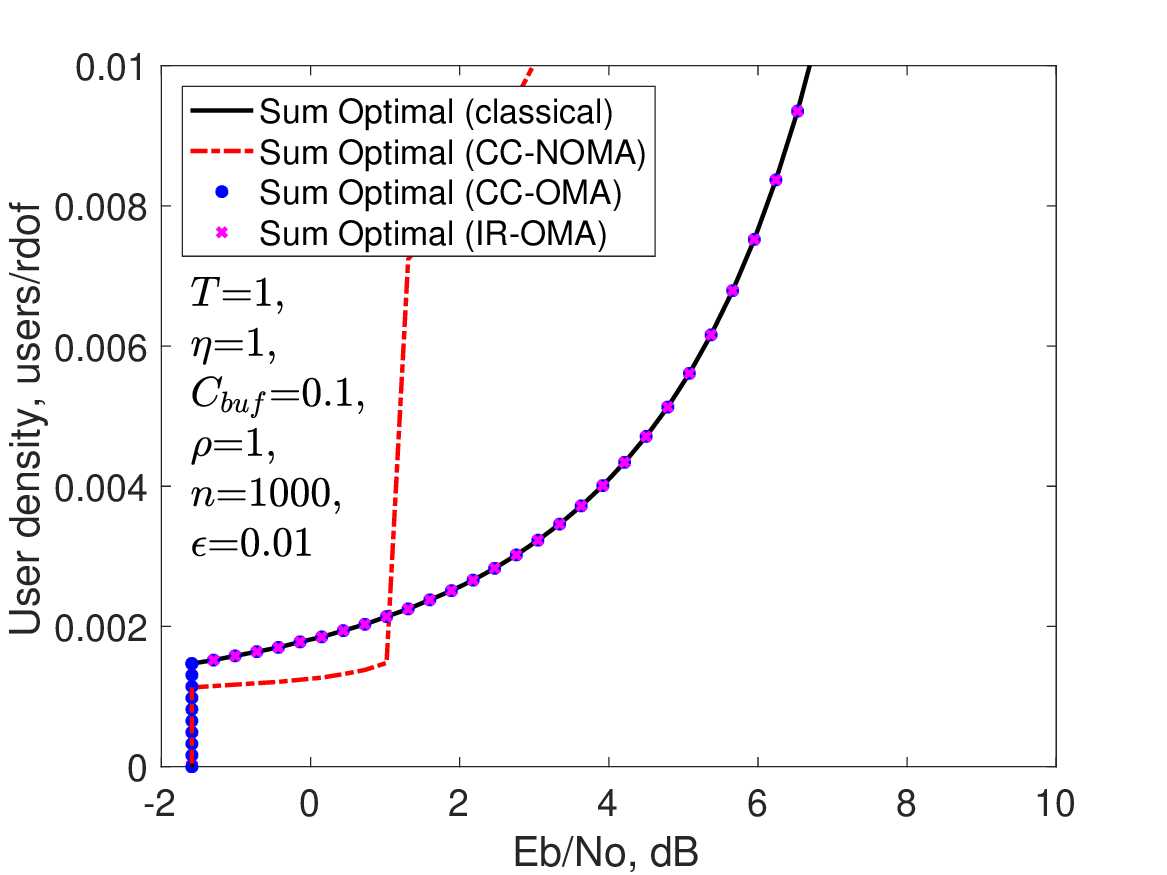}
\includegraphics[width=0.32\textwidth]{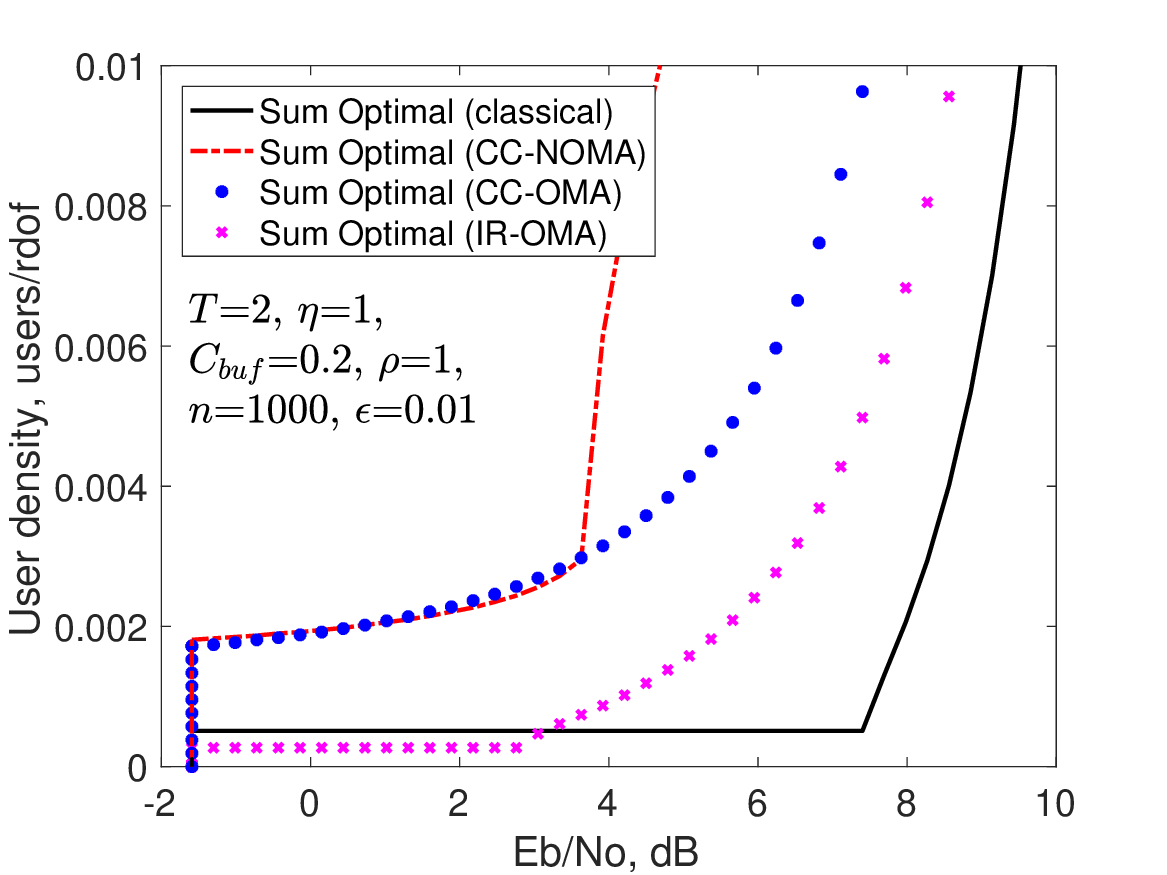}
\includegraphics[width=0.32\textwidth]{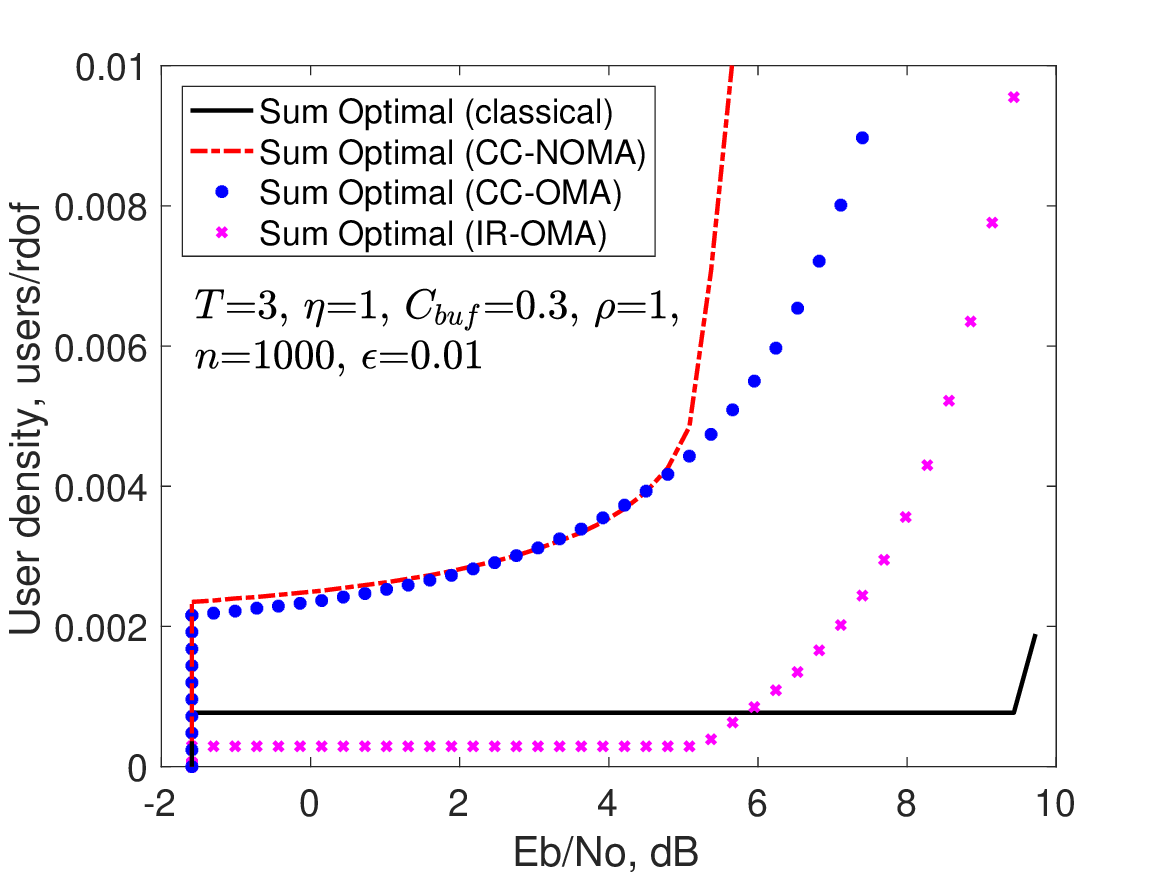}
\end{center}
\caption{\small{(Sum-rate optimal) Scaling of $J/n$ versus $E_b/N_0$ and $\Cb=0.1T$ for various $T$. (Row I) $\rho=0.1$, (Row II) $\rho=1$.}}
\label{fig:SO_UserDensity_lowC}
\end{figure}

\begin{figure}[t!]
\begin{center}
\includegraphics[width=0.32\textwidth]{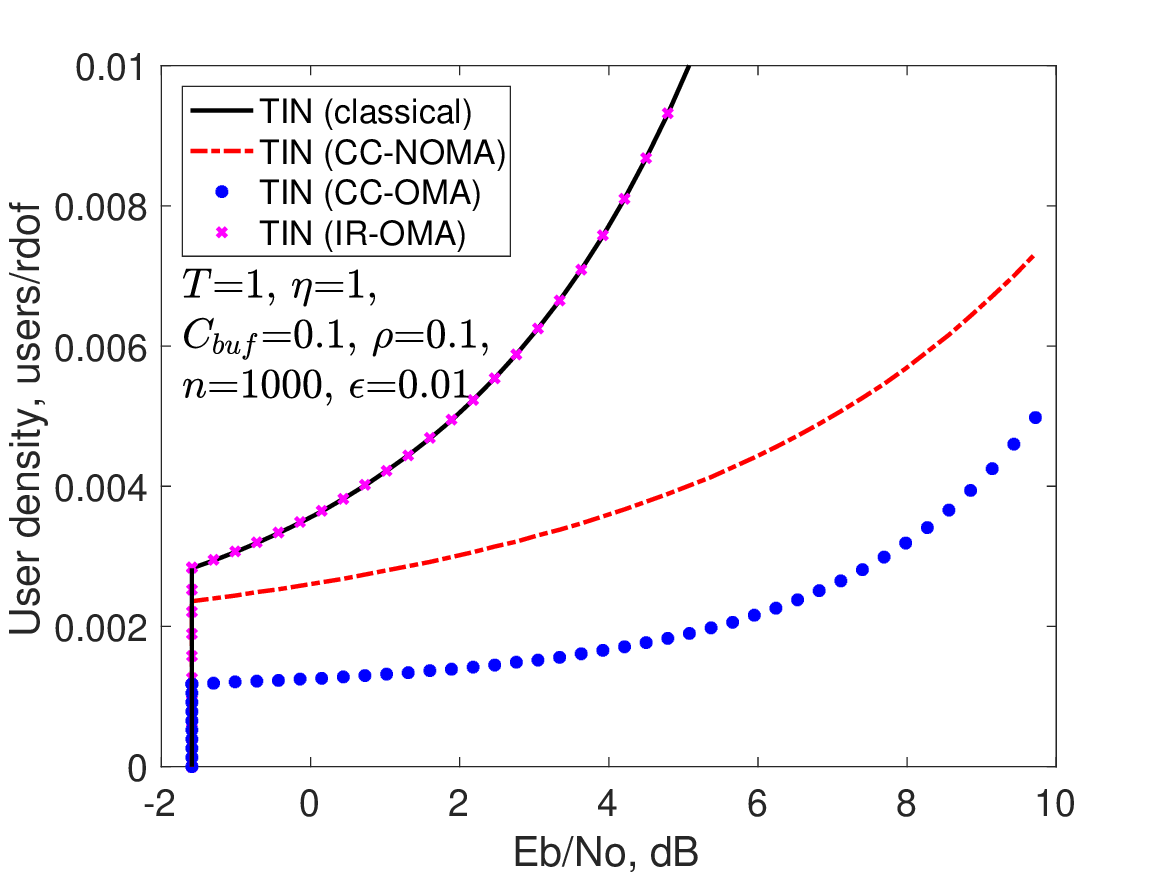}
\includegraphics[width=0.32\textwidth]{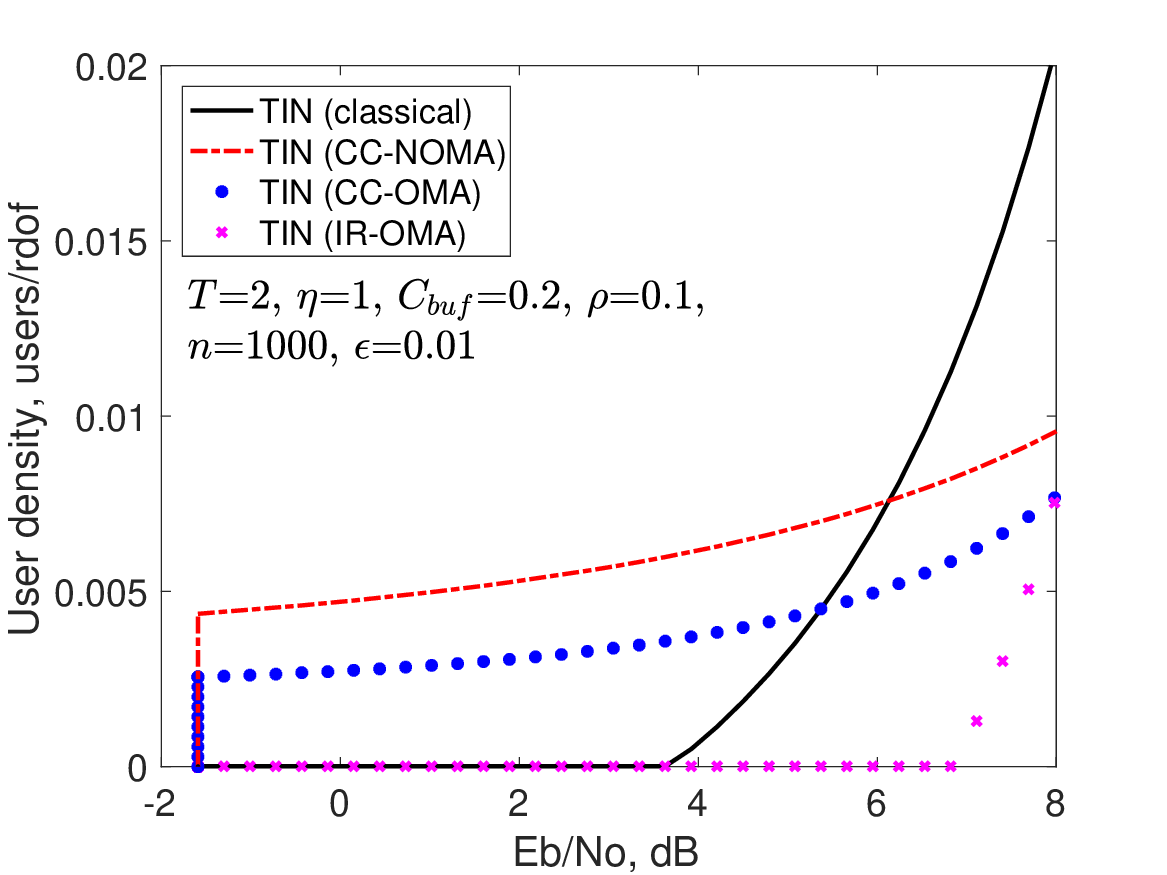}
\includegraphics[width=0.32\textwidth]{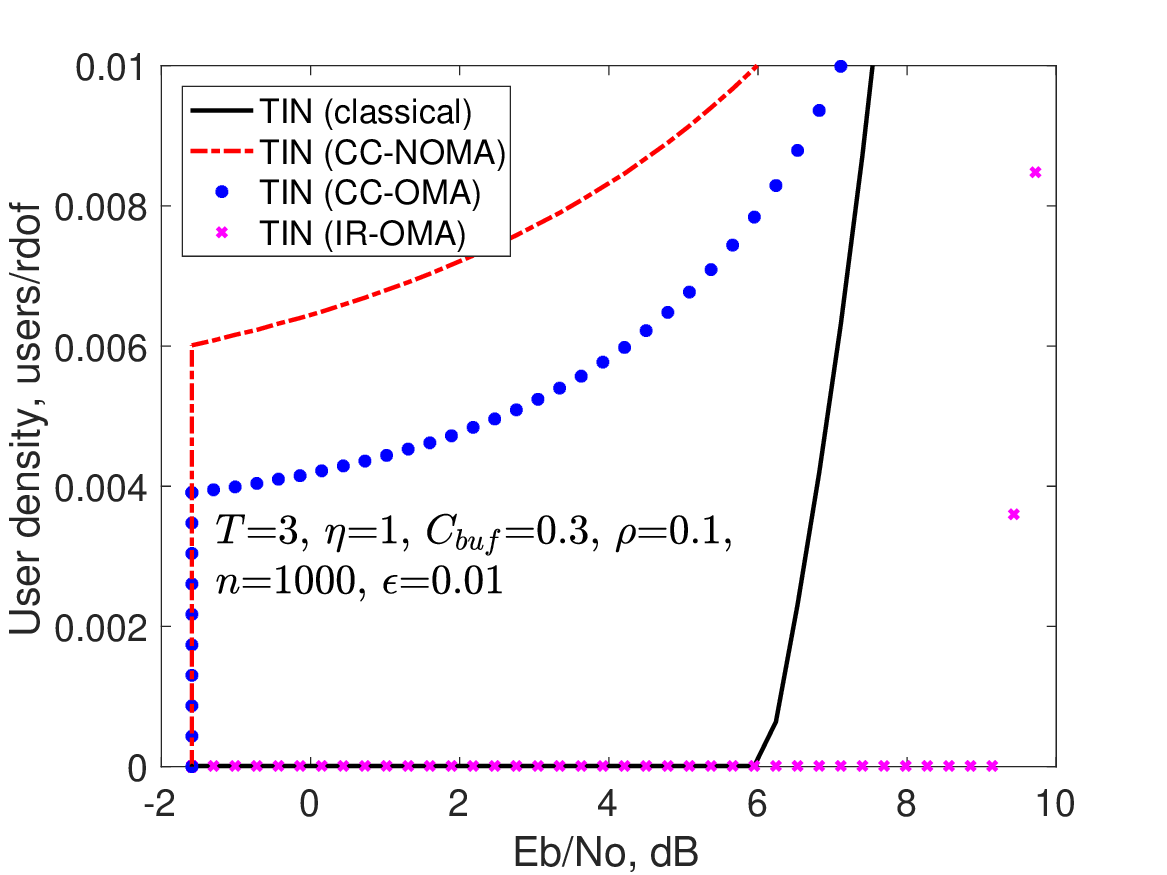}
%%%
\includegraphics[width=0.32\textwidth]{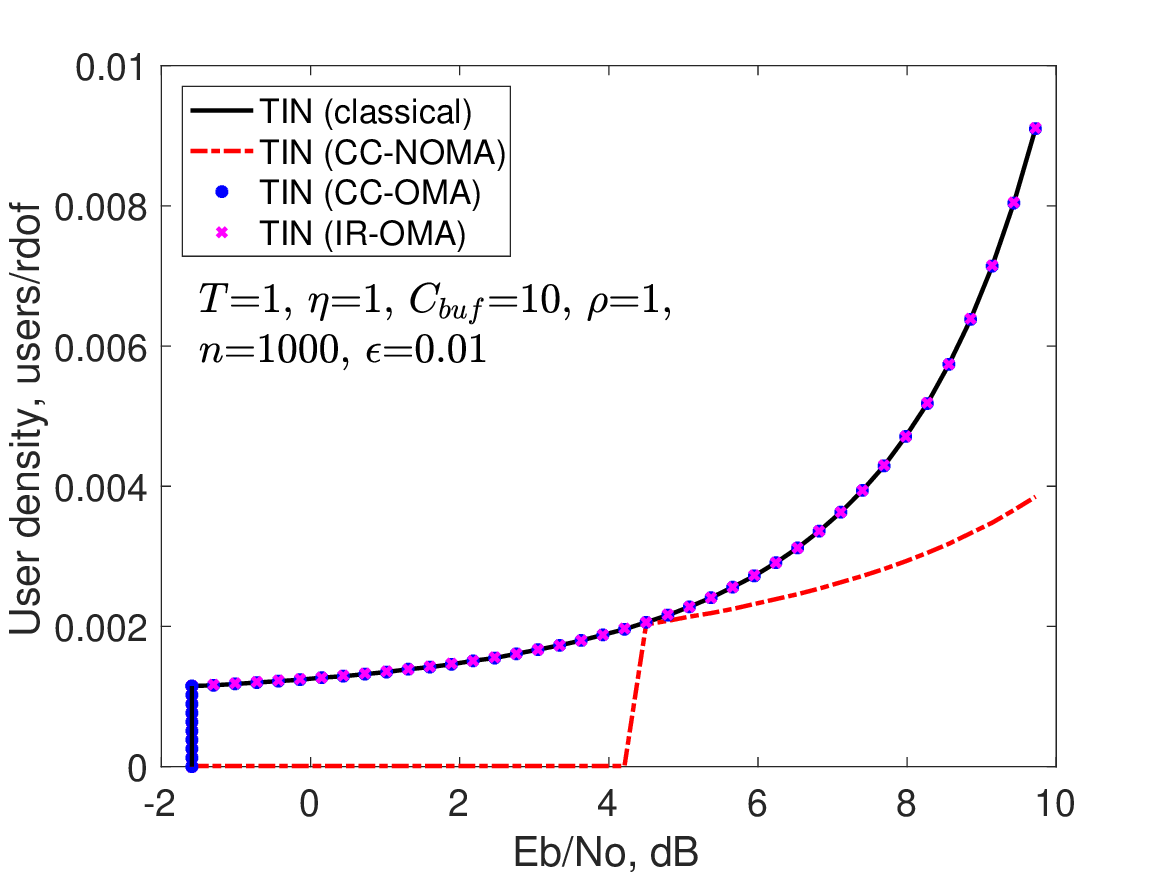}
\includegraphics[width=0.32\textwidth]{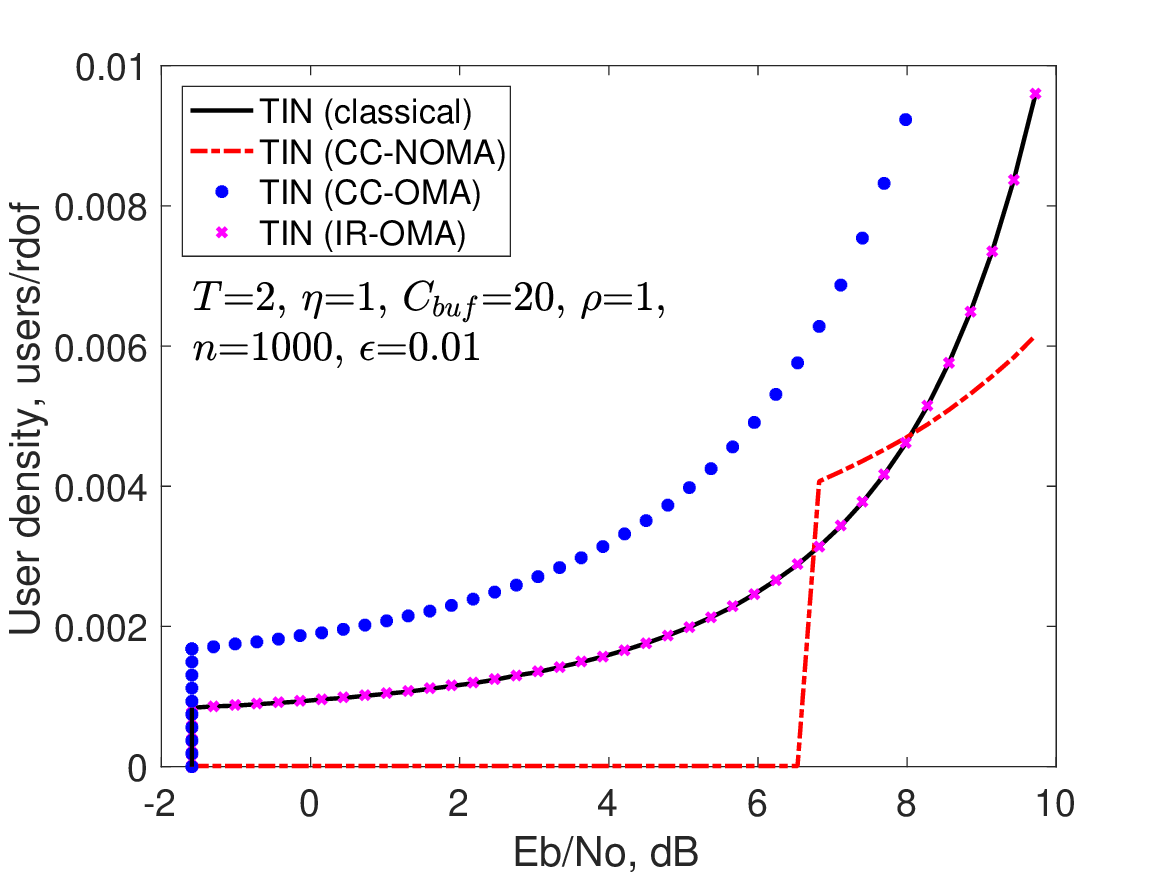}
\includegraphics[width=0.32\textwidth]{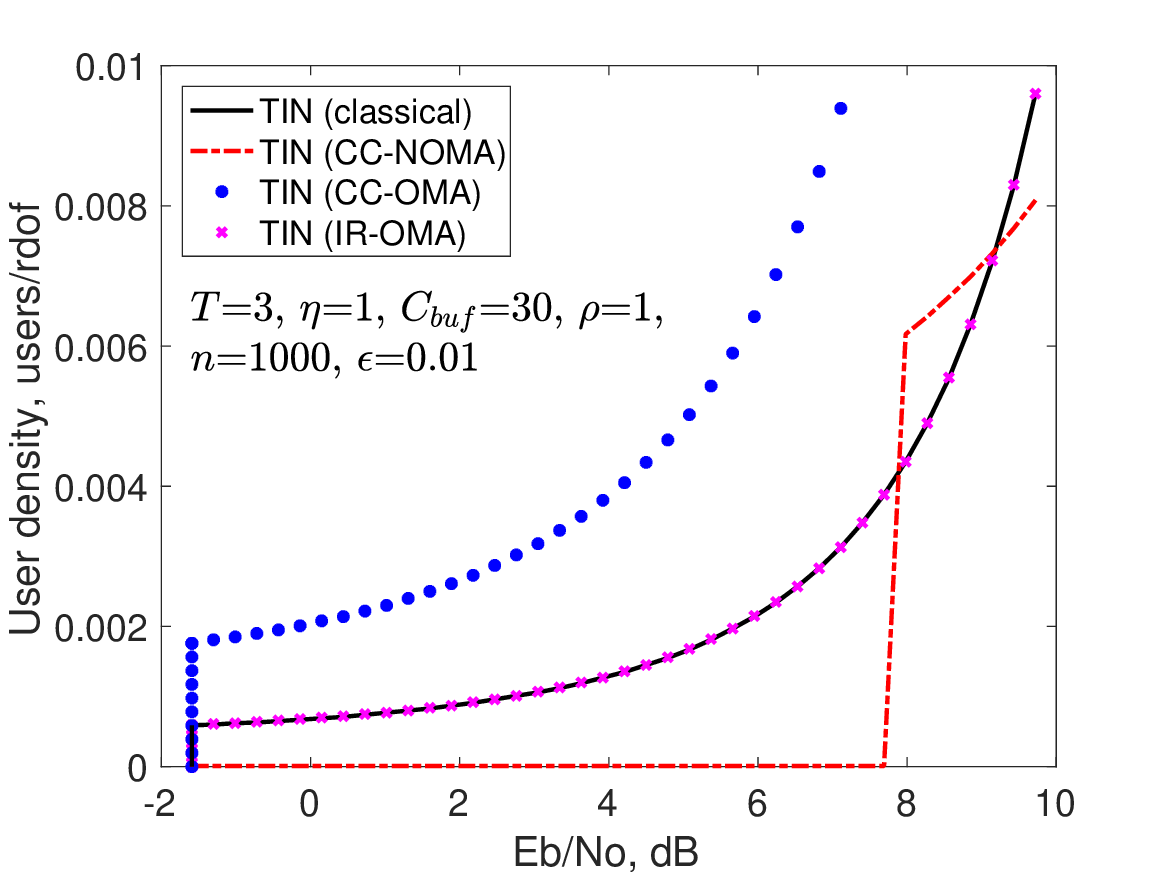}
\end{center}
\caption{\small{(TIN) $J/n$ versus $E_b/N_0$ for varying $\rho$ and $\Cb$. (Row I) $\rho=0.1$, $\Cb=0.1T$, (Row II) $\rho=1$, $\Cb=10T$.  
}}
\label{fig:TIN_UserDensity}
\end{figure}

In this section, exploiting our findings in Sections \ref{CommsModel} and \ref{NOMA_retransmission_no_fading}, we first study the SE (bits/rdof) versus the $E_b/N_0$ (dB) tradeoff (see Figures \ref{fig:SE_EbN0_eta1_J10} and \ref{fig:SE_EbN0_eta_J_varying}) for the different HARQ-based retransmission combining models for the sum-rate optimal and the TIN schemes detailed in Section \ref{NOMA_retransmission_no_fading}, as function of the design parameters $T$, $\eta$, $\Cb$, and $J$. Our numerical results (in Figures \ref{fig:SE_EbN0_eta1_J10} and \ref{fig:SE_EbN0_eta_J_varying}) are for the FBL regime (with $n=1000$, $\epsilon=0.01$), approximating the actual scaling behaviors for the SE models for each strategy. We then focus on the scaling behavior of the user density $J/n$ with respect to $E_b/N_0$ (dB) (see Figures \ref{fig:SO_UserDensity}, \ref{fig:SO_UserDensity_lowC}, and 
\ref{fig:TIN_UserDensity}) as function of $T$, $\eta$, $\Cb$, and $\rho$ under unity channel power gain. For various regimes of interest, we indicate the set of chosen parameters of the sum-rate optimal and the TIN schemes in the legend on each plot. Based on the numerical experiments run for different values of $T$, $\eta$, $\Cb$, and $J$ (for fixed values of $n$, $J$, and $\epsilon$), we next present our observations (see \Cref{fig:SE_EbN0_eta1_J10,fig:SE_EbN0_eta_J_varying,fig:SO_UserDensity,fig:SO_UserDensity_lowC,fig:TIN_UserDensity}).

\paragraph{Number of retransmissions $T$} From Cor. \ref{EbN0_no_multiplexing}, we note that both $\clasicalsumoptimal$ and $\clasicaltin$ are constant for fixed $J/T$. 
From (\ref{SE_CC_NOMA_sum_fading}) and Cor. \ref{Sensitivity_EbN0_CC_NOMA_sum}, increasing $T$ causes degradation in $\chasenoma$.  From Prop. \ref{Chase_SINR_nonorthogonal_TIN} and Cor. \ref{Sensitivity_EbN0_CC_NOMA_TIN}, $\chasetin$ improves with $T$. 
From (\ref{SE_OMA_fading}) $\chaseoma$ also decays with $T$, which we can observe from Figures \ref{fig:SE_EbN0_eta1_J10} and \ref{fig:SE_EbN0_eta_J_varying}. 
A competitor strategy in terms of SE is $\chasetinort$, which improves with $T$ and can even outperform $\chaseoma$, which follows from Prop. \ref{Chase_SINR_orthogonal_TIN}. 
By looking at the partial derivatives of $\chaseoma$ in (\ref{SE_OMA_fading}) and $\chasetinort$ in (\ref{SE_CC_OMA_TIN_fading}) with respect to $\rho$, 
at low $E_b/N_0$ values, TIN-based CC-OMA can perform better than sum-rate optimal CC-OMA because the interference term in $\chasetinort$ becomes low whereas $\chaseoma$ decreases in $T$ for large $J$, as can be observed from Figure \ref{fig:SE_EbN0_eta1_J10}. 
From Figures \ref{fig:SE_EbN0_eta1_J10} and \ref{fig:SE_EbN0_eta_J_varying}, at high $E_b/N_0$ values, $\chaseoma$ is better than $\chasetinort$, and $\chasetinort$ is higher than $\IRtinort$ with increasing gains in $T$. 
Overall, these scaling results show that for any given value of $E_b/N_0$, the best performance is attained by $\chasenoma$, in general for small $T$. Furthermore, at low $E_b/N_0$, the gap between $\chasetinort$ and $\chasenoma$ becomes smaller, and hence, the performance of $\chasetinort$ can approach or outperform $\chasenoma$, and $\chasetinort$ also outperforms the other TIN-based approaches because it exploits Chase combining and unlike $\chasetin$ it has lowered interference.

\paragraph{Finite buffer size $\Cb$ at the decoder} The SNR per bit values under TIN for the classical model %via TIN given 
in (\ref{EbN0_no_retransmissions_TIN}) and the IR-OMA model %under TIN %given in 
(Prop. \ref{cor:IR_OMA_TIN}) have a matching fundamental $E_b/N_0$ limit when $\Cb$ is sufficiently large for $\rho=0$ for any given $T$. 
For large $\Cb$, while $\IRtinort$ is higher than $\clasicaltin$, the behavior of $\clasicalsumoptimal$ and $\IRoma$ schemes are similar, and similarly, for $\clasicaltin$ and $\IRtinort$, see e.g., Figure \ref{fig:SE_EbN0_eta1_J10} (Row I). 
As $\Cb$ decays, implying a lower SE,  
the performance of IR-OMA degrades both for the sum-rate optimal and TIN scenarios, as explained in Section \ref{IR_retransmissions}, see e.g., Figure \ref{fig:SE_EbN0_eta1_J10} (Row II). At small $\Cb$, the SNR per bit of IR-OMA under TIN is higher than the classical TIN model. 
Intuitively, for any given $T$, at small $\Cb$, the curves for $\IRtinort$ and $\IRoma$ start to overlap. 
For large $\Cb$, when $\sigma_{q}(t,T)$  
becomes negligible, $\IRtinort$ is approximately the same as $\clasicaltin$, and $\IRoma$ approaches that of the $\clasicalsumoptimal$, as expected. 
The evaluations %numerical results 
indicate that while $\chasenoma$ outperforms the other strategies almost in all regimes, and $\chaseoma$ is less sensitive to $\Cb$,  
at high $\Cb$, $\IRoma$ competes with $\chasenoma$ and $\chaseoma$, yet $\IRtinort$ is only slightly above $\clasicaltin$.

\paragraph{Contrasting SE versus SNR per bit and non-orthogonality of transmissions measured via $\eta$} 
For small $\Cb$, i.e., under high quantization noise, as $T$ increases, we expect the SE of IR-OMA (both the sum-rate optimal and TIN models), where each retransmission successively refines the information, to be a lower bound to CC-NOMA and CC-OMA and classical models for $T>1$. 
On the other hand, for large $\Cb$, $\IRtinort$ can perform superior to $\chasetin$ when interference is high, e.g., if  
$\eta=1$ or $J$ is high, from (\ref{SE_CC_NOMA_TIN_fading}) and (\ref{SNR_per_bit_IR_OMA_TIN}), and $\IRtinort$ can perform similarly to $\clasicaltin$, from the equivalence of (\ref{SE_no_retransmissions_TIN_fading}) to (\ref{SE_IR_OMA_TIN_fading}) as $\Cb\to\infty$. 
For large $\Cb$, $\IRoma$ performs inferior to $\chasenoma$ and, in general, better than $\chaseoma$, and could outperform $\chasenoma$ for small $\eta$, which follows from contrasting (\ref{SE_CC_NOMA_sum_fading}) and (\ref{SE_IR_sum}). 
Decreasing $\eta$ in CC-NOMA reduces  
the interference and improves $\chasetin$ via combining retransmissions, as illustrated in Figure \ref{fig:SE_EbN0_eta_J_varying}. On the other hand, $\chasenoma$ degrades.
For larger $T$ and  
$\Cb$, which requires a higher $\rho$ as could be inferred from (\ref{SE_OMA_fading}) and (\ref{SE_IR_sum}), $\IRoma$ could be lower than $\chaseoma$ (and similarly for $\IRtinort$ versus $\chasetinort$), and %the performance of 
$\chasetinort$ can be superior to $\clasicaltin$. 
In general, for the sum-rate optimal models, in the bandwidth-limited regime (high SNR), the SE is less sensitive to the changes in $\rho$ versus the power-limited (low SNR) region \cite[Ch. 5]{tse2005fundamentals}.

\paragraph{Scaling of the SE with $J$} Increasing $J$ moves the curve of $\chasenoma$ to the left (see Figure~\ref{fig:SE_EbN0_eta_J_varying}), which coincides with the discussion after Cor.~\ref{Sensitivity_EbN0_CC_NOMA_sum}. Similarly, the SEs of the other sum-rate optimal models also become steeper. On the other hand, for the TIN models, the SE drops in $J$, e.g.,  
with unitary channel gains, from (\ref{SE_CC_NOMA_TIN_fading}), $\chasetin$ scales as $T/(J\log 2)$ as $J\to\infty$, and $J/T\log_2(1+\rho T)$ as $J\to T$, i.e., no interference. The trend of SE is similar for $\clasicaltin$, $\chasetinort$, and $\IRtinort$.

While retransmissions are inevitable in HARQ-based protocols, they  
generally degrade the performances of SE and $J/n$ versus $E_b/N_0$. 
Our numerical results on $J/n$ versus $E_b/N_0$ for the sum-rate optimal models in Figures \ref{fig:SO_UserDensity} and \ref{fig:SO_UserDensity_lowC} and the TIN models in Figure \ref{fig:TIN_UserDensity} 
at FBLs, provide approximations of the actual scaling behaviors. We next detail them,  
reminding the reader that  
$n=1000$ and $\epsilon=0.01$.

\paragraph{Scaling of user density $J/n$ versus $E_b/N_0$} We investigate this scaling behavior in Figures \ref{fig:SO_UserDensity}-\ref{fig:TIN_UserDensity} 
as a function of $\rho$. A stricter average probability of error requirement at a decoder is equivalent to a high $\rho$ value, which yields a higher $E_b/N_0$ to achieve the same $J/n$.  
As $T$ increases, the supported density $J/n$ drops. As $\rho$ increases, the user density scalings of different models for the sum-rate optimal strategy become more similar under high $\Cb$. This situation is because the growth of $E_b/N_0$ is not much sensitive to $\rho$ at low $\rho$ and the approximate growth rate for the sum-rate optimal models is $\frac{\rho}{\log_2 \rho}$ for high $\rho$, which causes a significant drop in $J/n$.  
With the conventional MFR, the optimal SE for the sum-rate optimal model cannot be accurately captured %when $\rho$ is high 
at high $\rho$ \cite{tse1999linear}. 
However, we might not observe this behavior for the TIN models under high $\Cb$ (see Figure \ref{fig:TIN_UserDensity} (Row II)). For CC-NOMA under TIN, from (\ref{EbN0_TIN}), $E_b/N_0$ roughly grows with $\eta^2(J-T)^2/T$ at low $\rho$, and the scaling is subquadratic at high $\rho$, for CC-OMA from (\ref{EbN0_CC_OMA_TIN}), $E_b/N_0$ grows with $J/T$ at low $\rho$, and the scaling becomes sublinear at high $\rho$. 
For the classical TIN model from (\ref{EbN0_no_retransmissions_TIN}), and similarly for IR-OMA with TIN at high $\Cb$, $E_b/N_0$ grows linearly with $J$ at low $\rho$, and $J/T$ is not much sensitive to $E_b/N_0$ at high $\rho$.  

We next compare different TIN models. At low $\rho$ values, the scalings of $J/n$ versus $E_b/N_0$ for CC-NOMA and CC-OMA improve similarly, whereas the schemes that do not promote combining do not perform well. However, at higher $\rho$ values, the scaling of the CC-NOMA scheme deteriorates due to high interference, whereas CC-OMA performs the best because it combines retransmissions while not being susceptible to interference. The TIN-based CC-NOMA and CC-OMA models improve the user density scaling by increasing $T$ that causes diminishing returns in gains, and scaling for the IR-OMA and the classical models are not robust to retransmissions. 
From SNR per bit of the classical sum-rate optimal model in (\ref{EbN0_no_retransmissions_sum_optimal}), and exploiting the SNR per bit for the other sum-rate optimal models, which are given for CC-NOMA in (\ref{Eb_N0_CC_NOMA_sum}), for CC-OMA in (\ref{EbN0_CC_OMA_sum}), and for IR-OMA in (\ref{EbN0_IR_sum_optimal}), when $T=1$, the classical model, CC-OMA, and IR-OMA behave the same, and CC-NOMA has a lower $E_b/N_0$ than the classical sum-rate optimal approach, CC-OMA, and IR-OMA for a given $J/n$. This behavior can be observed in Figure \ref{fig:SO_UserDensity} for $T=1$. 
%In Figure \ref{fig:SO_UserDensity}, 
For $T=2$ and $T=3$, for the classical sum-rate optimal approach, the effective $J/n$ for a given $E_b/N_0$ decays as a function of $T$, following from (\ref{EbN0_no_retransmissions_sum_optimal}), which is similar for the other sum-rate optimal models, namely CC-NOMA, CC-OMA, and IR-OMA. However, the scaling of these three models is less sensitive than the classical model, indicating %which indicates 
their robustness. From the most to the least sensitive as an increasing function of $T$, the ordering for the user density scalings of the models in the sum-rate optimal regime is classical, CC-OMA, IR-OMA, and CC-NOMA.

We observe from Figure \ref{fig:TIN_UserDensity} for TIN-based models that the different models we considered in this paper perform the best at low SE.   
Increasing $\rho$ decreases the scaling performance of $J/n$ for IR-OMA, CC-NOMA, and CC-OMA. To compensate for the loss of CC-NOMA, even though we can incorporate better coding signatures to enable lower $\eta$, this model still requires a higher minimum SNR per bit versus the other models with a higher sensitivity to $\rho$. 
From Part (ii) of Cor. \ref{Sensitivity_EbN0_CC_NOMA_TIN}, for CC-NOMA with the TIN model, the SNR per bit $E_b/N_0$ limit to ensure a nonzero user density increases with $\rho\eta^2$, and it is easy to notice from Figure \ref{fig:TIN_UserDensity} (Row II) that this limit could indeed be very high (i.e., $>8$ dB). The conventional MFR approach is ideal for the low SINR regime, and with MFR, the characterization for the TIN-based might be suboptimal at high $\rho$ \cite{tse1999linear}. 
As $T$ increases, achieving a superior number of users per rdof and a better scaling for IR-OMA via increasing $\Cb$ is possible. At higher $\Cb$ (or when $\rho\geq 10$), IR-OMA yields a better performance over CC-OMA, where CC-OMA scales better due to the combining of transmissions as given by the SNR per bit in the first step of (\ref{EbN0_CC_OMA_TIN}) than IR-OMA with an SNR per bit in (\ref{EbN0_IR_TIN}) versus vice versa for lower 
$\Cb$ (or when $\rho\leq 1$).

%%%%%%%%%%%%%%%%%%%%%%%%%%%%%%%%%%%%%%%%
\section{Conclusions}
\label{sect:conclusion}

In this paper, we proposed eight HARQ-based \MA models for 5G wireless communication networks: (i) the classical scheme with no retransmissions, and the retransmission-based schemes using different combining techniques at the receiver, namely (ii) CC-NOMA, (iii) CC-OMA, and (iv) IR-OMA, both for the sum-rate optimal and TIN-based strategies. 
For each model, we characterized the tradeoffs for SE versus SNR per bit, and the user density versus SNR per bit, and demonstrated through numerical simulations that retransmissions can improve the scaling behaviors of SE and the user density.  
Our results indicate that sum-rate optimal CC-NOMA provides the best scaling almost in all regimes, and at low SNR per bit, the performance of TIN-based CC-OMA, which outperforms  
the TIN-based classical and IR-OMA approaches 
with increasing $T$ via exploiting CC and providing reduced interference, can attain  
the best performance.
Furthermore, at high $\Cb$, the SE performance of IR-OMA approximates CC-NOMA and CC-OMA under the sum-rate optimal models. 
At low $\rho$ values, the user densities of CC-NOMA and CC-OMA improve similarly, whereas the schemes that do not promote combining do not perform well. 
At high $\rho$ values, as interference is %becomes 
more dominant, the scaling of the CC-NOMA-based scheme deteriorates, whereas CC-OMA performs the best because it combines retransmissions and has %experiences 
less interference. 
The ordering of the $J/n$ versus $E_b/N_0$ performances of the models, from the most to the least sensitive as an increasing function of $T$ -- which degrade in $T$ -- in the sum-rate optimal regime, is classical, CC-OMA, IR-OMA, and CC-NOMA.  
Comparing different TIN models at low $\rho$ values, the scalings of $J/n$ versus $E_b/N_0$ for CC-NOMA and CC-OMA improve similarly, whereas the schemes without combining do not perform well.

Critical future directions include incorporating feedback and optimizing the number of retransmissions $T$ and the number of frequency bins $B$.  
From a resource-allocation perspective, handling the issues of identification of user IDs, asynchrony, and traffic burstiness are of critical importance and left as future work. 
Another direction is the joint design of the physical and network layer aspects. It is crucial to support heterogeneous traffic type requirements on one platform where distinct classes of users are under different SINR requirements. Power allocation for the cell edge versus cell center users could be different to mitigate the interference, and capacity model could be revisited under general power control mechanisms. Furthermore, the MMSE receiver is superior than the conventional MFR over a wide range of SIRs \cite{tse1999linear}, which makes it more suitable under multiple traffic types.

The generalization of the classical capacity models is of primary interest through incorporating path loss, and SINR distribution, and investigating the outage capacity exploiting the fading distribution for the asymptotic (IBL) and FBL models, as well as techniques to achieve optimal performance for the SU and the MU settings, and for joint decoding of users. This will pave the way for understanding the 3-way tradeoff between SE, $E_b/N_0$, and $L$. In addition, the BER performance for the MU NOMA model depends on the modulation and coding scheme. The study of the probability of error (per-user or for all users) achieved for a given user density, payload, and energy, is left as future work.

%%%
\appendix

%$a_{tj}$ denotes the complex amplitude of the product of the values of the transmitted symbol $b_{tj}$ and the channel gain $H_{tj}$ of user $j$ at slot $t$.

We recall that $|a_{tj}|^2=m\sigma_t^2\rho_{tj} |H_{tj}|^2$, which is the product of the values of the transmitted symbol power $|b_{tj}|^2$ of user $j$ and the channel power gain $|H_{tj}|^2$ at slot $t$, and $m\sigma_t^2$ is the total noise power across the number of frequency bins, which is $B$. With unit $|H_{tj}|$, it holds that $|a_{tj}|^2=m\sigma_t^2\rho_{tj}$.

%%%%%%%%%%%%%%%%%%%%%%%%
\subsection{Proof of Proposition \ref{Chase_SINR_nonorthogonal_sumoptimal}}
\label{App:Chase_SINR_nonorthogonal_sumoptimal}

Here we compute $\chasenoma$ as well as $\frac{E_b}{N_0}$ for the sum-rate optimal model, for HARQ with Chase Combining of NOMA transmissions (CC-NOMA). 
We recall (\ref{retransmissions_combining_chase}) and combining $T$ transmissions yields
\begin{align}
\sum\limits_{t=1}^T a^*_{tj} {\bf Y}_t&=\sum\limits_{t=1}^T |a_{tj}|^2 {\bf S}_{j}+\sum\limits_{t=1}^T a^*_{tj}\Big(\sum\limits_{j'\in \mathcal{S}_{t,-j}}a_{tj'}{\bf S}_{j'}+{\bf Z}_t\Big)\ .\nonumber%\\
%&=\sum\limits_{t=1}^T |a_{tj}|^2 {\bf S}_{j}+\sum\limits_{t=1}^T a^*_{tj} \sum\limits_{j'\in \mathcal{S}_{t,-j}}a_{tj'}{\bf S}_{j'}+\sum\limits_{t=1}^Ta^*_{tj}{\bf Z}_t \ .\nonumber
\end{align}

Decoding via matched filter and SUDs, we obtain 
\begin{align}
\Big\langle {\bf S}_{j},\,\sum\limits_{t=1}^T a^*_{tj} {\bf Y}_t\Big\rangle=\sum\limits_{t=1}^T |a_{tj}|^2+\sum\limits_{t=1}^T a^*_{tj} \sum\limits_{j'\in \mathcal{S}_{t,-j}}a_{tj'}\langle {\bf S}_{j},{\bf S}_{j'}\rangle+\sum\limits_{t=1}^Ta^*_{tj}\langle {\bf S}_{j},{\bf Z}_t\rangle\ ,\nonumber
\end{align}
where the total received signal power from user $j\in\mathcal{J}$
is given as
\begin{align}
\label{signalpower}
P_S=\Big(\sum\limits_{t=1}^T |a_{tj}|^2 \Big)^2,
\end{align}
and %the expected value of 
the total noise plus interference power, where interference is treated as noise (TIN), equals
\begin{align}
\label{TIN_totalnoisepower}
P_N=&\mathbb{E}\Big[\Big|\sum\limits_{t=1}^T a^*_{tj} \sum\limits_{j'\in \mathcal{S}_{t,-j}}a_{tj'}\langle {\bf S}_{j},{\bf S}_{j'}\rangle+\sum\limits_{t=1}^Ta^*_{tj}\langle {\bf S}_{j},{\bf Z}_t\rangle\Big|^2\Big]\nonumber\\
&=\Big|\sum\limits_{t=1}^T a^*_{tj} \sum\limits_{j'\in \mathcal{S}_{t,-j}}a_{tj'}\langle {\bf S}_{j},{\bf S}_{j'}\rangle\Big|^2
+\Big(\sum\limits_{t=1}^T a^*_{tj} \sum\limits_{j'\in \mathcal{S}_{t,-j}}a_{tj'}\langle {\bf S}_{j},{\bf S}_{j'}\rangle\Big)a_{tj}\langle {\bf S}_{j},{\bf Z}_t\rangle\nonumber\\
&+\Big(\sum\limits_{t=1}^T a_{tj} \sum\limits_{j'\in \mathcal{S}_{t,-j}}a^*_{tj'}\langle {\bf S}_{j},{\bf S}_{j'}\rangle^*\Big)a^*_{tj}\langle {\bf S}_{j},{\bf Z}_t\rangle^*
+\sum\limits_{t=1}^T|a_{tj}|^2 |\langle {\bf S}_{j},{\bf Z}_t\rangle|^2\nonumber\\
&=\sum\limits_{t=1}^T\sum\limits_{t'=1}^T  a_{tj}a^*_{t'j} \sum\limits_{j'\in \mathcal{S}_{t,-j}}\sum\limits_{k'\in \mathcal{S}_{t',-j}}a^*_{tj'}a_{t'k'}\langle {\bf S}_{j'},{\bf S}_{j}\rangle
\langle {\bf S}_{j},{\bf S}_{k'}\rangle\nonumber\\
&+2\sum\limits_{t=1}^T |a_{tj}|^2 \sum\limits_{j'\in \mathcal{S}_{t,-j}}Real(a_{tj'}\langle {\bf S}_{j},{\bf S}_{j'}\rangle \langle {\bf S}_{j},{\bf Z}_t\rangle)
+\sum\limits_{t=1}^T|a_{tj}|^2 |\langle {\bf S}_{j},{\bf Z}_t\rangle|^2\ .
\end{align}

From (\ref{signalpower}), (\ref{TIN_totalnoisepower}), and using $|\langle{\bf S}_{j},{\bf S}_{j'}\rangle|=\eta$ for $j\neq j'$, and $|\langle {\bf S}_{j},{\bf Z}_t\rangle|=0$, 
$\sigma_t^2=\sigma^2$, and $\rho_{tj}=\rho=
\frac{|a|^2}{m\sigma^2}$, $\forall$  $t\in \mathcal{T}$ and $j\in\mathcal{J}_t$, the total received %power from all $j\in\mathcal{J}$, i.e., 
signal plus interference power from all $j\in\mathcal{J}$ is
\begin{align}
\label{total_received_power_CC_NOMA_sumoptimal}
\Big(\sum\limits_{t=1}^T |a_{tj}|^2 \Big)^2+\eta^2\Big|\sum\limits_{t=1}^T \sum\limits_{j'\in \mathcal{S}_{t,-j}} a_{tj}a^*_{tj'}\Big|^2 
=T^2|a|^4+\eta^2 \Big(\sum\limits_{t=1}^T(J_t-1)|a|^2\Big)^2 \ ,%\nonumber\\
%&=(Tm\sigma^2\rho)^2+\eta^2(J-T)^2(m\sigma^2\rho)^2 \ ,
\end{align}
%$a^*_{tj}\cdot a_{t'j}=0$ for $t'\neq t$ for all $j$, 
where $\sum\limits_{t=1}^T J_t=J$. We can derive $\chasenoma$, via  rescaling (\ref{total_received_power_CC_NOMA_sumoptimal}) %the total received power 
by the noise power $\sum\limits_{t=1}^T |a|^2 |\langle {\bf Z}_{t},{\bf Z}_{t}\rangle|=T(m\sigma^2)^2\rho$ % (excluding the interference power)
, yielding %which yields 
the following effective sum-rate optimal SINR for HARQ with CC-NOMA: %under the sum-rate optimal model: 
\begin{align}
\label{total_received_SINR_CC_NOMA_sum}
{\rm SINR}={\Big((Tm\sigma^2\rho)^2+\eta^2(J-T)^2(m\sigma^2\rho)^2\Big)}\Big/{(T(m\sigma^2)^2\rho)}=
T\rho\Big[1+\eta^2\Big(\frac{J}{T}-1\Big)^2\Big] \ .   
\end{align}

Incorporating (\ref{total_received_SINR_CC_NOMA_sum}) with (\ref{PowerConstraint}), the total capacity of the $J=\sum\limits_{t\in\mathcal{T}} J_t$ user Gaussian MAC is given as
\begin{align}
\sum\limits_{t=1}^T\sum\limits_{j\in\mathcal{J}_t}R_{tj} \leq \frac{1}{2}\log_2\Big(1+\rho T\Big[1+\eta^2\Big(\frac{J}{T}-1\Big)^2\Big]\Big)\,\, bit/rdof \ .\nonumber    
\end{align}

We next incorporate the channel power gains, and order the users from the one with the largest SINR (say user $j$) to the one with the lowest SINR, which is then followed by decoding via SIC. Then, from (\ref{signalpower}), (\ref{TIN_totalnoisepower}), and (\ref{total_received_SINR_CC_NOMA_sum}), effective SINR as a result of $T$ transmissions given as
\begin{align}
{\rm SINR}&={\Big(\Big(\sum\limits_{t=1}^T |a_{tj}|^2 \Big)^2+\eta^2\Big|\sum\limits_{t=1}^T \sum\limits_{j'\in \mathcal{S}_{t,-j}} a_{tj}a^*_{tj'}\Big|^2\Big)}\Big/{\Big(\sum\limits_{t=1}^T|a_{tj}|^2 m\sigma_t^2\Big)}\nonumber\\
&={\Big((m\sigma^2\rho \sum\limits_{t=1}^T|H_{tj}|^2)^2+\eta^2\Big|\sum\limits_{t=1}^T \sum\limits_{j'\in \mathcal{S}_{t,-j}}H_{tj}H^*_{tj'}\Big|^2(m\sigma^2\rho)^2\Big)}\Big/{\Big(\sum\limits_{t=1}^T|H_{tj}|^2(m\sigma^2)^2\rho\Big)}\nonumber \\
&=\rho \Big(\sum\limits_{t=1}^T|H_{tj}|^2\Big)\Big[1+\eta^2\Big(\sum\limits_{t=1}^T|H_{tj}|^2\Big)^{-2}\Big|\sum\limits_{t=1}^T \sum\limits_{j'\in \mathcal{S}_{t,-j}}H_{tj}H^*_{tj'}\Big|^2\Big] \ ,\nonumber 
\end{align}
where different from the constant gain scheme, we have $|a_{tj}|^2=m\sigma_t^2\rho_{tj} |H_{tj}|^2=m\sigma^2\rho |H_{tj}|^2$.

%\chasenoma
By expanding the energy and the SE terms and evaluating their ratio, and using (\ref{EbN0_defn}) we obtain (\ref{Eb_N0_CC_NOMA_sum}).

%%%%%%%%%%%%%%%%%%%%
\subsection{Proof of Proposition \ref{Chase_SINR_nonorthogonal_TIN}}
\label{App:Chase_SINR_nonorthogonal_TIN}
Here we detail how to obtain $\chasetin$ as well as $\frac{E_b}{N_0}$ for the TIN model, for HARQ with CC-NOMA. 
Using (\ref{TIN_totalnoisepower}), the expected value of the total noise plus interference power (by TIN) equals 
\begin{align}
P_N=\sum\limits_{t=1}^T\sum\limits_{t'=1}^T  a_{tj}a^*_{t'j} \sum\limits_{j'\in \mathcal{S}_{t,-j}}\sum\limits_{k'\in \mathcal{S}_{t',-j}}a^*_{tj'}a_{t'k'}\eta^2+\sum\limits_{t=1}^T|a_{tj}|^2 m\sigma_t^2%\nonumber\\
=\eta^2\Big|\sum\limits_{t=1}^T \sum\limits_{j'\in \mathcal{S}_{t,-j}} a_{tj}a^*_{tj'}\Big|^2+\sum\limits_{t=1}^T|a_{tj}|^2 m\sigma_t^2 \ ,\nonumber
\end{align}
using which we will determine $\chasetin$, given by (\ref{SE_CC_NOMA_TIN_fading}).

The total received power from user $j\in\mathcal{J}$ divided by the total noise plus interference power (by TIN) as a result of $T$ transmissions is given as
\begin{align}
\label{SINR_NOMA_retransmission}
{\rm SINR}&={\Big(\sum\limits_{t=1}^T |a_{tj}|^2 \Big)^2}\Big/{\Big(\eta^2\Big|\sum\limits_{t=1}^T \sum\limits_{j'\in \mathcal{S}_{t,-j}} a_{tj}a^*_{tj'}\Big|^2+\sum\limits_{t=1}^T|a_{tj}|^2 m\sigma_t^2\Big)}\\  
\label{noise_index_simplification}
&= {(\sum\limits_{t=1}^T\rho_{tj}m\sigma^2)^2}\Big/{\Big(\eta^2\Big|\sum\limits_{t=1}^T \sum\limits_{j'\in \mathcal{S}_{t,-j}}\sqrt{\rho_{tj}\rho_{tj'}}m\sigma^2\Big|^2+\sum\limits_{t=1}^T\rho_{tj}m\sigma^2 m\sigma^2\Big)} 
\end{align}
\begin{align}
%\label{division_simplification}
%&={\Big(\sum\limits_{t=1}^T\rho_{tj}\Big)^2}\Big/{\Big(\eta^2\Big|\sum\limits_{t=1}^T \sum\limits_{j'\in \mathcal{S}_{t,-j}}\sqrt{\rho_{tj}\rho_{tj'}}\Big|^2+\sum\limits_{t=1}^T\rho_{tj}\Big)} \\
\label{t_index_simplification}
&={(T\rho)^2}\Big/{\Big(\eta^2\Big(\sum\limits_{t=1}^T(J_t-1)\rho\Big)^2+T\rho\Big)}={\rho T^2}\Big/{\Big(T+\rho\eta^2\Big(\sum\limits_{t=1}^T J_t-T\Big)^2\Big)} \ ,
\end{align}
where (\ref{noise_index_simplification}) follows from $\sigma_t=\sigma$, %(\ref{division_simplification}) from dividing both the numerator and the denominator terms by $(m\sigma^2)^2$, 
(\ref{t_index_simplification}) by letting $\rho_{tj}=\rho$ and $\sum\limits_{j'\in \mathcal{S}_{t,-j}} 1_{a_{tj'}\neq 0}=J_t-1$, with $\eta\approx \frac{1}{\sqrt{m}}$  \cite{polyanskiy2018information}. 

To determine $\chasetin$, we incorporate the channel power gains, and use (\ref{SINR_NOMA_retransmission}). From these the total received power from user $j\in\mathcal{J}$ divided by the noise power as a result of $T$ transmissions is 
\begin{align}
{\rm SINR}&={\Big(\sum\limits_{t=1}^T m\sigma^2\rho |H_{tj}|^2 \Big)^2}\Big/{\Big(\eta^2\Big|\sum\limits_{t=1}^T \sum\limits_{j'\in \mathcal{S}_{t,-j}} m\sigma^2\rho H_{tj}H^*_{tj'}\Big|^2+\sum\limits_{t=1}^Tm\sigma^2\rho |H_{tj}|^2 m\sigma_t^2\Big)}\nonumber\\
&={\rho\Big(\sum\limits_{t=1}^T  |H_{tj}|^2 \Big)^2}\Big/{\Big(\sum\limits_{t=1}^T |H_{tj}|^2+\rho\eta^2 \Big|\sum\limits_{t=1}^T \sum\limits_{j'\in \mathcal{S}_{t,-j}}  H_{tj}H^*_{tj'}\Big|^2\Big)} \ . \nonumber
\end{align}

%%%%%%%%%%%%%%%%%%%%
\subsection{Proof of Proposition \ref{Chase_SINR_orthogonal_sumoptimal}}
\label{App:Chase_SINR_orthogonal_sumoptimal}

Here we detail how to obtain $\chaseoma$ as well as $\frac{E_b}{N_0}$ for the sum-rate optimal model, for HARQ with Chase Combining of OMA transmissions, namely CC-OMA. Combining $T$ transmissions results in a total noise plus interference power (by TIN):
\begin{align}
&P_N=\Big(\sum\limits_{t=1}^T a_{tj} \sum\limits_{j'\in \mathcal{S}_{t,-j}}a^*_{tj'}{\bf S}^*_{j'}+\sum\limits_{t=1}^Ta_{tj}{\bf Z}^*_t\Big) \Big(\sum\limits_{t=1}^T a^*_{tj}\sum\limits_{j'\in \mathcal{S}_{t,-j}}a_{tj'}{\bf S}_{j'}+\sum\limits_{t=1}^Ta^*_{tj}{\bf Z}_t\Big)\nonumber\\
&=\sum\limits_{t=1}^T\sum\limits_{t'=1}^T a_{tj}a_{t'j}^* \sum\limits_{j'\in \mathcal{S}_{t,-j}}\sum\limits_{k'\in \mathcal{S}_{t,-j}} a^*_{tj'}a_{t'k'} \langle {\bf S}_{j'},{\bf S}_{k'} \rangle + \sum\limits_{t=1}^T |a_{tj}|^2 m\sigma_t^2 \nonumber\\
&=\sum\limits_{t=1}^T \sum\limits_{j'\in \mathcal{S}_{t,-j}} |a_{tj}|^2 |a_{tj'}|^2 \langle {\bf S}_{j'},{\bf S}_{j'} \rangle %\nonumber\\ 
+\sum\limits_{t=1}^T\sum\limits_{t'\neq t} \sum\limits_{j'\in \mathcal{S}_{t,-j}}\sum\limits_{k'\in \mathcal{S}_{t,-j}} (a^*_{tj} a_{t'j})^* (a^*_{tj'} a_{t'k'})\langle {\bf S}_{j'},{\bf S}_{k'} \rangle +\sum\limits_{t=1}^T |a_{tj}|^2 m\sigma_t^2 \nonumber\\
\label{Expansion_orthogonality_j_jprime}
&=\sum\limits_{t=1}^T \sum\limits_{j'\in \mathcal{S}_{t,-j}} |a_{tj}|^2 |a_{tj'}|^2 + \sum\limits_{t=1}^T |a_{tj}|^2m\sigma_t^2 \ ,
\end{align}
where  (\ref{Expansion_orthogonality_j_jprime}) follows from that $\langle {\bf S}_{j},{\bf S}_{j} \rangle=1$ for all $j\in\mathcal{J}$, and $\langle {\bf S}_{j},{\bf S}_{j'} \rangle=0$ for $j\neq j'$, which yields $(a_{tj'}^*\cdot a_{t'k'}) \langle {\bf S}_{j'},{\bf S}_{k'} \rangle=0$ for $j'\neq k'$ and $t,t'\in\mathcal{T}$. Furthermore, $a^*_{tj}\cdot a_{t'j}=0$ for $t'\neq t$ for all $j\in\mathcal{J}$.

Incorporating the channel gains, and via ordering the users to select the one with the largest SINR (say user $j$) to the one with the lowest SINR, which is then followed by decoding via SIC, yields
\begin{align}
{\rm SINR}&={\Big(\Big(\sum\limits_{t=1}^T \rho|H_{tj}|^2m\sigma_t^2\Big)^2+\Big(\sum\limits_{t=1}^T \sum\limits_{j'\in \mathcal{S}_{t,-j}} \rho H_{tj}m\sigma_t^2 \rho H^*_{tj'} m\sigma_t^2\Big)\Big)}\Big/{\Big(\sum\limits_{t=1}^T \rho|H_{tj}|^2 m\sigma_t^2 m\sigma_t^2 \Big)}\nonumber \\
&=\rho \sum\limits_{t=1}^T |H_{tj}|^2\Big[1+\Big(\sum\limits_{t=1}^T |H_{tj}|^2\Big)^{-2} \Big(\sum\limits_{t=1}^T \sum\limits_{j'\in \mathcal{S}_{t,-j}} H_{tj} H^*_{tj'} \Big)\Big] \ ,\nonumber
\end{align} 
which gives the SE for the sum-rate optimal capacity model, which is given by (\ref{SE_OMA_fading}).

%%%%
\subsection{Proof of Proposition \ref{Chase_SINR_orthogonal_TIN}}
\label{App:Chase_SINR_orthogonal_TIN}
Here we detail how to obtain $\chaseoma$ as well as $\frac{E_b}{N_0}$ for the TIN model, for HARQ with CC-OMA. 
Letting $\rho_{tj}=\frac{\mathbb{E}[{\bf X}_{tj}^{\intercal}{\bf X}_{tj}]}{m\sigma^2}=\frac{|a_{tj}|^2}{m\sigma^2}=\frac{1}{m\sigma^2}\sum\limits_{i=1}^m \mathbb{E}[X_{tji}^2]$, the effective SINR as a result of $T$ transmissions is %given by the following ratio:
\begin{align}
{\rm SINR}={\Big(\sum\limits_{t=1}^T \rho_{tj}m\sigma_t^2\Big)^2}\Big/{\Big(\sum\limits_{t=1}^T \sum\limits_{j'\in \mathcal{S}_{t,-j}} \rho_{tj}m\sigma_t^2 \rho_{tj'}m\sigma_t^2 + \sum\limits_{t=1}^T \rho_{tj}m\sigma_t^2 m\sigma_t^2 \Big)} %= \frac{\Big(\sum\limits_{t=1}^T \rho_{tj}\Big)^2}{\sum\limits_{t=1}^T \sum\limits_{j'\in \mathcal{S}_{t,-j}} \rho_{tj} \rho_{tj'} + \sum\limits_{t=1}^T \rho_{tj} } 
\ .\nonumber
%\\
%{\rm SINR}={\rho\Big(\sum\limits_{t=1}^T  |H_{tj}|^2\Big)^2}\Big/{\Big(\rho\sum\limits_{t=1}^T \sum\limits_{j'\in \mathcal{S}_{t,-j}}  |H_{tj}|^2  |H_{tj'}|^2 + \sum\limits_{t=1}^T |H_{tj}|^2 \Big)}
\end{align}
If $\rho_{tj}=\rho$, then letting $K_t=\sum\limits_{j'\in \mathcal{S}_{t,-j}} 1_{a_{tj'}\neq 0}=J_t-1$, the SU decoder sees an effective SINR, given as 
\begin{align}
\label{Chase_SINR_orthogonal_eqn_TIN}
{\rm SINR}={\rho T^2}\Big/{\Big(\rho\sum\limits_{t=1}^T K_t +T\Big)}={\rho T^2}\Big/{\Big(T+\rho\big(\sum\limits_{t=1}^T J_t-T\big)\Big) }\ .
\end{align}

Reordering the terms in (\ref{Chase_SINR_orthogonal_eqn_TIN}), we can obtain the SE for the sum-rate optimal capacity model in (\ref{SE_CC_OMA_TIN_fading}).
%\begin{align}
%R_{sum} 
%\chaseoma=\frac{1}{2}\log_2\Big(1+\rho T\Big[1+\frac{1}{T}\Big(\frac{J}{T}-1\Big)\Big]\Big) \ .\nonumber
%\end{align} 

Unlike (\ref{t_index_simplification}), where the SINR is obtained via Chase combining of non-orthogonal transmissions over $T$ slots, we evaluate the SINR in (\ref{Chase_SINR_orthogonal_eqn_TIN}) via Chase combining of orthogonal transmissions over $T$ slots.

%%%%%
\subsection{Proof of Proposition \ref{IR_SINR}}
\label{App:IR_SINR}
%Each time slot accommodates the transmission of a packet of length $m$ symbols. 
The buffer capacity is $m\Cb$, where $\Cb$ is the buffer size normalized with the packet lengths. Given $\Cb$, the transmit rate is given by $\frac{m\Cb}{n}=\frac{\Cb}{T}$. We can relate the quantization noise $\sigma_{q}^2(t,T)$ to the number of allocated bits $m\Cb/T$ via the rate-distortion theory \cite{cover2012elements}. Combining the retransmissions, each providing an addendum to the first transmission so that combining $T$ transmissions achieves the desired distortion, we obtain (\ref{SE_IR_sum}). 
Using (\ref{retransmissions_combining_chase}) we infer, at transmission attempt $t$, $t<T$, it holds that 
\begin{align}
\label{buffer_size_sum_rate_IR}
\frac{\Cb}{T}=I({\bf Y}_t;{\bf \hat{Y}}_t)= \frac{B}{2}\log_2\Big(1+\frac{J\rho m\sigma^2/B+m\sigma^2}{\sigma_{q}^2(t,T)/B}\Big),   \quad t<T\ ,
\end{align}
which leads to (\ref{distortion_IR_sum_optimal}). We also have the notational convention $\sigma_{q}^2(T,T-1)=0$. 

Incorporating the channel gains, and via ordering the users to select the one with the largest SINR (say user $j$) followed by SIC yields the following SINR for the sum-rate optimal approach for IR-OMA:
\begin{align}
{\rm SINR}={\rho \sum\limits_{j\in\mathcal{J}_{t}}|H_{tj}|^2/B}\Big/{\Big(1+\sigma_{q}^2(t,T-1)/(Bm\sigma^2)\Big)} \ ,\,\, t<T \ ,\quad {\rm SINR}=\rho \sum\limits_{j\in\mathcal{J}_{T}}|H_{Tj}|^2/B \ , \,\, t=T\ , \nonumber
\end{align} 
where the quantization noise is given as 
\begin{align}
\label{distortion_IR_sum_optimal_fading_app}
\sigma_{q}^2(t,T)=B\big(\sum\limits_{j\in\mathcal{J}_{t}}|H_{tj}|^2\rho/B+1\big)m\sigma^2\cdot \frac{1}{2^{\frac{2\Cb}{TB}}-1} \ ,\, \quad t<T \ , \quad \sigma_{q}^2(T,T)=0 \ .
\end{align}

%%%%%
The SNR per bit of IR-OMA for sum-rate optimal case is given by (\ref{EbN0_IR_sum_optimal}), where (\ref{buffer_size_sum_rate_IR}) and (\ref{distortion_IR_sum_optimal_fading_app}) yield $$\IRoma=\frac{B}{2}\log_2(1+\rho J/B)+\sum\limits_{t=1}^{T-1}\frac{B}{2}\log_2\Big(1+\frac{\rho J/B}{1+\sigma_{q}^2(t,T-1)/(Bm\sigma^2)}\Big) \ .$$ 

In the case of no fading, using (\ref{distortion_IR_sum_optimal}) yields $\sigma^2_q(t,T-1)=B(J\rho/B+1)m\sigma^2 \cdot\frac{1}{2^{\frac{2\Cb}{(T-1)B}}-1}$ for $t<T-1$, and using the notational convention $\sigma^2_q(T,T-1)=0$. 
Note that as $\Cb\to\infty$, the $E_b/N_0$ result ($B=1$) is the same as the SNR per bit for the classical sum-rate optimal model in (\ref{EbN0_no_retransmissions_sum_optimal}). 

%%%%%%
\subsection{Proof of Proposition \ref{SNR_IR_OMA_TIN}}
\label{App:SNR_IR_OMA_TIN}

The SE for IR-OMA with TIN is given as
\begin{align}
\IRtinort%&=\frac{JB}{2T}\log_2\Big(1+\frac{\rho/B}{\rho/B(J/T-1)+1}\Big)\nonumber\\
%&+
=\sum\limits_{t=1}^{T}\frac{JB}{2T}\log_2\Big(1+\frac{\rho/B}{\rho/B(J/T-1)+1+\sigma_{q}^2(t,T-1)/(Bm\sigma^2)}\Big)\,\, bit/rdof/(T\mbox{ slots})\ , \nonumber
\end{align}
%For TIN, total transmit power cannot be $J\sigma^2\rho$ it needs to be divided by $T$. This is indeed the case as there are $J_t$ users and the power is $J_t\sigma^2\rho$.
where the buffer size normalized with respect to the packet lengths, i.e., $\Cb$, more precisely the transmit rate, provided that $\Cb<\frac{BT}{2}\log_2\Big(1+\frac{B(\rho/B+1)m\sigma^2}{(J/T-1)\rho m\sigma^2}\Big)$, satisfies:
\begin{align}
\label{TIN_Cb}
\frac{\Cb}{T}=\frac{B}{2}\log_2\Big(1+\frac{\rho m\sigma^2/B+m\sigma^2}{(J/T-1)\rho m\sigma^2/B+\sigma_{q}^2(t,T)/B}\Big)\ , 
\end{align}
which implies that as $\sigma_{q}^2(t,T)\to 0$ for $\Cb\geq\frac{BT}{2}\log_2\Big(1+\frac{\rho\sigma^2/B+\sigma^2}{(J/T-1)\rho\sigma^2/B}\Big)$, where the size of $\Cb$ could be tuned to the channel capacity (\ref{FBL_capacity}) in the FBL regime. The expression (\ref{TIN_Cb}) leads to (\ref{distortion_IR_TIN}).

%%%
Using (\ref{SE_IR_OMA_TIN_fading}), we can express the SNR per bit of IR-OMA for TIN as (\ref{EbN0_IR_TIN}), where
$$\IRtinort=\frac{JB}{2T}\log_2\Big(1+\frac{\rho/B}{\rho/B(J/T-1)+1}\Big)+\frac{JB(T-1)}{2T}\log_2\Big(1+\frac{\rho/B}{1+\rho/B\zeta_t}\Big)\ ,$$ 
where $\zeta_t=\big(J/T-J/(T-1)+1/(2^{\frac{2\Cb}{(T-1)B}}-1)\big)+B/\rho\cdot 1/(2^{\frac{2\Cb}{(T-1)B}}-1)$ for $t<T$, i.e., $\zeta_t$ is fixed for $t<T$, and $\zeta_T=(J/T-1)$ for $t=T$. 
We note that this step follows from incorporating (\ref{distortion_IR_TIN}) for $\IRtinort$. 
This yields the $E_b/N_0$ result of the proposition given in (\ref{EbN0_IR_TIN}).

We also note that the SE of IR-OMA with TIN $\IRtinort$ {\em incorporating the channel power gains} is given in (\ref{SE_IR_OMA_TIN_fading}) using the same intuition for $\IRtinort$ given at the beginning of this proof. %(\ref{SE_TIN_IR})

%%%%%
\begin{spacing}{1}
\bibliographystyle{IEEEtran}
\bibliography{references}
\end{spacing}

\end{document}